# Improvements in Three-Dimensional Automated Shimming Techniques in High-Resolution Nuclear Magnetic Resonance

A thesis submitted to the University of Manchester

for the degree of Doctor of Philosophy

in the Faculty of Science and Engineering

2004

Vladimir V. Korostelev

Department of Chemistry

# Contents

Figures

## Chapter 2

## Chapter 3

## Chapter 4

## Chapter 5

# Chapter 6

Tables

# Chapter 1

# Chapter 2

# Chapter 5

# Chapter 6

THE UNIVERSITY OF MANCHESTER

ABSTRACT OF THESIS submitted by Vladimir V. Korostelev

for the Degree of **Doctor of Philosophy** and entitled

# Improvements in Three-Dimensional Automated Shimming Techniques for High-Resolution Nuclear Magnetic Resonance Spectroscopy

July 2004

Traditionally the improvement of static magnetic field homogeneity ("shimming") of the magnet in Nuclear Magnetic Resonance (NMR) spectroscopy is performed manually, which has many limitations. However, in recent years a number of automated shimming techniques based on Fourier Imaging Technique have been proposed. Existing 3D automated shimming methods require special, Pulsed Field Gradient (PFG) hardware, which is not available on majority of high-resolution NMR spectrometers. The modified technique, presented in this thesis uses the normal NMR hardware provided with the majority of high-resolution NMR spectrometers.

The 3D shimming technique described was optimised for use with Varian UNITY INOVA spectrometers and successfully tested with both protonated and deuterated solvents. A method for calibrating linear transverse shim field gradients and correcting any non-orthogonality and imbalance of strengths is proposed. The effect of thermal convection on field mapping was observed and is reported here for the first time.

# ABSTRACT

The improvement of static magnetic field homogeneity ("shimming") of the magnet in high-resolution Nuclear Magnetic Resonance (NMR) spectroscopy has proved to be a necessary routine and, in many cases, a difficult task. Traditionally, shimming is performed manually, which has many limitations. It is often tedious, and does not always yield satisfactory results. This is because it is difficult for a user to see from the lock signal normally used in manual shimming how the strength of the magnetic field varies across the sample, and thus which shims need correction. For these reasons other methods of shimming have been sought; one such method, reported by van Zijl in 1994, is based on 3D Fourier imaging (in particular, on a field mapping technique proposed by Maudsley in 1979) and subsequent numerical calculations of the shim corrections. This type of shimming is known as 3D automated shimming.

Originally, 3D automated shimming was described solely for use with special Pulsed Field Gradient (PFG) hardware, which is not typically available on the majority of high-resolution NMR spectrometers. The use of normal hardware, available on all commercial NMR spectrometers, was therefore investigated. 3D shimming with normal hardware uses shim coils and the homospoil facility for the production of transverse and $z$ field gradients respectively. A method for calibrating linear transverse shim field gradients and correcting any non-orthogonality and imbalance of strengths is proposed. The experimental parameters for this calibration method were optimised, and it has been integrated into standard routines for 3D shimming.

The speed of 3D shimming has been optimised by reduction of the minimum number of transverse phase-encoding increments, and the transverse mapping digitization. This was thoroughly tested using both proton and deuterium observation and has proven to be very effective. Since the shim gradients used have long switching and slow recovery times, the timing of the Pulsed Field Gradient Stimulated Echo (PFGSTE) sequence was investigated in detail and optimised for the use with Varian UNITY INOVA spectrometers.

The experimental limitations of the 3D automated shimming were thoroughly investigated. The effect of thermal convection on field mapping was observed and is reported here for the first time. Low-viscosity samples are particularly vulnerable to thermal convection, and hence successful 3D shimming with these requires certain precautions.

No portion of this work referred to in the thesis has been submitted in support of an application for another degree or qualification of this or any other university or other institute of learning.

# Acknowledgements

First, I want to thank my supervisor Professor Gareth A. Morris, for his guidance and help throughout this project.

Funding for this project was provided by Manchester University and Varian Inc.

I want to thank gratefully the people whom I have had the pleasure of working with in the Department of Chemistry.

I must also thank Dr. Frank Heatley for helpful teaching and lectures, and also for giving me the opportunity to teach undegraduate students in the Department of Chemistry during the first year of the project.

I am very thankful to the various people at Varian who made this project possible and, especially to Dr. Paul Bowyer for his cooperation.

My sincerest thanks to my family and friends for their support and patience.

# 1

# Introduction and Literature Review

Spectroscopy is the branch of science concerned with investigation of the physical properties of matter by analysing the energy that it absorbs at characteristic wavelengths of the electromagnetic spectrum. Nuclear Magnetic Resonance (NMR) is the study of the absorption of radiofrequency (RF) irradiation by nuclear spins in matter. NMR may be used for the investigation of many nuclei in all states of matter, liquid, gaseous and solid and has grown nowdays into large interdisciplinary study, with a range spread over such diverse fields as quantum physics, chemistry, medicine and biology.

The history of the ideas and concepts on which modern NMR is based can be traced as far back as the early days of quantum physics, when Pauli had introduced the concept of nuclear mechanical moment, named nuclear spin(1.1). This quantity represents the angular momentum of the nucleus and plays a key role in NMR. By 1936 Rabi and co-workers(1.2) had measured the values of magnetic moments for proton and deuterium, and in 1937 the equilibrium nuclear paramagnetic susceptibility of protons in bulk material was directly detected(1.3). These early works provided the experimental basis for the first attempts to detect NMR, which soon followed in 1942-45(1.4, 1.5). However, these were either unsuccessful(1.4), or unreliable because of imperfections in the apparatus (1.5).

The first condensed phase NMR signals were observed in 1945-46, in experiments undertaken by two groups working independently in Harvard(1.6) and Stanford(1.7) . These experiments also suffered from experimental limitations, but finally were a success when proton NMR signals were detected in solid and liquid samples. The supervisors of the groups,

Purcell and Bloch respectively, shared the Nobel Prize for Physics in 1952 as a sign of recognition for the discovery of NMR in bulk materials.

This thesis is concerned with the development of experimental techniques in liquid-state NMR for correction of the static magnetic field inhomogeneity in NMR magnets. The significance of NMR instrument stability and, particularly, maintenance of high homogeneity of the static magnetic field during experiments has been clear since the first attempts to detect NMR signals(1.5). The rest of this chapter reviews some of the major developments in experimental NMR since its discovery and, especially, the techniques that have led to improvements in static magnetic field homogeneity optimisation, which is especially important in high-resolution NMR.

## 1.1 Fourier Transform NMR Spectroscopy

Generally, in order to observe NMR signals from spins in matter a sample has to be irradiated by an electromagnetic field whose frequency corresponds to the Larmor frequency of the spins. As the NMR spectrum has finite width, in order to observe the whole spectrum the frequency range of electromagnetic field must at least cover the width of the spectrum. In early NMR experiments, this was achieved by sweeping either the frequency of RF field or the magnitude of the static magnetic field over the spectrum width and recording the response of the spins at every frequency(1.8). This technique, called continuous wave (CW) spectroscopy, requires considerable experimental time because, in order to observe NMR signal in such way, every frequency in the spectrum must be irradiated, one at time, until the entire spectrum has been scanned.

CW NMR spectroscopy has been succeeded by a different approach, proposed by R.R.Ernst and W.A. Anderson(1.9), that allows the nuclear spins to be irradiated simultaneously over the full NMR spectral width. This was named Fourier Transform (FT) NMR Spectroscopy after the French scientist J.B. Fourier, who first showed that general mathematical functions can be represented by series of trigonometric functions[1.10]. The implication of this finding for NMR is that a pulsed RF irradiation of some frequency actually includes a variety of frequencies, distributed within some range about this frequency. The application of a pulsed RF field to a sample of matter allows observation of the whole NMR spectrum at once. However, the FT requires complex calculations and therefore called for introduction of computers in the field.

## 1.2 Computerization of NMR Spectroscopy

The first computers combined with NMR spectrometers were used for time averaging of low signal-to-noise ratio signals, and also for the calculation of spectra by a numerical Fourier Transform routine[1.11]. The computational efficiency of the latter was significantly improved by the invention of the Fast Fourier Transform (FFT) algorithm in 1965[1.12]. This advance, together with developments in computer memory and computational power of processors, has stimulated a remarkable increase in the application of data processing techniques in NMR since 1966[1.11].

Early computers in NMR were used mostly for hard-wired calculations, with modest facilities for programming[1.11, 1.13]. However, significant progress has been made since then and nowdays the modern computers in NMR are provided with diverse software, including that for phase correction, apodization, digital filtering, advanced techniques of spectrum calculation, multidimensional FFT and visualization[1.14]. Another consequence of the introduction of computers to NMR was that it assisted the automation of many routine operations, including a limited form of automated shimming (static magnetic field optimisation) by the use of the simplex search algorithm in 1968[1.15].

## 1.3 High-Resolution NMR

NMR was of little use for Chemistry until 1949-50, when it was found that the NMR frequency of a nucleus depends on its chemical enviroment[1.16]. However, NMR signals are intrinsically weak and spectral resolution was also insufficient for many NMR applications, even in the liquid state, at that time. It was known that greater sensitivity and spectral resolution could be achieved by the use of stronger static magnetic fields, stimulating developments in NMR magnet design.

The other requirements for NMR magnets in high-resolution applications include high stability and homogeneity of the static magnetic field[1.6]. In order to control the former, the field/frequency feedback circuit was introduced by R. Varian in 1965[1.17]; this stabilizes the ratio between the magnitude of static magnetic field and the frequency of the RF field.

## 1.4 NMR Magnets

The magnet, generating a static magnetic field over a sample of matter, has been an essential part of NMR since its discovery in 1945. Generally, the NMR magnet may be one of three types: a permanent magnet, an electromagnet, or a superconducting magnet. The two former are simpler and cheaper than the latter, however, the maximum strength of their magnetic fields is limited to 1.4 T and 2.35 T, respectively[1.1, 1.8], because of the magnetic saturation of the iron materials used in their construction. The saturation effect restricts the maximum NMR frequency of such magnets to 100 MHz in the case of proton resonance.

Further improvements in sensitivity and resolution came with the introduction of superconducting solenoids, which can produce static magnetic fields in the range up to 25T[1.18]. Superconducting magnets have exceptional advantages over permanent and electromagnets in high resolution NMR, both in the terms of field strength and in terms of short- and long-time stability.

## 1.5 Superconducting Magnets

The phenomenon of superconductivity was first observed by H.K. Onnes in 1911, when he investigated the behaviour of the temperature-dependent component of metal resistivity at absolute temperatures near 0 K. He found a sharp drop in the resistivities of metals near some critical temperature, and gave to this phenomenon the name superconductivity[1.18]. It was soon established, that the critical temperature, which determines transition from the resistive to the superconductive state depends strongly on magnetic field. It was found that the critical magnetic fields for metallic superconductors are limited to about 2T, comparable to the fields produced by permanent magnets and electromagnets. Subsequently, the possibility of producing stronger magnetic fields by the use of superconductive alloys was theoretically proposed in 1957 by Ginzburg, Landau, Abrikosov and Gorkov[(1.19)]. These alloys, called type 2 superconductors, are used in many fields nowadays, including experimental NMR spectroscopy. For their contribution, Ginzburg and Abrikosov shared the Nobel Prize for Physics in 2003.

The first superconducting magnets for 200 MHz proton NMR, using type 2 superconductors, were built in several groups under the supervision of Weaver in 1964[(1.20)]. Further development came with the introduction of a high-resolution NMR spectrometer with a superconducting magnet operating at 300 MHz for proton NMR in 1970, which allowed spectral resolution better than 0.1 Hz[(1.21)].

## 1.6 Shim Coils

Generally, the homogeneity of the raw static magnetic field of NMR magnets is poorer than 0.1 ppm over a typical sample volume[(1.18)], and this is not sufficient for high-resolution NMR. A variety of different techniques for compensation of the static magnetic field inhomogeneity have been tried, but the most efficient so far was proposed by Golay in 1957[1.22]. He realized that the error in static magnetic field may be compensated by the use of a set of electric coils of specific geometries, which generate compensating fields. These coils, called shim coils (or shims for short) allow the inhomogeneity of the raw static magnetic field of the magnet (the magnetic field produced by the magnet alone, without use of shim coils) to be compensated from 0.1 to about 0.001 ppm. The calculation of the static magnetic field gradients produced by coils of different shapes was carried out in 1961 by Anderson[1.23].

Some of the field error components can be effectively averaged by spinning the sample in the transverse plane of the magnet.This technique, proposed by Bloch[1.24] and demonstrated by Anderson and Arnold[1.25], enables averaging of the static magnetic field in the plane perpendicular to the axis of spinning, which partly reduces the effect of the static magnetic field inhomogeneity experienced by spins. As a result, field homogeneity optimisation was often reduced, in practice, to a one-dimensional problem, where only inhomogeneity along the axis of spinning needs compensation.

## 1.7 Field Gradients

Static field gradients have been used in NMR since its early days. The classic work of Hahn on spin echoes described the relationship between static field inhomogeneity, molecular diffusion and echo intensity. Gabillard had described in 1952[1.26] the effect of static magnetic field gradients on NMR signals. Carr and Purcell introduced a static field gradient coil for the production of gradients and measurement of diffusion[1.27]. Static field gradients were soon augmented by pulsed field gradients, proposed by Anderson et al[1.28]. The scope of applications of field gradients has gradually grown, especially after Tanner and Stejskal proposed the use of pulsed field gradients for the measurement of molecular diffusion in 1965[1.29]. Another seminal work, which described the idea of Magnetic Resonance Imaging was published by Lauterbur in 1973[1.30]. He showed the possibility of obtaining information about the spatial distribution of spins in the sample by the use of linear static field gradients produced by shim coils[1.31].

The use of pulsed field gradients for Fourier Imaging was first proposed by Kumar et al[1.32] in 1975, in collaboration with Ernst, using gradients in two or three dimensions with Fourier Transform NMR. The pulsed field gradients in this work were produced by switching on and off the direct currents passing through linear shim coils.

Many original imperfections in static and pulsed field gradients have been known since their first applications in NMR. The need for improvements gradually led to the introduction of new hardware: improved shim power supply modules, and pulsed field gradient (PFG) modules. The latter development allows the production of reproducible, stable and fast rise- and fall-time field gradient pulses[1.33]. Another important improvement to gradient coils was the introduction of actively shielded gradients[1.34], which shield gradient coils from the influence of eddy currents in metallic parts of the NMR probe and magnet.

## 1.8 Magnetic Resonance Imaging (MRI)

The idea to use static magnetic field gradients for obtaining images by NMR was first proposed by Lauterbur in 1971. After working out the technique, he got the first 2D images of a water sample and described the results and general approach in 1973[1.30]. He introduced a frequency-encoding technique, which uses a static field gradient to encode nuclear spins spatially along the direction of the applied gradient so that the Larmor frequency of the spins becomes a function of their position in the sample. The frequency-encoding gradients are referred to in the terminology developed later, as 'read' gradients for their ability to encode the spatial distribution of spins in the sample into the free induction delay signal.

The idea of using NMR for non-destructive imaging of humans in medicine was patented by Damadian[1.35]. He also observed the remarkable phenomenon that water relaxation times in tumor tissues are longer than those in healthy ones[1.36]. His observation was one of the main impetuses for the introduction of MRI into medicine. Further progress came with the introduction in 1974 of a technique for imaging of specific volumes of a sample by the application of selective radio-frequency pulses in the presence of a field gradient[1.37].

In 1975 Kumar et al[1.32] obtained NMR images using pulsed field gradients. This method exploits three orthogonal linear gradients applied in succession after a 90 degree radio-frequency pulse: x- and y- gradient pulses, used for phase encoding, and then a z 'read' gradient. The FID was sampled during the 'read' gradient and the durations of the x- and y- gradient pulses were successively incremented. The complex data matrix obtained was then processed by 3D Fourier transformation to yield a 3D image of the sample. Application of this technique had brought 3D image of the sample. This experiment became the prototypical imaging pulse sequence, used later with many variations. The basic experiment includes the preparation of the spin system by application of a radio-frequency pulse (or pulses), spin evolution during the phase encoding interval, when x- and y- gradients are applied and successively incremented, and a detection period, when the FID is sampled while the applied 'read' gradient is on.

It soon became apparent that the phase encoding used in this technique leads to distortions in the image, caused by the inhomogeneity of the static magnetic field over the volume of sample[1.38]. This observation led Maudsley et al[1.39] in 1979 to modify the imaging technique to allow 3D phase mapping of the static magnetic field inhomogeneity of magnet. An improvement to the basic technique, which eliminated the distortions caused by field inhomogeneity, was proposed by Edelstein et al[1.40], who incremented the magnitudes of phase encoding gradients instead of their durations.

The obtaining of 3D images is time-consuming and called for faster methods better suited to practical applications. In 1977, Mansfield[1.41] modified the slice selection technique to extract a single slice from a 3D image. This improved method allowed obtaining images from particular parts of a volume, and is now widely used, with modifications, in medical and biological applications of MRI. For their contributions to the development of MRI, Lauterbur and Mansfield shared the Nobel Prize for Medicine in 2003.

## 1.9 Development of Shimming Techniques

The improvement of the static magnetic field homogeneity of the magnet has been an important problem since the discovery of NMR. Early NMR magnets were permanent, with pole pieces produced from high quality steel and polished to optical flatness in order to improve the homogeneity of the static magnetic field. To obtain high field homogeneity, the pole faces needed to be parallel to each other. Thus, small pieces of thin sheet steel, called shims, were placed between the pole faces and the steel core of the magnet for adjustment of their position relative to each other. This method of shimming, later called passive shimming, was not precise, and far from easy for the user.

Passive shimming was replaced by active shimming after the introduction of Golay coils[1.22]; each of these coils, also called shims, is designed to produce a specific shape of weak magnetic field. Multiple shim coils are able together to produce the total shape of static magnetic field required to compensate the residual gradients in the static magnetic field. Shim coils are easier to use than passive shimming, and offer more possibilities for automation of the procedure. The basic technique, also called manual shimming, relies on manual adjustment of each of the shims by the user[1.42].

R.R. Ernst proposed in 1968 the first automated shimming technique[1.15], based on a simplex search algorithm. This approach uses the peak amplitude of the NMR signal to estimate field inhomogeneity without specifying its spatial distribution in the sample volume.

This limitation makes shimming of transverse and high order shims, whose optimisation is important for high-resolution NMR, especially difficult.

A further improvement in automated shimming came in 1975 after the measurement of the first 3D images of samples with 3D pulsed field gradients[(1.32)]. It was soon noticed that the phases of spins, spatially encoded by transverse pulsed gradients, are also influenced by static magnetic field inhomogeneities and can be used for mapping these. Maudsley et al[(1.39)] modified this technique in 1979 for the 3D mapping of static magnetic field inhomogeneity in MRI. This 3D field mapping technique was adapted for shimming with high-resolution NMR samples and coupled with an algorithm for 3D automated shimming by van Zijl et al[(1.43)]. The gradient-recalled echo used in this work relies on the use of phase-encoding gradient pulses with short rise and fall times, requiring the use of triple axis PFG modules, although these are not generally available. The method was later adapted for use with deuterated solvents[(1.44)]. An offspring of 3D field mapping, 1D gradient shimming, is now routinely used commercial spectrometers, and is exploited here as part of the 3D shimming procedure.

A further improvement in 3D automated shimming came with an application of the principle, proposed in reference (1.43), for NMR spectrometers equipped only with standard hardware[(1.45)]. This technique is applicable to samples in either protonated or deuterated solvents, and requires for the production of field gradients only the normal hardware, which includes standard shim coils and the homospoil facility for the 'read' gradient. The latter, of course, could use PFG hardware.

## 1.10 The Work Presented in this Thesis

The thesis is concerned with work aimed at improving 3D automated shimming with the use of normal hardware[(1.45)] in high-resolution, and particularly, liquid state NMR. The improvements include optimisation of the parameters for different experimental conditions, and development of the software and pulse sequence.

As has been shown in this chapter, much has been done over the years to improve shimming procedures, from the introduction of new hardware to novel methods for field mapping and shimming. Nevertheless, despite the efforts of spectroscopists and instrument designers, optimisation of the static magnetic field homogeneity is still a very significant problem in high-resolution NMR.

The contents of this thesis can be divided into three main parts:

Part 1 (Chapters 2, 3) contains a theoretical description of NMR and, in particular, of 3D phase shimming.

Part 2 (Chapters 4, 5) is concerned with NMR instrumentation and the software used for 3D automated shimming.

Chapter 6 describes the results achieved. A discussion of these is presented in Chapter 7, which completes the thesis.

# 2

# Pulsed NMR in the presence of static magnetic field inhomogeneity

The aim of this part of the thesis is to give a mathematical description "to represent as simply, as completely, and as exactly as possible a whole group of experimental laws"[(2.1)], of NMR, particularly as relates to 3D automated shimming. The theory of modern NMR describes the behaviour of nuclear spins by the use of quantum mechanics, statistical physics and classical electrodynamics.

Quantum mechanics is a theory of physical phenomena at a microscopic scale. As NMR actually occurs at a microscopic level and relates to the absorption and emission of electromagnetic energy by a spin, the use of a quantum mechanical description of NMR is appropriate. However, the NMR signals observed in experiments build up from a large number of spins, whose states may differ from each other and therefore, the concept of an ensemble[(2.2)], originally introduced in statistical physics, is also used for description of the spin ensembles in modern NMR. Both statistical physics and quantum mechanics had been found useful for the formulation of a new concept – the quantum density matrix[(2.3)], which is used widely in theory of modern NMR[(2.4)].

Classical electrodynamics is the study of the phenomena associated with charged particles in motion, and is concerned with effects such as magnetism, electromagnetic induction and electromagnetic radiation[(2.5)]. Static and radiofrequency (rf) magnetic fields are used in NMR in order to exchange energy with nuclear spins during an NMR experiment, by the interaction of the electromagnetic field with nuclear magnetic moments.

## 2.1 Magnetic properties of nuclear spins

Atomic nuclei possess mechanical spin angular momentum - spin, for short. The nuclear magnetic moment $\vec{\mu}$ and spin $\vec{I}$ relate as(2.6):

$$\vec{\mu} = \gamma\hbar\vec{I} \tag{2.1}$$

where $\hbar$ is Planck's constant divided by $2\pi$ and $\gamma$ is the magnetogyric ratio, measured in the units $rad\ s^{-1}\ T^{-1}$. In this thesis, a positive sign of magnetogyric ratio is assumed. The magnetic moment unit is the Bohr magneton, denoted $\mu_B$. Magnetogyric ratio is an intrinsic characteristic of a nucleus, defined as the ratio of its magnetic moment and spin angular momentum. The properties of the proton and deuterium (denoted $^1$H and $^2$H, respectively) whose NMR is observed in shimming experiments, are presented in Table 2.1(2.7).

**Table 2.1 The properties of proton and deuterium**

| Nucleus | Spin quantum number | Magnetogyric ratio, $rad\ s^{-1}\ T^{-1}$ | Magnetic moment, $\mu_B$ |
|---|---|---|---|
| $^1$H | $1/2$ | $26.75221280\times10^7$ | 4.837353570 |
| $^2$H | 1 | $4.10662791\times10^7$ | 1.21260077 |

Since magnetic moment is proportional to magnetogyric ratio, according to Eqn.(2.1), the magnetic moment of proton is larger than that of deuterium. Typically either proton or deuterium is present in the solvents used in the great majority of samples for high-resolution NMR, and therefore, either of these can be used for shimming.

## 2.2 Quantum mechanics of a spin ½

In classical physics, the physical states of an object of interest can be defined exactly, to a degree which is mainly limited by experimental factors such as random and systematic errors. The measurement of a physical state in quantum mechanics is different as it includes intrinsic uncertainty, which cannot be influenced by improvements in experimental techniques[(2.8)]. This concept, originally proposed by Heisenberg[(2.9)], is called the Uncertainty Principle, which states that the uncertainties of measurement of energy $\Delta E$ and interval of time $\Delta t$, during which a microscopic particle possesses that energy, relate as:

$$\Delta E \Delta t \geq h \tag{2.2}$$

where $h$ is Planck's constant. Hence, only the probabilities of getting particular results can be obtained in quantum mechanical experiments, fundamentally distinguishing them from the classical ones. The uncertainties in quantum-mechanical measurements stem from the disturbance which measurement itself causes to the measured state at a microscopic scale. In NMR, the life times of spin states do not generally exceed the spin-lattice relaxation time $T_1$, and therefore the half-widths of NMR lines in spectra must be at least of the order of $1/T_1$ [(2.10)].

The uncertainties featured in quantum-mechanical measurements lead to a probability interpretation of phenomena, where the quantum-mechanical states are described by wave functions. In the Dirac notation, used in this thesis, the wave function, denoted $|\Psi\rangle$ is called a ket and its complex conjugate, denoted $\langle\Psi|$, is called a bra. The wave function describes the likelihood of a particle originally in state $|\Psi_1\rangle$, to be in a different state $\langle\Psi_2|$ when measured. This likelihood is denoted $\langle\Psi_2|\Psi_1\rangle$, and called a probability amplitude[(2.11)]. According to quantum mechanics, its square gives the probability of finding the particle in the state $\langle\Psi_2|$ at time of measurement[(2.12)]:

$$P = \left| \langle \Psi_2 | \Psi_1 \rangle \right|^2 \qquad (2.3).$$

In the matrix version of quantum mechanics[2.13, 2.14], kets and bras are represented by vectors and columns, respectively, with their elements indexed by number $n$ and given by[2.13]:

$$|\Phi\rangle = \begin{pmatrix} C_1 \\ C_2 \\ . \\ . \\ C_n \end{pmatrix} \qquad (2.4),$$

$$\langle\Phi| = \begin{pmatrix} C_1{}^* & C_2^* & . & . & C_n{}^* \end{pmatrix} \qquad (2.5).$$

The measured quantum-mechanical state, represented by a ket, can be expanded into a linear combination of its basis states $|\Psi_n\rangle$:

$$|\Phi\rangle = \sum_{n=1}^{N} C_n |\Psi_n\rangle \qquad (2.6)$$

where $|\Psi_n\rangle$ is a set of basis states, which are indexed by integer $n = 1,2,...N$. The basis states are special kets, which are normally orthogonal functions and represent the states in which a quantum-mechanical system can be found when its state is measured. The coefficients of the expansion $C_n$ determine the probabilities of a quantum-mechanical system being in the $n-$th state. The particular basis states can be chosen for description of a quantum mechanical state, like a frame of reference. However, any chosen set of basis states must be normalised and complete, in order to be appropriate for representation of a quantum mechanical state. A basis state is called normalised, when its length (defined as a scalar product of its ket and bra) is unity[2.15]:

$$\langle \Psi_n | \Psi_n \rangle = 1 \qquad (2.7).$$

The condition of orthogonality for different basis states of the same basis set is given by[(2.15)]:

$$\langle \Psi_i | \Psi_j \rangle = 0 \qquad (2.8).$$

When a particle is certainly present in the basis states, the set is called complete and the following condition of completeness[(2.16)] is satisfied:

$$\sum_{n=1}^{N} | \Psi_n \rangle \langle \Psi_n | = 1 \qquad (2.9).$$

This is the sum of the diagonal elements of matrix, given by products of bras and kets, and represents the total probability of finding a particle in the basis states. This condition, according to the Copenhagen Interpretation[(2.17)] of quantum mechanics, ensures a complete description of a quantum system in terms of probability amplitudes.

Another new concept, specific to quantum mechanics, relates to the way of calculating the values of dynamical variables measured in an experiment. In classical physics, the dynamical variables are normally described by functions, as opposed to quantum mechanics, where dynamical variables are represented in terms of expectation values of their operators.

An operator is a symbolic representation of a mathematical operation accomplished on one or more functions. Only linear operators are used in quantum mechanics to represent dynamical variables. The operator $\hat{A}$ is linear when the following properties hold[(2.18)]:

$$\hat{A}\left(|\Psi_1\rangle + |\Psi_2\rangle\right) = \hat{A}|\Psi_1\rangle + \hat{A}|\Psi_2\rangle \qquad (2.10)$$

$$\hat{A}\left(c|\Psi_1\rangle\right) = c\hat{A}|\Psi_1\rangle \qquad (2.11)$$

where $c$ and $|\Psi_1\rangle$, $|\Psi_2\rangle$ are a constant and the kets respectively. When linear operator is applied to a ket or bra it results in a new ket or bra, respectively. This can be written for operator $\hat{A}$ and kets $|\Psi_1\rangle$ and $|\Psi_2\rangle$ in the form:

$$\hat{A}|\Psi_1\rangle = |\Psi_2\rangle \qquad (2.12)$$

where $|\Psi_1\rangle$ and $|\Psi_2\rangle$ are the kets, which represent the original and the new quantum-mechanical states respectively. The linear operator $\hat{A}$ is Hermitian when it is equal to its own adjoint(2.19):

$$\hat{A} = \hat{A}^* \qquad (2.13).$$

In quantum mechanics, only Hermitian operators represent dynamical variables.

In summary, linear operators are used for the quantum-mechanical description of physical interactions and dynamical variables, while quanum-mechanical states are represented by wave functions. A result of the measurement of a dynamical variable is one of the eigenvalues of the operator which represents this dynamical variable(2.20). The eigenvalues of a linear operator $\hat{A}$ can be found solving its eigenvalue equation(2.21):

$$\hat{A}|\Psi_n\rangle = A_n|\Psi_n\rangle \qquad (2.14),$$

which links the known operator $\hat{A}$ and two unknowns – its eigenvalues and their corresponding eigenfunctions. For example, the Hamiltonian operator (denoted $\hat{H}$) represents the total energy of the quantum mechanical system and is used for description of the physical interactions and time evolution of the system. Its eigenvalues are found by solving the time-independent Schrödinger equation(2.22), given by:

$$\hat{H}|\Psi_n\rangle = E_n|\Psi_n\rangle \qquad (2.15).$$

The unknowns in (2.15) are the eigenvalues $E_n$ and the eigenfunctions $\left|\Psi_n\right\rangle$. When a time-independent Hamiltonian is applied to a ket $\left|\Phi\right\rangle$, the latter changes according to the time-dependent Schrödinger equation(2.12):

$$\imath\hbar\frac{\partial\left|\Phi(t)\right\rangle}{\partial t}=\hat{H}\left|\Phi(t)\right\rangle \qquad (2.16).$$

The solution is:

$$\left|\Phi(t)\right\rangle=U(t)\left|\Phi(0)\right\rangle \qquad (2.17),$$

where $\left|\Phi(0)\right\rangle$ is the ket at $t=0$, and

$$U(t)=e^{-\imath\frac{\hat{H}t}{\hbar}} \qquad (2.18)$$

is the propagator, which describes the time evolution of a quantum mechanical system. Propagators are linear operators, and generally can be non-Hermitian.

In the matrix version of quantum mechanics(2.13, 2.14), operators are represented by matrices, whose elements correspond to the different states. For example, the dimensionless spin angular momentum operators $\hat{I}_x$, $\hat{I}_y$, $\hat{I}_z$ for spin 1/2 are represented by the matrices(2.23):

$$I_x=\begin{pmatrix}0 & \frac{1}{2}\\ \frac{1}{2} & 0\end{pmatrix} \qquad I_y=\begin{pmatrix}0 & -\frac{i}{2}\\ \frac{i}{2} & 0\end{pmatrix} \qquad I_z=\begin{pmatrix}\frac{1}{2} & 0\\ 0 & -\frac{1}{2}\end{pmatrix} \qquad (2.19).$$

Any ket of spin $1/2$ can be represented as a linear combination of the basis states $\left|\Psi(+1/2)\right\rangle$ and $\left|\Psi(-1/2)\right\rangle$ (denoted $\left|\alpha\right\rangle$ and $\left|\beta\right\rangle$, respectively) written in the form(2.24):

$$\left|\Phi\right\rangle=C_1\left|\Psi(+1/2)\right\rangle+C_2\left|\Psi(-1/2)\right\rangle \qquad (2.20)$$

where $C_1$ and $C_2$ satisfy the completeness condition: $\left|C_1\right|^2+\left|C_2\right|^2=1$ and the basis states of spin 1/2 are given by(2.24):

$$\left|\Psi(+1/2)\right\rangle = \begin{pmatrix}1\\0\end{pmatrix} \quad \text{and} \quad \left|\Psi(-1/2)\right\rangle = \begin{pmatrix}0\\1\end{pmatrix} \qquad (2.21\text{-}22).$$

Quantum-mechanical states and operators relate to experimental results through observables. These are defined as expectation values of operators of dynamical variables whose values are measured in experiments. In quantum mechanics, "any result of a measurement of a real dynamical variable is one of its eigenvalues. Conversely, every eigenvalue is a possible result of a measurement of the dynamical variable for some state of the system"(2.25). The expectation value of the observable for an operator $\hat{A}$ can be represented in the form(2.26):

$$\left\langle \hat{A} \right\rangle = \sum_j \sum_k C_j C_k \left\langle \Psi_j \right| \hat{A} \left| \Psi_k \right\rangle \qquad (2.23).$$

Generally, a quantum mechanical state can be described by the quantum mechanical density operator, which is given in matrix form by(2.27, 2.28):

$$\rho = \sum_j \sum_k C_j C_k^* \left|\Psi_j\right\rangle\left\langle\Psi_k\right| \qquad (2.24).$$

The diagonal elements of the density matrix represent populations of the corresponding quantum states(2.29). Wave functions and density matrices are both describing quantum-mechanical systems. The density matrix of a single particle provides exactly the same information as its wave function(2.30), and therefore either is equally accurate for description of the single particle. However, the use of the density matrix is especially convenient for the description of ensembles of many spins, as shown in Section 2.5.

The time evolution of the density operator $\hat{\rho}$ is described by the Liouville-von Neumann equation(2.31):

$$\frac{d\hat{\rho}}{dt} = -i\left[\hat{H}, \hat{\rho}\right] \qquad (2.25).$$

When $\hat{H}$ is independent of time, the solution is given by(2.32):

$$\hat{\rho}(t) = e^{-i\hat{H}t}\hat{\rho}(0)e^{i\hat{H}t} \qquad (2.26).$$

The expectation value of dynamical variable for operator $\hat{A}$ can be represented by[2.33]:

$$\langle \hat{A}(t) \rangle = Tr(\hat{A}\hat{\rho}(t)) \qquad (2.27).$$

This description of the time evolution of the density operator, in which operators for dynamical variables remain time independent[2.33] is referred to as 'the Schrödinger representation'.

## 2.3 Static magnetic field and its inhomogeneity

Static magnetic fields are described by a magnetic induction vector $\vec{B}(x, y, z)$, given at position $\vec{r}$ as[2.34]:

$$\vec{B}(\vec{r}) = -\nabla\varphi(\vec{r}) \qquad (2.28),$$

where $\varphi(\vec{r})$ is the magnetic scalar potential, $\nabla\varphi(\vec{r})$ is its gradient, and the vector $\vec{r}$ is specified in a suitable frame of reference. A magnetic moment $\vec{\mu}(O)$, placed at the origin of a reference frame, produces a static magnetic field whose scalar potential is given at point $P(\vec{r})$ by[2.35]:

$$\varphi(\vec{r}) = \frac{\vec{\mu}(O)\cdot\vec{r}}{\left|r^3\right|} \qquad (2.29).$$

The scalar potential of the static magnetic field is found by solving the Laplace equation[2.36]:

$$\nabla^2\varphi(\vec{r}) = 0 \qquad (2.30).$$

The solution in spherical coordinates is given by a sum of spherical harmonics[2.37]:

$$\varphi(x, y, z) = -\sum_{n=1}^{\infty}\sum_{m=0}^{n} r^n P_n^{\ m}(\cos\theta)\left[A_n^{\ m}\cos(m\phi) + B_n^{\ m}\sin(m\phi)\right] \qquad (2.31),$$ where $n$ is order of spherical harmonic function. In NMR, shim coils are designed to produce field distributions, which follow the spherical harmonics of certain orders. These fields are used in 3D shimming, discussed in detail in Chapter 3.

Static magnetic fields whose magnetic induction magnitudes do not vary with the coordinate variables are called homogeneous[(2.37)] (or uniform – a term widely used in the technical literature[(2.38)]). Technically, the improvement of field homogeneity is by compensation of the coordinate variation of the induction vector, which can be achieved by application of appropriate static magnetic field variations of opposite sign.

Generally, NMR magnets are designed to produce static magnetic fields of high homogeneity. In practice, their fields are inhomogeneous due to imperfections in design and manufacturing. Since the magnetic susceptibilities of the sample and NMR probe are different from those of air, these also contribute to the total field inhomogeneity over sample volume. The total inhomogeneity of the magnetic field generated by a magnet may be defined in terms of a ratio $\Delta B/B_0$, where $B_0$ is the nominal value of magnetic induction and $\Delta B$ is the maximum deviation from the nominal value within a given volume. Modern NMR magnets produce static magnetic fields with $\Delta B/B_0 \leq 10^{-5}$ over a sphere of about 20 per cent of the bore diameter of magnet[(2.38)]. This may be improved to about $10^{-9}$, which is sufficient for most NMR experiments[(2.39)] by the use of additional correcting coils[(2.38)].

## 2.4 A spin in a static magnetic field

In quantum mechanics, the Zeeman Hamiltonian describes the interactions of spins and static magnetic fields. For a spin placed in a static magnetic field with induction $B_0$, the Zeeman Hamiltonian is given by[(2.40)]:

$$\hat{H}_z = \omega_{0L}\hat{I}_z \qquad (2.32),$$

where

$$\omega_{0L} = -\gamma B_0 \qquad (2.33)$$

is the Larmor frequency, which is the frequency of spin precession in the static magnetic field.

The electron surroundings induce on nuclei the static shielding fields opposite to the externally applied field $B_0$. Hence, the field $B_0$ is altered by the shielding fields and nuclei experience a net field, which is given by[2.41]:

$$B(nucleus) = B_0(1-\sigma) \qquad (2.34),$$

where inhomogeneity of the field $B_0$ and anisotropy of the shielding are neglected at the moment. Thus, the frequency of spin precession is altered by the value of shielding constant and can be given by chemically shifted Larmor frequency[2.41, 2.42]:

$$\omega_{csL} = -\gamma B_0(1-\sigma) \qquad (2.35).$$

When the effects of the shielding fields and field inhomogeneity are neglected, the energy of a spin placed in a static magnetic field is given by[2.44]:

$$E_m = -\gamma\hbar m B_0 \qquad (2.36).$$

For a spin1/2, the energy $E_0$, defined in absence of the static magnetic field, splits due to the Zeeman interaction into two energies, described as[2.45] (Fig.2.1):

$$E_{|\alpha\rangle} = E_0 - \frac{\gamma\hbar B_0}{2} \qquad (2.37)$$

and

$$E_{|\beta\rangle} = E_0 + \frac{\gamma\hbar B_0}{2} \qquad (2.38),$$

where $|\alpha\rangle$ and $|\beta\rangle$ refer to the eigenstates with $m = +1/2, -1/2$, respectively[2.46].

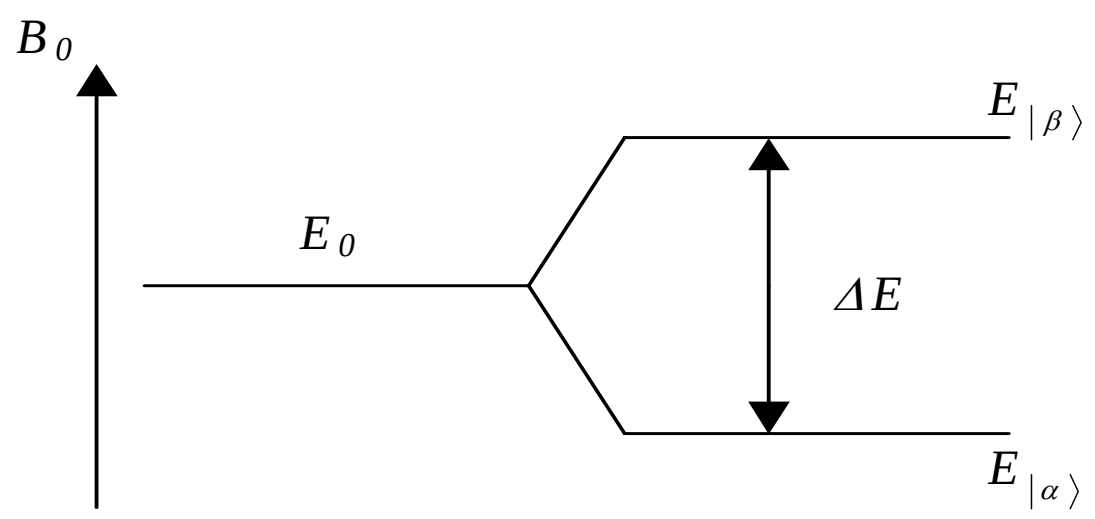

**Fig.2.1** Splitting of $E_0$ energy level in the static magnetic field $B_0$

The energy difference between the $|\beta\rangle$ and $|\alpha\rangle$ states is:

$$\Delta E = E_{|\beta\rangle} - E_{|\alpha\rangle} = \gamma\hbar B_0 \qquad (2.39).$$

The behaviour of a spin in a static magnetic field is described by the time-dependent Schrödinger equation[2.22, 2.47]:

$$\frac{\partial}{\partial t}|\Phi(t)\rangle = -i\hat{H}_z|\Phi(t)\rangle \qquad (2.40).$$

The solution is given by:

$$|\Phi(t)\rangle = e^{-i\omega_{0L}t\hat{I}_z}|\Phi(0)\rangle \qquad (2.41).$$

It can be shown that $\langle \hat{I}_z \rangle$ is time-independent for the states given by Eqns.(2.37) and (2.38) but $\langle \hat{I}_x \rangle$, $\langle \hat{I}_y \rangle$ are time-dependent and can be represented as vectors rotating about $z$ axis[2.48]. This corresponds to Larmor precession of the spin through an angle $\omega_{0L}t$ in a time $t$.

## 2.5 Ensemble of spins 1/2

The density matrix describes state of an ensemble of spins. It is a convenient approach since the individual states of a large number of spins in an ensemble cannot be defined, in practice[2.49, 2.50]. Generally, each of the spins in an ensemble can be in a different state, described by its own ket. The proportion of spins in a state with energy $E_i$ for an ensemble in thermal equilibrium is given by the Boltzmann distribution[2.28]:

$$P_i = \frac{e^{-\frac{E_i}{kT}}}{\sum_m e^{-\frac{E_m}{kT}}} \qquad (2.42).$$

The density matrix for an ensemble of spins is given by[2.28]:

$$\rho = \sum_i P_i \sum_j \sum_k C_j{}^i C_k{}^{*i} \left|\Psi_j\right\rangle\left\langle\Psi_k\right| = \\ = \sum_j \sum_k \overline{C_j C_k{}^*} \left|\Psi_j\right\rangle\left\langle\Psi_k\right| \qquad (2.43).$$

where overbar indicates an averaging over all the spins in the ensemble.Thus, the expectation value of an operator $\hat{A}$ for ensemble is represented by[(2.49)]:

$$\left\langle\left\langle\hat{A}\right\rangle\right\rangle = \sum_j \sum_k \overline{C_j C_k{}^*} \left\langle\Psi_j\right|\hat{A}\left|\Psi_k\right\rangle \qquad (2.44),$$

where the double bracket denotes a double averaging; "the first averaging is due to statistical interpretation inherent in quantum mechanics whereas the second averaging is of a classical nature and would be necessary if we were treating the system classically"[(2.49)]. After the second averaging over $N$ spins in ensemble, the observable of operator $\hat{A}$ can be approximately expressed by[(2.51)]:

$$\left\langle\hat{A}\right\rangle \cong N Tr\left(\hat{\rho}\hat{A}\right) \qquad (2.45),$$

which is accurate within a factor $N^{-1/2}$ for large $N$ . For the samples normally used in NMR this factor is smaller than $10^{-7}$, which is an excellent approximation[(2.51)].

## 2.6 Ensemble of spins 1/2 in a static magnetic field

In the absence of a static magnetic field at thermal equilibrium, the spins are isotropically oriented. However, as will be shown here, when a field is applied (along the $z$ axis, by convention) at thermal equilibrium the spins in ensemble are oriented along the field with a higher probability than against it[(2.52)].

Formally, the state of the spin ensemble at equilibrium in a static magnetic field may be represented by a density operator[(2.53)]:

$$\hat{\rho}_0 = \frac{e^{-\frac{\hat{H}_z}{kT}}}{Z} \quad (2.46),$$

where $Z$ is the partition function, given by:

$$Z = Tr\left( e^{-\frac{\hat{H}_z}{kT}} \right) \quad (2.47).$$

In the high-temperature limit ( $\hat{H}_z << kT$ ) the density operator can be approximately written as[(2.53)]:

$$\hat{\rho}_0 \approx \frac{1 - \hat{H}_z /(kT)}{Tr\left(1 - \hat{H}_z /(kT)\right)} \quad (2.48).$$

It is convenient to express the density operator in the form[(2.53)]:

$$\hat{\rho}_0 \approx a + b\hat{I}_z \quad (2.49),$$

where the constant $a = 1/Z$ , and represents the density operator in the absence of the external static magnetic field; $b\hat{I}_z$ is the part of density operator proportional to the strength of the applied static magnetic field, given by:

$$b = \frac{\gamma\hbar B_0}{ZkT} \quad (2.50).$$

The ratio of the spin state populations for $|\alpha\rangle$ and $|\beta\rangle$ states is described by[(2.54)]:

$$\frac{n_{|\alpha\rangle}}{n_{|\beta\rangle}} = e^{\frac{\Delta E}{kT}} \quad (2.51),$$

where $\Delta E$ is the difference between the energies of the states.

Since this ratio changes as a function of $B_0$, the net magnetization of spin ensemble builds up along the applied static magnetic field[(2.55)]. This is called the equilibrium longitudinal spin magnetization $M_0$, which is given for an ensemble of $N$ spins 1/2 by the Curie-Weiss Law[2.56]:

$$\vec{M}_0 = \frac{N\mu^2 \vec{B}_0}{kT} \qquad (2.52),$$

where $\mu$ is the magnitude of the magnetic moment of a spin and $T$ is the temperature.

## 2.7 Pulsed RF magnetic fields

When a magnetic field oscillating with frequency equal to the Larmor frequency is applied perpendicular to $B_0$, its energy can be absorbed by spins. The absorbed energy changes the polarization of the spins and can transform spin states[2.57]. The Larmor frequencies of nuclei in NMR experiment are in the radiofrequency (rf) range, typically tens or hundreds of MHz (or tens and hundreds millions of precession cycles per second).

RF pulses produce oscillating flux density, which can be mathematically represented by a vector with magnitude $B_1$ which oscillates with angular frequency $\omega_0$ and phase $\phi$[2.58]:

$$\vec{B}_{RF} = 2B_1 \cos(\omega_0 t + \phi)\vec{r}_{rf} \qquad (2.52),$$

where $\vec{r}_{rf}$ is the unit vector, which points in direction of vector $\vec{B}_{RF}$.

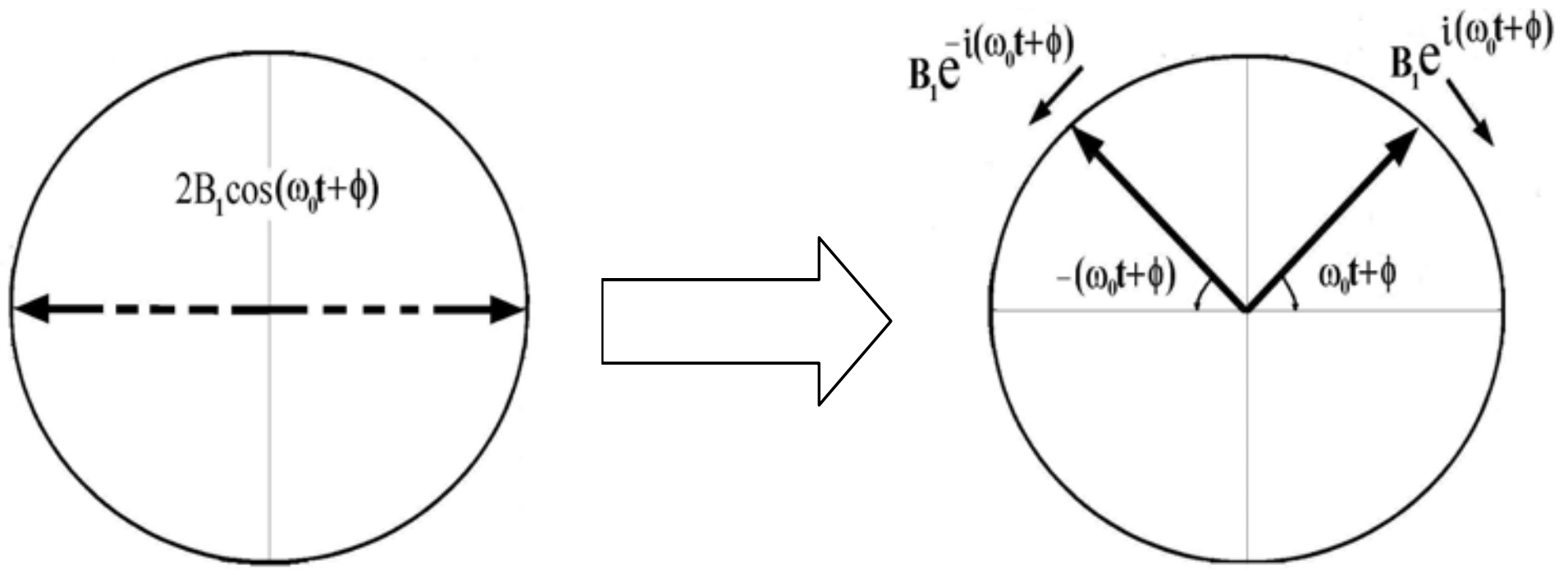

**Fig.2.2** Decomposition of an oscillating rf field into two counter-rotating components.

Since

$$\cos\varphi = \frac{e^{\varphi} + e^{-\varphi}}{2} \qquad (2.54)$$

the rf field can be represented as the sum of two components, rotating with angular frequency $\omega_0$ in opposite directions, represented by (see Fig.2.2):

$$\vec{B}_{RF} = B_1\left(e^{i(\omega_0 t+\phi)} + e^{-i(\omega_o t+\phi)}\right)\vec{r}_{rf} \qquad (2.55).$$

A rf field, applied along the $x$ axis in a Cartesian frame of reference is described by:

$$\vec{B}_{xRF} = 2B_{1x}\cos(\omega_0 t + \phi_x)\vec{x} \qquad (2.56).$$

The Hamiltonian of rf pulse, applied along $x$ axis is given by[2.58]:

$$\hat{H}_{xRF} = -\hat{\mu}_x \vec{B}_{xRF} = -2\gamma\hbar B_{1x}\hat{I}_x\cos(\omega_0 t + \varphi_x) \qquad (2.57).$$

This Hamiltonian can be simplified by transformation into frame of reference rotating about the $z$ axis of the laboratory frame with angular frequency $\omega_0$ in the same sense as the Larmor precession. The Hamiltonian given by (2.57) becomes time independent in this frame and can be described by[2.59]:

$$\hat{H}_{xRF}^{\ rot} = -\gamma\hbar B_{1x}\hat{I}_x \qquad (2.58),$$

where only one rotating field component contributes. The counter-rotating component, rotating in the opposite direction to the Larmor precession of the spin, is $2\omega_0$ away from the Larmor frequency of the nuclei. Thus, the energy of counter-rotating component is not absorbed by the spins and its effect can be neglected[2.60]. The other component interacts with the spins and its energy is absorbed.

## 2.8 Observation of NMR signal

The longitudinal static paramagnetism of nuclear spins is not only very weak, it is also obscured by the much larger electron diamagnetism, which prevents the direct study of nuclear

magnetism[2.61]. Therefore, longitudinal magnetization of spins is transferred into transverse plane (which is perpendicular to the direction of the static magnetic field), where the NMR signal is detected. This is achieved by application of pulsed rf fields, whose strength and duration are adjusted as required in experiment.

When an rf pulse with a frequency close to the Larmor frequency of the spins and of magnitude $B_1$ is applied to the spins for a time $t_p$, the spin magnetization is rotated through an angle[2.62]:

$$\alpha = \gamma B_1 t_p \qquad (2.59).$$

It is convenient to introduce a quantity $\omega_{nut} = |\gamma B_1|$, called the nutation frequency[2.63]. This describes frequency of the rotation due to the rf field. An rf pulse which rotates the spins through 90 degrees is called a 90 degree pulse; its frequency, magnitude and duration depend on the nucleus involved and on the instruments used in particular experiments. Application of a 90 degree pulse transfers maximum longitudinal magnetization into the transverse plane, which allows the observation of NMR signal of maximal amplitude.

## 2.9 NMR signal in presence of static magnetic field inhomogeneity

This section describes the effect of the static magnetic field inhomogeneity on an NMR signal. The analysis is carried out in the rotating frame of reference with a 90 degree pulse applied along the $x$ axis. Relaxation times are assumed to be much longer than the time of spin evolution so that relaxation effects can be neglected.

The strength of the static magnetic field induction can be represented as the sum of a spatially independent part $B_0$, which corresponds to the nominal strength of magnetic induction generated by magnet, and its local deviations as a function of position $\vec{r}$:

$$B(\vec{r}) = B_0 + \delta B(\vec{r}) \qquad (2.60).$$

When 90 degree pulse is applied along $x$ axis, the density matrix evolution is represented in the rotating frame of reference by[2.32]:

$$\rho(\vec{r},t_{90^0}) = e^{-\frac{i}{\hbar}\hat{H}_{xRF}{}^{rot}t_{90^0}} \rho_0(\vec{r}) e^{\frac{i}{\hbar}\hat{H}_{xRF}{}^{rot}t_{90^0}} \qquad (2.61),$$

where $\hat{H}_{xRF}{}^{rot}$ represents the Hamiltonian for the rf pulse and $t_{90}$ is its duration. The rf pulse is assumed to be applied at the Larmor frequency with the static magnetic field $B_0$, so the frame of reference rotates at $\omega_{0L} = -\gamma B_0$. The density matrix for spins ½ at equilibrium in the static magnetic field can be given by:

$$\rho_0(\vec{r}) = a + b(\vec{r})I_z \qquad (2.62).$$

After the 90 degree pulse, the density matrix becomes (see Appendix A for detail):

$$\rho(\vec{r},t_{90^0}) = a + b(\vec{r})I_y \qquad (2.63).$$

The sign of rotation, used here is consistent with that used in reference (2.64). The Zeeman Hamiltonian of the spins, in the inhomogeneous static magnetic field is[2.64]:

$$\hat{H}_z{}' = -\gamma\hbar\hat{I}_z\delta B(\vec{r}) \qquad (2.64),$$

where $\delta B(\vec{r})$ represents the difference between the strength of the static magnetic field at a position $\vec{r}$ and its nominal value $B_0$. Spin evolution after the 90 degree pulse is determined by inhomogeneous magnetic field, which leads to new state of the spin ensemble[2.64]:

$$\hat{\rho}(\vec{r},t_1) = a + b(\vec{r})\left(\hat{I}_y \cos\varepsilon_1(\vec{r},t_1) + \hat{I}_x \sin\varepsilon_1(\vec{r},t_1)\right) \qquad (2.65),$$

where $\varepsilon_1(\vec{r},t_1)$ represents the local phase, given by:

$$\varepsilon_1(\vec{r},t_1) = \gamma\delta B(\vec{r})t_1 \qquad (2.66).$$

This can be expressed through the local Larmor frequency $\Omega(\vec{r})$, which varies with $\vec{r}$ due to the field inhomogeneity. Hence, the local phases can be given by

$$\varepsilon_1(\vec{r},t_1) = \Omega(\vec{r})t_1 \qquad (2.67).$$

The average of the density matrix over all $\Omega(\vec{r})$ in the spin ensemble is represented by:

$$\left\langle \hat{\rho}(\vec{r},t_1)\right\rangle_\Omega = a + b(\vec{r})\left(\hat{I}_y\left\langle \cos\varepsilon_1(\vec{r},t_1)\right\rangle_\Omega + \hat{I}_x\left\langle \sin\varepsilon_1(\vec{r},t_1)\right\rangle_\Omega\right) \quad (2.68)$$

The complex transverse spin magnetization maybe written as:

$$M_{xy}(\vec{r},t_1) = Tr\left(\left\langle \hat{\rho}(t)\right\rangle_\Omega \left(\hat{I}_x + i\hat{I}_y\right)\right) \quad (2.69),$$

thus, the NMR signal can be finally written, after computation, in the form[(2.64)]:

$$M_{xy}(\vec{r},t_1) = M_0\left\langle e^{-i\varepsilon_1(\vec{r},t_1)}\right\rangle_\Omega \quad (2.70),$$

where phase factor $\left\langle e^{-i\varepsilon_1(\vec{r},t_1)}\right\rangle_\Omega$ represents the damping of transverse magnetization due to spin dephasing caused by the static magnetic field inhomogeneity. The longitudinal magnetization $M_0$, given by Eqn. (2.52) determines maximal intensity of NMR signal.

It is useful to analyze the frequency content of a signal by application of Fourier Transformation (FT). The FT converts a time-domain signal into the frequency domain by:

$$S(\omega) = \int_{-\infty}^{\infty} M(t)e^{-i\omega t}dt \quad (2.71),$$

where $M(t)$ is a signal in the time domain, and $S(\omega)$ is a signal in the frequency domain, called a spectrum. The spectrum can be represented as a result of decomposition of a signal into its components, which are described by corresponding amplitudes and phases at each of the frequencies in spectrum.

The NMR signal can also be represented as a sum of spin isochromats:

$$M_{xy}(\vec{r},t_1) = \sum_{k=1}^{N} m_k e^{i\varepsilon_{1k}(\vec{r},t_1)} \quad (2.72),$$

where $m_k$ is transverse magnetization of the $k-$th spin isochromat, $N$ is the number of isochromats and $\sum_{k=1}^{N} m_k$ is the net magnetization equal to $M_0$. Hence, Eqn. (2.72) simplifies:

$$M_{xy}(\vec{r},t_1) = M_0\sum_{k=1}^{N} e^{i\varepsilon_{1k}(\vec{r},t_1)} \quad (2.73).$$

Transverse magnetization also decays exponentially with a time constant $T_2$, called the spin-spin relaxation time. This effect is described by:

$$M^*_{xy}(\vec{r},t_1) = M_0 \sum_{k=1}^{N} e^{i\varepsilon_{1k}(\vec{r},t_1)} e^{-\frac{t_1}{T_2}} \quad (2.74).$$

where $\varepsilon_{1k}(\vec{r},t_1) = \Omega_k(\vec{r})t_1$. The FT of the $k-$th isochromat is expressed by:

$$S_k(\omega) = \int_0^{\infty} \left[ e^{i\Omega_k(\vec{r})t_1} e^{-\frac{t_1}{T_2}} \right] e^{-i\omega t_1} dt_1 \quad (2.75).$$

This can be rewritten in the form:

$$S_k(\omega) = \int_0^{\infty} e^{-a_k t_1} dt_1 \quad (2.76),$$

where

$$a_k = \frac{1}{T_2} + i(\omega - \Omega_k(\vec{r})) \quad (2.77).$$

The whole spectrum is given as a sum of Fourier transformations for $N$ isochromats:

$$S(\omega) = M_0 \sum_{k=1}^{N} S_k(\omega) \quad (2.78).$$

After integration, the result of FT can be expressed as a complex Lorenzian:

$$\Im(\omega) = \mathrm{A}(\omega) + i\mathrm{B}(\omega) \quad (2.79),$$

where

$$\omega = \Omega_k(\vec{r}) - \Omega_0(\vec{r}) \quad (2.80).$$

Its real and imaginary parts, called respectively absorption and dispersion signals are represented by:

$$\mathrm{A}(\omega) = \frac{T_2}{1 + T_2^{\,2}\omega^2} \quad (2.81),$$

$$\mathrm{B}(\omega) = -\frac{T_2^{\,2}\omega}{1 + T_2^{\,2}\omega^2} \quad (2.82).$$

The absorption and dispersion signals are illustrated in Fig.2.3.

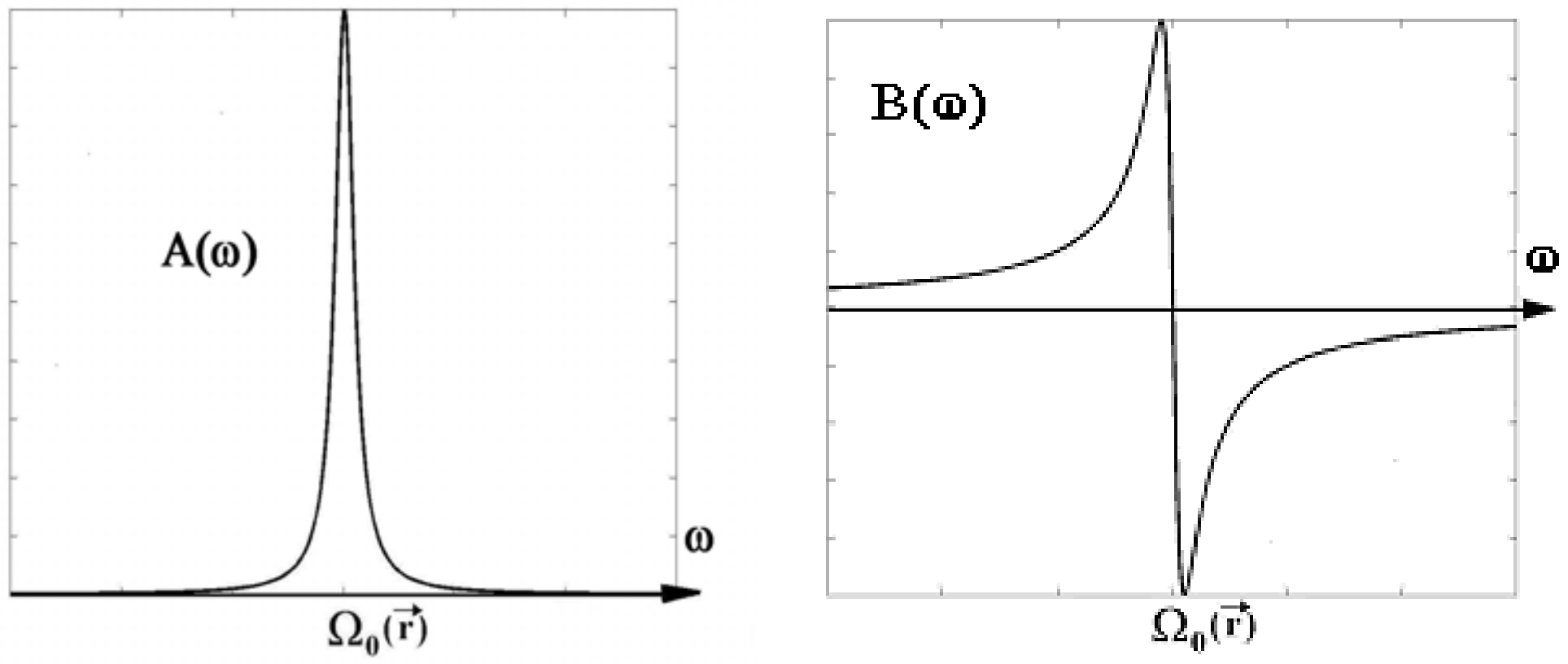

**Fig.2.3** Absorption and dispersion signals (from left to right).

The absorption signal is normally used to represent the spectrum of the NMR signal. The line-width of absorption signal of an isochromat, measured at half of its height, is called the natural linewidth and is given in Hertz units by:

$$\Delta\nu_{1/2} = \frac{1}{\pi T_2} \qquad (2.83).$$

The natural line width is determined by the inverse of relaxation time $T_2$. The resonance frequency $\Omega_0(\vec{r})$ is the Larmor frequency shifted proportionally to the field inhomogeneity at the positions of the spins, which contribute to a given isochromat.

The NMR signal observed in an experiment is represented as a sum of the complex Lorenzians of different isochromats, whose resonance frequencies are shifted by the field inhomogeneities in the sample volume. Thus, NMR spectrum is broader than the Lorenzian of single isochromat. This broadening is caused by inhomogeneity, i.e. the different Larmor frequencies in different parts of the sample volume. The broadening may be empirically described by a decay constant $T_2^*$, sometimes called the effective instrumental relaxation time. The half-width of a line in NMR spectrum in Hz units is given by:

$$\Delta \nu_{1/2}^{*} = \frac{1}{\pi T_2^{*}} \qquad (2.84).$$

The NMR line shape observed due to field inhomogeneity, NMR spectrum does not uniquely specify how the static magnetic field inhomogeneity is distributed within a sample volume. It will be shown in the next chapter how the 3D distribution of field inhomogeneity can be mapped by the use of Fourier imaging techniques.

# 3

# Theory of 3D shimming

The aim of this chapter is to describe in general terms the basic stages of 3D shimming. These include 3D mapping of field inhomogeneity, 3D mapping of the fields, generated by shim coils, and optimisation of shim settings. Calibration of linear transverse field gradients is also included in the 3D shimming procedure described; this testing of the $x1$ and $y1$ shim gradient strengths and balance is needed for imaging experiments, in which shim gradients are used. This chapter forms a preface to Part 2, where experimental aspects of 3D automated shimming and the results achieved by its applications to $^1$H and $^2$H shimming are presented.

## 3.1 3D mapping of static magnetic field inhomogeneity

It was shown in Chapter 2 that the information about the static magnetic field inhomogeneity, coded into the NMR signal with the averaged phase factor does not directly specify the spatial distribution of local magnetic fields in the sample volume. In the absence of special hardware directly measuring the magnetic fields, as described in reference (3.1), a Magnetic Resonance Imaging (MRI) method can be used. It maps the static magnetic field inhomogeneity by measuring the rate of spin phase evolution as a function of position[(3.2)].

### *3.1.1 Applications of linear pulsed field gradients*

It was demonstrated by Lauterbur that the application of a static magnetic field which varies with position within a sample makes the Larmor frequency dependent on the position[(3.3)]. Consider the magnetic field created by application of a linear field gradient along $z$ axis during time $t_z$. In a Cartesian reference frame with origin in the centre of NMR magnet, this produces a magnetic field with strength given at position $z$ by:

$$B(z) = B_0 + G_z z \quad (3.1),$$

where $G_z$ is the strength of the gradient. Then Larmor frequency varies along $z$, and can be described by:

$$\omega_L(z) = \omega_{0L} + \gamma G_z z \quad (3.2),$$

where $\omega_{0L}$ is the Larmor frequency in the absence of the field gradient. A single line in the NMR spectrum, whose Larmor frequencies are encoded by application of linear field gradient represents a projection of the sample along the direction of the gradient, and is called a profile. The width of the profile, denoted $\Delta\omega_L$, relates to the sample $\Delta z$ by:

$$\Delta\omega_L = \gamma G_z \Delta z \quad (3.3).$$

In experiments the spectral width is typically set larger than $\Delta\omega_L$ in order to avoid aliasing of the spectrum.

The minimum distinguishable separation between two adjacent data points in a spectrum is called spectral resolution and is denoted $\delta\omega$. Spatial resolution is the ability to distinguish two points in coordinate space. In MRI, the spatial resolution of sample, $\delta z$ is limited by the spectral resolution, and given by[3.4]:

$$\delta\omega = \gamma G_z \delta z \quad (3.4).$$

It can be seen from Eqn.(3.4) that increase of gradient strength allows imaging of smaller details of a sample with a given spectral resolution. Thus, very large gradient strengths are required to get images of very small objects, for example in NMR microscopy.

*3.1.2 Frequency and phase encoding gradients*

Frequency encoding uses a field gradient during acquisition of the NMR signal to define the direction along which the spatial position is encoded into Larmor frequency.

If the field gradient is applied during acquisition of NMR signal then the Larmor frequency of spins given by Eqn.(3.3) becomes spatially dependent. The Fourier Transform decomposes the NMR signal into individual frequency components, where each frequency corresponds to a given position[3.5]. When several gradients are applied at the same time, their net effect is the same as a single gradient at an intermediate angle, given by:

$$\vec{G} = \vec{G}_x + \vec{G}_y \quad (3.5).$$

In phase encoding, originally proposed by Kumar et al[3.6], field gradients are applied between the periods of excitation and acquisition of NMR signal. If the gradient is applied for a fixed period of time before acquisition of NMR signal, then the phase of the signal, instead of frequency, becomes spatially encoded. The phase acquired by spins at position $z$ and time $t$ due to the application of a linear gradient is given by:

$$\varphi(z,t) = \gamma(B_0 + G_z z)t \text{ , for } 0 < t < t_z \quad (3.6).$$

This process is repeated with incremented gradient strengths (or durations) in order to cause phase shifts of spins by these increments. In this method, the phases of spins become a function of their position in the sample volume. Unlike frequency encoding, the application of several phase encoding gradients affects the spin system as if these were applied independently, allowing simultaneous application of several phase encoding gradients without increasing the duration of pulse sequence.

### *3.1.3 Mapping of static magnetic field inhomogeneity*

The mapping of static magnetic field inhomogeneity by the MRI technique is a very effective way to find the local strength of the static magnetic field experienced by spins in the sample. It exploits the fact that the spin phases depend on the local magnetic field strengths, which alter in different parts of a sample.

In practice, the form of the pulse sequence for mapping of the static magnetic field is determined by experimental considerations. For consistency with Part 2 of the thesis, the modified stimulated echo with 3-dimensional pulsed field gradients is presented here(3.7). Fig.3.1 shows a simplified pulse sequence based on the modified pulsed field gradient stimulated echo (PFGSTE), used for field mapping in this thesis.

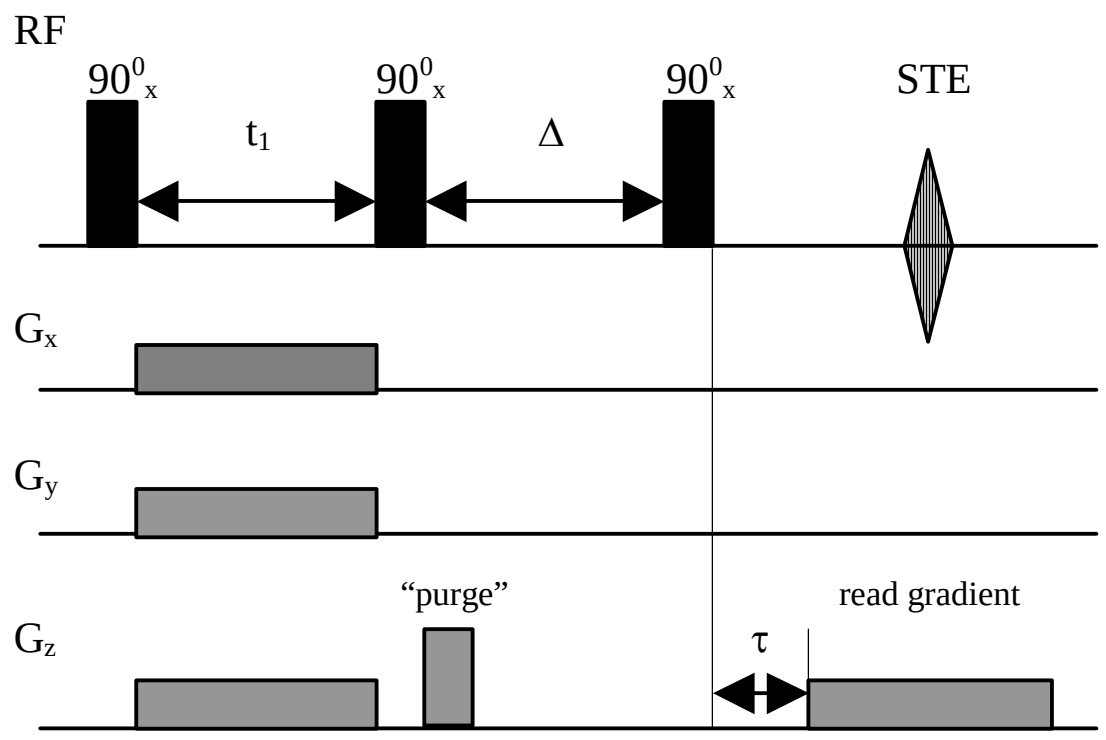

**Fig.3.1** PFGSTE pulse sequence, modified for 3-dimensional field mapping.

It has been shown that a number of echoes are produced by this sequence(3.8). The stimulated echo is a signal with maximum intensity at $t = 2t_1 + \Delta + \tau$. The density matrix formalism can be used to describe the formation of the stimulated echo in time domain. Before application of the first rf pulse, the spins are polarized along the $z$ axis. The analysis is carried out in the rotating frame of reference, described in Chapter 2. All rf pulses are 90 degree pulses applied along the $x$ axis in the rotating frame. The chosen sign convention for rotations is consistent with that used in reference (3.8). It is assumed that relaxation effects are negligible during the sequence, and that the phase- and frequency-encoding gradient strengths are larger than the static magnetic field inhomogeneity.

After the first rf pulse the density operator can be presented by:

$$\hat{\rho}\left(\vec{r}, t_{90^0}\right) = a + b(\vec{r})\hat{I}_y \qquad (3.7).$$

Evolution during the time $t_1$ due to the phase-encoding gradient $\vec{G}\left(G_x, G_y, G_z\right)$ gives:

$$\hat{\rho}\left(\vec{r}, t_{90^0} + t_1\right) = a + b(\vec{r})\left(\hat{I}_y \cos\varphi(\vec{r}, t_1) + \hat{I}_x \sin\varphi(\vec{r}, t_1)\right) \qquad (3.8),$$

where $\varphi(\vec{r})$ is a local phase, given by:

$$\varphi(\vec{r}) = \gamma \vec{G} \cdot \vec{r} t_1 \qquad (3.9).$$

The second 90 degree pulse rotates $\hat{I}_y$ magnetization to the $-z$ axis:

$$\hat{\rho}\left(\vec{r}, t_{90^0} + t_1 + t_{90^0}\right) = a + b(\vec{r})\left(-\hat{I}_z \cos\varphi(\vec{r}) + \hat{I}_x \sin\varphi(\vec{r})\right) \qquad (3.10).$$

This is followed by strong $z$ ”purge” gradient, dephasing transverse magnetization, which can be subsequently neglected. The remaining density operator is:

$$\hat{\rho}\left(t_{90^0} + t_1 + t_{90^0} + \Delta\right) = a + b(\vec{r})\left(-\hat{I}_z \cos\varphi(\vec{r})\right) \qquad (3.11).$$

The third 90 pulse rotates $z$ magnetization into the $xy$ plane:

$$\hat{\rho}\left(t_{90^0} + t_1 + t_{90^0} + \Delta + t_{90^0}\right) = a + b(\vec{r})\left(-\hat{I}_y \cos\varphi(\vec{r})\right) \qquad (3.12).$$

Evolution during the period $\tau$ is caused by the static magnetic field inhomogeneity, and the density operator becomes:

$$\begin{aligned} &\hat{\rho}\left(t_{90^0} + t_1 + t_{90^0} + \Delta + t_{90^0} + \tau\right) = \\ &a + b(\vec{r})\left(-\hat{I}_y \cos\Omega(\vec{r})\tau - \hat{I}_x \sin\Omega(\vec{r})\tau\right)\cos\varphi(\vec{r}) \end{aligned} \qquad (3.13),$$

where $\Omega(\vec{r})$ is an angular frequency representing the local difference between Larmor frequencies in inhomogeneous and homogeneous fields, given at position $\vec{r}$ by:

$$\Omega(\vec{r}) = \gamma \delta B(\vec{r}) \qquad (3.14).$$

The read-out $z$ gradient applied during acquisition of NMR signal gives:

$$\begin{aligned} &\hat{\rho}\left(\vec{r}, t_{90^0} + t_1 + t_{90^0} + t_2 + t_{90^0} + \tau + t_z\right) = \\ &a + b(\vec{r})\cos\varphi(\vec{r})\left[-\hat{I}_y \cos\left(\Omega(\vec{r})\tau + \varphi_z(z)\right) - \hat{I}_x \sin\left(\Omega(\vec{r})\tau + \varphi_z(z)\right)\right] \end{aligned} \qquad (3.15),$$

where $\varphi_z$ is the phase acquired by spins during the read-out gradient time $t_z$:

$$\varphi_z(z) = \gamma G_z^{read} t_z z \qquad (3.16).$$

The local NMR signal may be written as:

$$M_{xy}(\vec{r}, t_z, \tau) = -M_0 \frac{i}{2} \cos\varphi(\vec{r}) e^{i(\Omega(\vec{r})\tau + \varphi_Z(z))} \quad (3.17).$$

After the substitution:

$$\cos\varphi = \frac{e^{\varphi} + e^{-\varphi}}{2} \quad (3.18).$$

This may be rewritten as:

$$M_{xy}(\vec{r}, t_z, \tau) = -M_0 \frac{i}{4} \left( e^{i(\Omega(\vec{r})\tau + \varphi_z(\vec{r}) + \varphi(\vec{r}))} + e^{i(\Omega(\vec{r})\tau + \varphi_z(\vec{r}) - \varphi(\vec{r}))} \right) \quad (3.19),$$

where the first signal in brackets continues to dephase, and gives in normal conditions a negligible signal (referred as an anti-echo[3.9]). The second signal refocuses forming a stimulated echo (STE) at time $t_z = t_1$ for equal strengths of $z$ gradient strengths during $t_1$ and $t_z$ intervals. The echo signal is given by:

$$M^{STE}(\vec{r}, t_z, \tau) = -M_0 \frac{i}{4} e^{i(\Omega(\vec{r})\tau + \varphi_z(\vec{r}) - \varphi(\vec{r}))} \quad (3.20).$$

The STE signal builds up while the difference between the phases $\varphi(\vec{r})$ and $\varphi_z(\vec{r})$ diminishes, and reaches a maximum when phase difference

$$\Delta\varphi(\vec{r}) = |\varphi(\vec{r}) - \varphi_z(\vec{r})| \quad (3.21)$$

is minimum. The three-dimensional Fourier Transform of this signal brings:

$$S(\vec{r}, \tau) = -M_0 \frac{i}{4} e^{i\Omega(\vec{r})\tau} \quad (3.22).$$

This represents the NMR signal as a function of position, with a phase modulated by an angle proportional to the variation of the static magnetic field inhomogeneity as a function of position. The complex ratio of images, acquired with and without the delay $\tau$ is given by:

$$R(\vec{r}, \tau) = \frac{S(\vec{r}, \tau)}{S(\vec{r}, 0)} = e^{i\Omega(\vec{r})\tau} \quad (3.23).$$

This ratio is a complex function, whose real and imaginary parts are given by:

$$\mathrm{Re}\{R(\vec{r}, \tau)\} = \mathrm{Re}\{e^{i\Omega(\vec{r})\tau}\} = \cos(\Omega(\vec{r})\tau) \quad (3.24),$$

$$\mathrm{Im}\{R(\vec{r},\tau)\} = \mathrm{Im}\{e^{i\Omega(\vec{r})\tau}\} = i\sin(\Omega(\vec{r})\tau) \qquad (3.25).$$

The phase difference between two images may be expressed by:

$$\Omega(\vec{r})\tau = Arc\tan\left(\frac{\mathrm{Im}[R(\vec{r},\tau)]}{\mathrm{Re}[R(\vec{r},\tau)]}\right) \qquad (3.26).$$

Then local inhomogeneity of the static magnetic field is(3.10):

$$\delta B(\vec{r}) = \frac{1}{\gamma\tau} Arc\tan\left(\frac{\mathrm{Im}[R(\vec{r},\tau)]}{\mathrm{Re}[R(\vec{r},\tau)]}\right) \qquad (3.27).$$

A plot of function, given by Eqn. (3.27) versus position represents a map of the static magnetic field inhomogeneity, or field map in short.

As the trigonometric function $Arc\tan$ has a period $\pi$ it is single-valued only for the angles within the period. When phase changes by more than $\pi$, phase unwrapping is used in order to avoid ambiguity in value of the function. It means that when a phase difference between adjacent points in an image is more than $\pi$, a $2\pi$ angle (or its multiple) is either added or subtracted to bring the phase difference into the $\pi$ radian range.

Experimentally, phase differences between profiles may also result from other causes, for example thermal convection. Field mapping can be affected by these simultaneously with the effect of field inhomogeneity(3.11). Hence, it is important to provide such experimental conditions that the spins will evolve during the time $\tau$ mainly due to the field inhomogeneity without interference with other effects.

## 3.2 3D mapping of the fields produced by the shim coils

The technique for 3D mapping of the fields produced by the shim coils (3D shim mapping) is described in this section. Together with the 3D field mapping technique, this forms a preface to Section 3.3, in which the calculation of corrected shim values is presented. Practical applications of these methods are presented in Part 2. The 3D shim mapping technique is applied only to the room temperature shims, whose adjustment is a routine operation usually carried out before to begin NMR experiment. Field and shim mapping may start either from all shim values set to zero (‘cold’ shims’), or from some already adjusted values of shims (‘warm’ shims).

It has been shown that the local static magnetic field strength is measured by mapping the phase change as a function of position during an evolution time. This technique can easily be extended by measuring field maps with and without a change in one shim settings, so that the difference gives a map of the field change produced by the shim.

*3.2.1 Magnetic field shapes produced by shim coils*

It was proposed by Golay that if the spatial variation of a static magnetic field can be described mathematically by a sum of spherical harmonics, then generating experimental fields corresponding to such harmonics with appropriate amplitude can correct that spatial variation[3.12]. Each shim coil or set of coils has a particular winding geometry, designed to produce a field shape representing one of the functions of the basis set used for expansion of the spatial variation of the field error. The field error, denoted $\delta B(\vec{r})$, is the difference between nominal $B_0$ and actual $B(\vec{r})$ values of the magnetic field at a position $\vec{r}$. The functions of the basis set are chosen to be independent, so that the fields produced by each of the shims can be changed independently (i.e. without interference with the fields produced by other shims). The strength of the field produced by the $i$-th shim can be given at a position $\vec{r}$ in the form:

$$\delta B_i^*(\vec{r}) = A_i \cdot f_i(\vec{r}) \qquad (3.29),$$

where $A_i$ represents amplitude of the field and $f_i(\vec{r})$ is a field shape function, which describes the spatial variation of the field produced by a shim. The field shape functions for the shim set used in this thesis are presented in the Table 3.1.

Table 3.1 **The field shape functions generated by the shim coils used.**

| Shim name | $z0$ | $z1$ | $z2$ | $z3$ | $z4$ | $z5$ | $x1$ | $y1$ | $xz$ | $yz$ | $xy$ | $x2-y2$ | $x3$ | $y3$ |
|---|---|---|---|---|---|---|---|---|---|---|---|---|---|---|
| Field shape function | 1 | $z$ | $z^2$ | $z^3$ | $z^4$ | $z^5$ | $x$ | $y$ | $xz$ | $yz$ | $xy$ | $x^2-y^2$ | $xz^2$ | $yz^2$ |

### *3.2.2 Mapping of the fields produced by shim coils*

The field shapes, generated by each of the shims are found as a difference between two field maps, the first acquired with the current shim values as described in Section 3.1, and the second with the same shim values except for one. The value of this shim is mis-set in order to produce the shim field shape of interest over a sample. The complete shim mapping experiment is performed by repetition of the technique described for each of the shims.

Ideally, the accuracy of the field mapping should not be disturbed by the field inhomogeneity. In practice, however, inhomogeneity can disturb field maps significantly. It is therefore desirable to adjust at least some of the shim settings before final shim mapping.

## 3.3 Optimisation of shim settings

The algorithm presented in this section, forms together with the 3D field and shim mappings described, the core of the 3D shimming technique. This can be applied in automation by the use of the software described in Chapter 5. The algorithm uses 3D field and shim maps as input data to find optimal corrections to shim settings in order to cancel 3D field inhomogeneity. This can be solved as an optimisation problem as described below.

### *3.3.1 Optimisation of N shims by linear least squares fitting*

3D shimming can be presented as an optimisation problem in $N$ dimensions (where $N$ is the number of shims) aimed at finding the set of values for shim setting corrections for $N$ shims which best cancel the field variations described by the field map. The optimal shim setting corrections can be calculated by solving the linear least squares fitting problem[3.13], which uses field and shim maps as input data and can be expressed as:

$$\chi^2 = \text{Min}\left(\left|\underline{\underline{A}} \cdot \underline{x} - \underline{b}\right|^2\right) \qquad (3.30),$$

where "chi-square" $\chi^2$ represents the statistical error of fitting. The shim maps are stored after the shim mapping experiment as *N* vectors of length *M*, and can be used to form a shim map matrix $\underline{\underline{A}}$ of size {*M* x *N*}. A vector $\underline{b}$ represents field map data. The problem is to find for each of $j = 1,...N$ shims the shim setting corrections $\underline{x}$, which minimise the square difference between field map and the shim map data multiplied by unknown $\underline{x}$. The least square difference can be expressed by a matrix equation:

$$\underline{\underline{A}} \cdot \underline{x} = \underline{b} \qquad (3.31).$$

This defines matrix $\underline{\underline{A}}$ as a linear mapping from the vector space $\underline{x}$ to the vector space $\underline{b}$, which can be given by a system of linear algebraic equations[3.14]:

$$\begin{aligned}
A_{11}x_1 + A_{12}x_2 + A_{13}x_3 + A_{14}x_4 + ... + A_{1N} &= b_1 \\
A_{21}x_1 + A_{22}x_2 + A_{23}x_3 + A_{24}x_4 + ... + A_{2N} &= b_2 \\
A_{31}x_1 + A_{32}x_2 + A_{33}x_3 + A_{34}x_4 + ... + A_{3N} &= b_3 \\
... \qquad &\quad ... \\
A_{M1}x_1 + A_{M2}x_2 + A_{M3}x_3 + A_{M4}x_4 + ... + A_{MN} &= b_M
\end{aligned} \qquad (3.32).$$

When *N*=*M*, the number of equations and unknown variables is the same and a single solution exists for the given system of equations[3.15]. For $N \neq M$, the set of equations is degenerate when either of the following occurs: some of the equations are linear combinations of each other, or all of the equations contain exactly the same linear combination of some unknown variables. A set of the degenerate linear equations, and the matrix, which these represent is called singular.

The matrices formed from experimental data can be degenerate, for example due to the influence of random or systematic errors. Degeneracy of the system of equations can result in several different "optimal" solutions, which correspond to local minima of the static magnetic field. Thus, it is important to solve the optimisation problem by a method, which is not mislead by local minima.

*3.3.2 Solution of linear least squares problem using SVD*

Singular Value Decomposition[3.16] (SVD) is a technique which is very effective for the solution of singular or almost singular sets of equations. The SVD method is based on the following theorem of linear algebra: a *M*x*N* matrix $\underline{\underline{A}}$ whose number of rows *M* is equal to or larger than its number of columns *N*, can be presented as a product of three matrices, a column-orthogonal matrix $\underline{\underline{U}}$ of size *M*x*N*, a diagonal matrix $\underline{\underline{W}}$ of size *N*x*N* with positive or zero elements called singular values, and the transposed orthogonal matrix $\underline{\underline{V}}$ of size *N*x*N*. This can be represented as:

$$\begin{bmatrix} \\ \underline{\underline{A}}\{M\text{x}N\} \\ \\ \end{bmatrix} = \begin{bmatrix} \\ \underline{\underline{U}}\{M\text{x}N\} \\ \\ \end{bmatrix} \cdot \begin{bmatrix} w_1 & & \\ & \cdot & \\ & & w_N \end{bmatrix} \cdot \begin{bmatrix} \underline{\underline{V}}^T\{N\text{x}N\} \end{bmatrix} \quad (3.32).$$

The matrices $\underline{\underline{U}}$ and $\underline{\underline{V}}$ are each orthogonal, i.e. $\underline{\underline{U}}^T \underline{\underline{U}} = \underline{\underline{I}}$ and $\underline{\underline{V}}^T \underline{\underline{V}} = \underline{\underline{I}}$, where $\underline{\underline{I}}$ is the identity matrix. When matrix $\underline{\underline{A}}$ is degenerate, it has one or more singular values of zero. The inverse of matrix $\underline{\underline{A}}$ is

$$\underline{\underline{A}}^{-1} = \underline{\underline{V}} \left[ \text{diag}\left( \frac{1}{\underline{\underline{W}}} \right) \right] \underline{\underline{U}}^T \quad (3.33).$$

In SVD the inverse singular values $\left(\frac{1}{\underline{\underline{W}}}\right)$ can be replaced by zeros[3.17] if $w_j \approx 0$, giving a non-singular inverse data matrix. The solution of Eqn.(3.31) can then be given directly as in the non-singular case[3.18]:

$$\underline{x} = \underline{\underline{A}}^{-1} \cdot \underline{b} \tag{3.34}$$

When SVD is used for $\underline{\underline{A}}$, the solution (written with the use of Einstein notation for some matrices and vectors) becomes[3.19]:

$$x_j = \mathbf{V} \cdot \left[ \mathrm{diag}\left( \frac{1}{w_j} \right) \right] \cdot \mathbf{U}^{\mathrm{T}} \cdot b_i \tag{3.35}$$

It is convenient to rewrite the solution in the following form:

$$x_j = \sum_{i=1}^{N} \left( \frac{U_{(i)} \cdot b_i}{w_i} \right) \cdot V_{(i)} \tag{3.36}$$

where $U_{(i)}$ and $V(i)$ are the vectors of length *M*, which, respectively, denote the columns of $\underline{\underline{U}}$ and $\underline{\underline{V}}$ for $i = 1,...M$.

The standard deviation in linear least squares fitting of the shim corrections $x_j$ by the use of SVD can be given by:

$$\sigma^2(x_j) = \sum_{i=1}^{M} \left( \frac{V_{ji}}{w_i} \right)^2 \tag{3.37}$$

The covariance matrix of the errors in the fitted parameters $x_j$ and $x_{j'}$ is:

$$\mathrm{Cov}(x_j, x_{j'}) = \sum_{i=1}^{M} \left( \frac{V_{ji} \cdot V_{j'i}}{w_i^2} \right) \tag{3.38}$$

The shim corrections $x_j$ (denoted in Einstein notation) calculated by the use of SVD are substracted from the current shim settings in order to get a set of corrected shim values. After the new shims are calculated, the remaining inhomogeneity of the static magnetic field can be estimated from the line width of a spectrum acquired with the corrected shim settings. If the line width of the spectrum is not small enough then homogeneity needs further improvement and therefore the field mapping and correction of the shim values must be repeated. This process can be run iteratively in automation by the use of the software presented in Chapter 5.

## 3.4 Calibration of x- and y- gradient strengths

In this thesis, transverse linear shim gradients are generated by $x1$ and $y1$ shim coils, which have similar design arranged at right angle to each other, in order to produce, in ideal, perpendicular field gradients of the same strength. In practice, the coils used are not ideal, due to imperfections in their design and manufacturing, and therefore the transverse gradient strengths are imbalanced to some extent. This imbalance can result in distorted images. Distortions can be corrected by the calibration procedure presented here and recommended for use before 3D shimming. The presentation given here is mainly theoretical, and describes the basis for the transverse gradient strength calibration programme presented in Chapter 5.

### *3.4.1 Technique for $x1$ and $y1$ gradient calibration*

Ideally, $x1$ and $y1$ shims produce linear field gradients with the same strength but perpendicular to each other, as illustrated in Fig.3.2 (a). Generally, when two field gradients $\vec{G}_x$ and $\vec{G}_y$ are applied, their total is a single gradient at an intermediate angle, defined as their vector sum[3.20]:

$$\vec{G} = \vec{G}_x + \vec{G}_y \qquad (3.39),$$

where $\vec{G}_x$ and $\vec{G}_y$ are vectors which represent the $x1$ and $y1$ gradients.

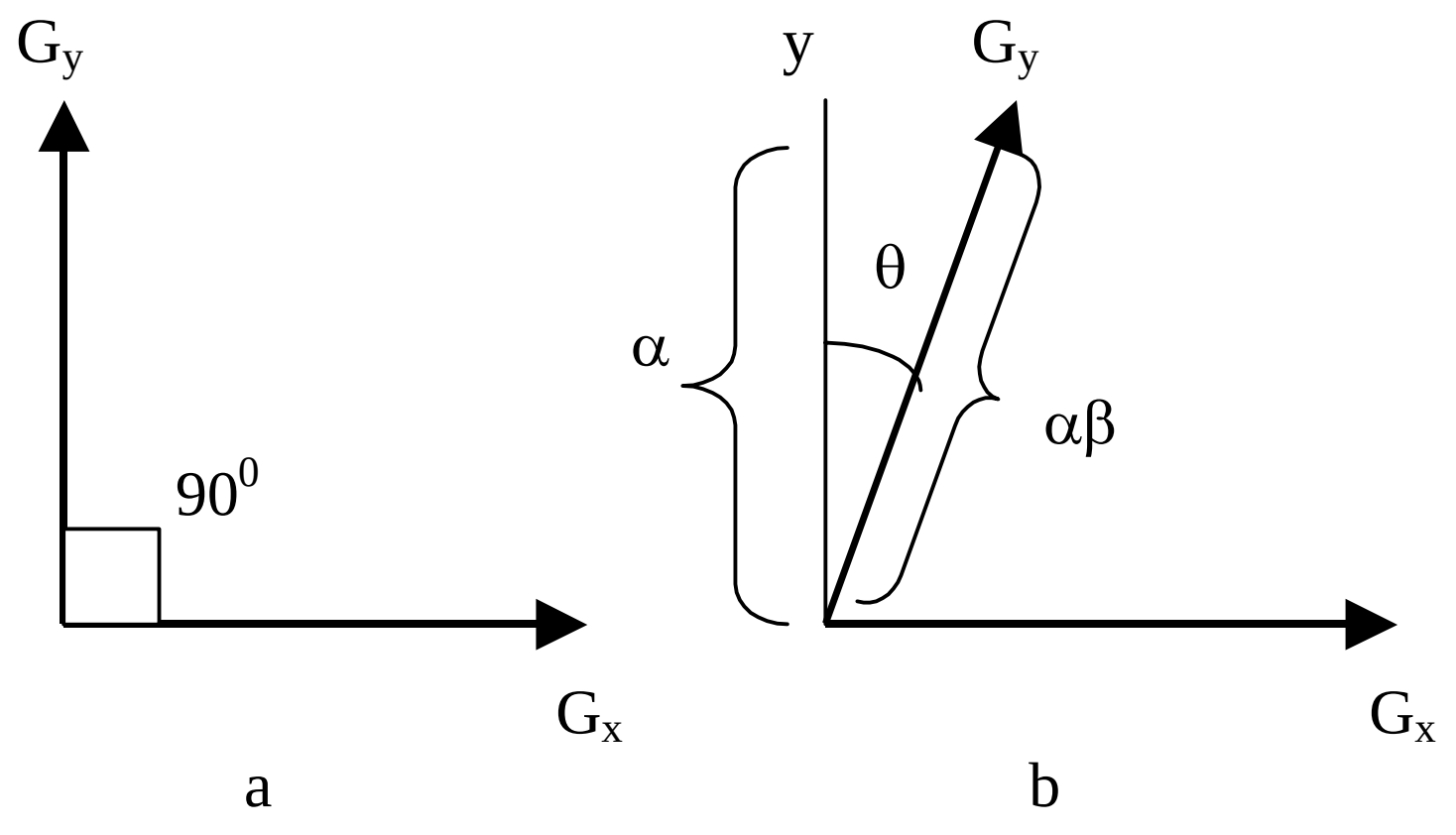

**Fig.3.2** Transverse gradients: (a) ideally balanced (b) misbalanced.

When the direction of the $y1$ gradient deviates from the true $y$ axis by an angle $\theta$, as shown in Fig.3.2 (b), the total gradient is:

$$\vec{G} = (\alpha x_1 + \alpha\beta y_1 \sin\theta)\vec{x} + (\alpha\beta y_1 \cos\theta)\vec{y} \quad (3.40),$$

where $\vec{x}$ and $\vec{y}$ are unit vectors for the $x$ and $y$ axes; and $x1$ and $y1$ are the nominal gradient strengths in the experimental units of Digital-to-Analog Converter (DAC) points. These are the signed integer numbers, which are used experimentally to define shim currents. When a shim setting (represented by the number of DAC points for a particular shim, $x1$ or $y1$ here) is changed for a particular shim, a corresponding shim field gradient of a strength proportional to the change is produced. The parameter $\alpha$ is the coefficient of proportionality between the nominal shim gradient strength $x1$ DAC points and the actual gradient strength $\vec{G}_x$ in $G\,cm^{-1}$. The parameter $\beta$ represents the transverse shim gradient strength imbalance, defined as the ratio of $y_1$ gradient strength to $\alpha$. The magnitude of the gradient given by Eqn.(3.40) can be written as:

$$\left|\vec{G}\right| = \alpha\sqrt{(x_1 + \beta \cdot y_1 \cdot \sin\theta)^2 + (\beta \cdot y_1 \cdot \cos\theta)^2} \quad (3.41).$$

When the gradients are perpendicular, as depicted in Fig.3.2 (a), $\theta = 0$, and the total gradient becomes:

$$\vec{G}_{\perp} = \alpha x_1 \vec{x} + \alpha\beta y_1 \vec{y} \quad (3.42).$$

The imbalance between the transverse shim gradient strengths and deviation from orthogonality can be determined by measuring the widths of the profiles of a cylindrical sample for different settings of the transverse shims. A simple method is to rotate a nominal applied linear transverse gradient through a succession of angles $\phi$, mapping out the variation in the width at base of the measured profile. For ideally balanced, orthogonal gradients this would simply yield a constant profile width at any value of angle $\phi$.

Consider the general case of measuring the signal profile width of a nucleus of magnetogyric ratio $\gamma$ for a cylindrical sample of internal diameter $d$. It is assumed that the shim gradients are mis-set by amounts $x_0$ and $y_0$ in $x1$ and $y1$ shims respectively and other gradients can be neglected. Then for the imperfect gradients defined above the effect of applying a nominal extra $x1$ gradient of $g_{inc}\cos\phi$ DAC points and a nominal $y1$ gradient of $g_{inc}\sin\phi$ DAC points will be to generate gradients in the $\vec{x}$ and $\vec{y}$ directions, given by:

$$G_x = \alpha\left(x_0 + g_{inc}\cos\phi + \beta\left(y_0 + g_{inc}\sin\phi\right)\sin\theta\right) \quad (3.43),$$

$$G_y = \alpha\left(y_0 + \beta g_{inc}\sin\phi\cos\theta\right) \quad (3.44).$$

Thus, the width in Hz of the base of the profile is given in Hz units by:

$$\Delta\nu^{Rot}\left(\alpha,\beta,\theta,x_0,y_0\right) = \frac{\gamma d\left|\vec{G}^{Rot}\right|}{2\pi} \quad (3.45),$$

where

$$\left|\vec{G}^{Rot}\right| = \alpha\sqrt{(x_0 + g_{inc}\cos\phi + \beta\sin\theta(y_0 + g_{inc}\sin\phi))^2 + \beta^2\cos^2\theta\left(y_0 + g_{inc}\sin\phi\right)^2}$$

(3.46).

As the nominal rotation angle $\phi$ of the gradient given by Eqn.(3.46) is varied from 0 to $2\pi$ radians, the measured width will vary in a manner determined by the parameters $\alpha$, $\beta$, $\theta$, $x_0$ and $y_0$, values for which may be found by least squares fitting.

The least squares problem is formulated as minimisation of the function:

$$F(\alpha,\beta,\theta,x_0,y_0;\phi)=\sum_{i=1}^{M}\left(\Delta\nu_i^{Rot.\exp}-\Delta\nu_i^{Rot.calc}\right) \qquad (3.47),$$

where $\Delta\nu_i^{Rot.\exp}$ is the width of profile measured for the $i$-th value of $\phi$, and $\Delta\nu_i^{Rot.calc}$ is the width calculated by the use of Eqn.(3.45). Nonlinear least-squares fitting[3.21] is used as described in detail in the next section.

### *3.4.2 Non-linear least squares fitting*

The fitting of parameters to a given function can be described as a search for values of parameters, which minimize the least squares error for the given function. Generally, it can be very difficult to find a global solution of this problem, i.e. a solution for the whole region where the function is defined. Thus, fitting methods are normally aimed at finding a local minimum inside of a small part (denoted here $\varepsilon$) of the region. The fitting of experimental data (which normally include errors) is a search for the best fit in a statistical sense. In the case of a normal error distribution, least squares fitting[3.22], described here, is used.

Least squares fitting is the problem of finding an argument (denoted here as $\underline{\tilde{b}_0}$) of a merit function, which corresponds to its minimum[3.21], given by:

$$F_{\mathrm{Min}}\left(\underline{\tilde{b}_0}\right)=\mathrm{Min}\sum_{i=1}^{M}\left(F\left(\underline{\tilde{b}}\right)\right)^2 \qquad (3.48),$$

within a region $\left|\underline{\tilde{b}}-\underline{\tilde{b}_0}\right|<\varepsilon$; the number $M$ represents a number of data point. The argument $\underline{\tilde{b}}$ can be described as a vector whose components are the fitting parameters (for example the described parameters of imbalance between $x1$ and $y1$ gradients). If the function $F(\underline{\tilde{b}})$ is differentiable within $\varepsilon$ then it can be represented in this region by a Taylor expansion:

$$F(\underline{\tilde{b}} + \delta\underline{\tilde{b}}) \approx F(\underline{\tilde{b}}) + \delta\underline{\tilde{b}}^T F'(\underline{\tilde{b}}) + \frac{1}{2}\delta\underline{\tilde{b}}^T F''(\underline{\tilde{b}})\delta\underline{\tilde{b}} + \ldots \qquad (3.49),$$

where $F'(\underline{\tilde{b}})$ and $F''(\underline{\tilde{b}})$ are respectively the first and second derivatives of the function $F(\underline{\tilde{b}})$, $\delta\underline{\tilde{b}}$ denotes an infinitesimal augment of vector $\delta\underline{\tilde{b}}$, defined inside of region $\varepsilon$ and $\delta\underline{\tilde{b}}^T$ is its transpose.

A search for the local minimum may be performed iteratively, where an optimal direction and step of $\delta\underline{\tilde{b}}$ leading towards local minimum of $F(\underline{\tilde{b}} + \delta\underline{\tilde{b}})$ are searched for each iteration. The choices of direction and step of the increment $\delta\underline{\tilde{b}}$ are the key to optimal fitting, as they determine how rapidly a minimum of the merit function will be found. They are also important in order that the fitting not mislead when there are several local minima and it is required to choose among them. These features are implemented in the technique proposed by Marquardt after a suggestion by Levenberg(3.23), where adjustment of the increment step and its direction is performed in each iteration more effectively than in other techniques. This method has became standard in nonlinear least-square fitting and is used in this thesis.

In the Marquardt-Levenberg technique, the optimal increment of the parameters $\delta\underline{\tilde{b}}$ can be found by solving the equation(3.21, 3.23):

$$(\underline{\underline{H}}(\underline{\tilde{b}}) + \mu\underline{\underline{I}})\delta\underline{\tilde{b}} = -\underline{J}(\underline{\tilde{b}}) \qquad (3.50),$$

where $\mu \geq 0$ and represents an adjustable parameter and the $\underline{\underline{I}}$ is the identity matrix. The derivatives $F'(\underline{\tilde{b}})$ and $F''(\underline{\tilde{b}})$ are represented by the vector $\underline{J}$ and the matrix $\underline{\underline{H}}$ (called Hessian), respectively. These are given (using Einstein notation for $\underline{\tilde{b}}$) by:

$$\underline{J}(\tilde{b}_i) = \sum_{i=1}^{M} \frac{\partial F(\tilde{b}_i)}{\partial \tilde{b}_i} \qquad (3.51),$$

$$\underline{\underline{H}}\left(\tilde{b}_i, \tilde{b}_j\right) = \sum_{i,j=1}^{M} \frac{\partial^2 F\left(\tilde{b}_i\right)}{\partial \tilde{b}_i \partial \tilde{b}_j} \quad (3.52),$$

The Marquardt-Levenberg technique works in the following way[3.23, 3.24]: after an initial guess $\underline{\tilde{b}}_{guess}$ is given, the merit function $F\left(\underline{\tilde{b}}_{guess}\right)$ can be calculated. Then some small value of $\mu$ is chosen for solving Eqn. (3.50) for $\delta\underline{\tilde{b}}$. After this, the merit function $F\left(\underline{\tilde{b}}_{guess} + \delta\underline{\tilde{b}}\right)$ is evaluated and if $F\left(\underline{\tilde{b}}_{guess} + \delta\underline{\tilde{b}}\right) \geq F\left(\underline{b}\right)$ then $\mu$ must be increased and Eqn.(3.50) has to be solved with the increased value of $\mu$. However, when $F\left(\underline{\tilde{b}}_{guess} + \delta\underline{\tilde{b}}\right) < F\left(\underline{\tilde{b}}\right)$, $\mu$ must be decreased and the parameter $\underline{\tilde{b}}$ has to be updated to $\underline{\tilde{b}}_{opt} = \underline{\tilde{b}}_{guess} + \delta\underline{\tilde{b}}$. In the next iteration Eqn. (3.50) is solved with $\underline{\tilde{b}}_{opt}$ and after it the merit function is evaluated as in the previous iteration. The solution $\delta\underline{\tilde{b}}$ is a signed quantity, which is positive when the direction of search is not a descent. Otherwise, the sign of solution $\delta\underline{\tilde{b}}$ is negative and indicates descent towards a minimum. The speed of approaching a minimum depends on the value of $\delta\underline{\tilde{b}}$ (step of parameter increment) and is also determined by $\mu$. This procedure is repeated iteratively until fitting converges to the required accuracy. The quality of the fitting can be estimated by calculation of the covariance matrix, given in Einstein notation by:

$$\mathrm{Cov}\left(\tilde{b}_j, \tilde{b}_k\right) = \frac{1}{H_{jk}} \quad (3.53)$$

The Marquardt-Levenberg technique is used in the software for transverse gradient calibration, presented in Chapter 5 of this thesis. The applications of the calibration procedure are thoroughly presented in Chapter 6.

# 4

# Hardware

## 4.1 Introduction

This chapter is intermediate between theory, presented in Chapter 3, and its experimental realisation, whose details and results will be described in the following chapters. An intermediate chapter is necessary in order to describe the hardware used for 3D shimming, including its practical limitations, before the presentation of experimental results. It starts with a general overview of a pulsed NMR spectrometer, and then describes in more detail the principal components of the NMR spectrometer, which was used for 3D automated shimming.

NMR instruments can be classified according to how they produce magnetic field gradients. Older NMR spectrometers are equipped with shims and a homospoil facility, while more modern spectrometers are also equipped with pulsed field gradient (PFG) modules. One of the aims of this thesis is to show that 3D shimming can be successfully accomplished to the point where the static magnetic field homogeneity achieved meets magnet specification without the use of any PFG hardware. This is particularly important for probes with 10 mm or greater sample diameters, as these rarely have a PFG coil.

## 4.2 General overview

Fig. 4.1 shows a simplified block diagram of a pulsed NMR spectrometer. Its main components are the magnet, transmitter, amplifier, probe, preamplifier, receiver and analog-to-digital converter[(4.1)]. Each of these accomplishes a specific function:

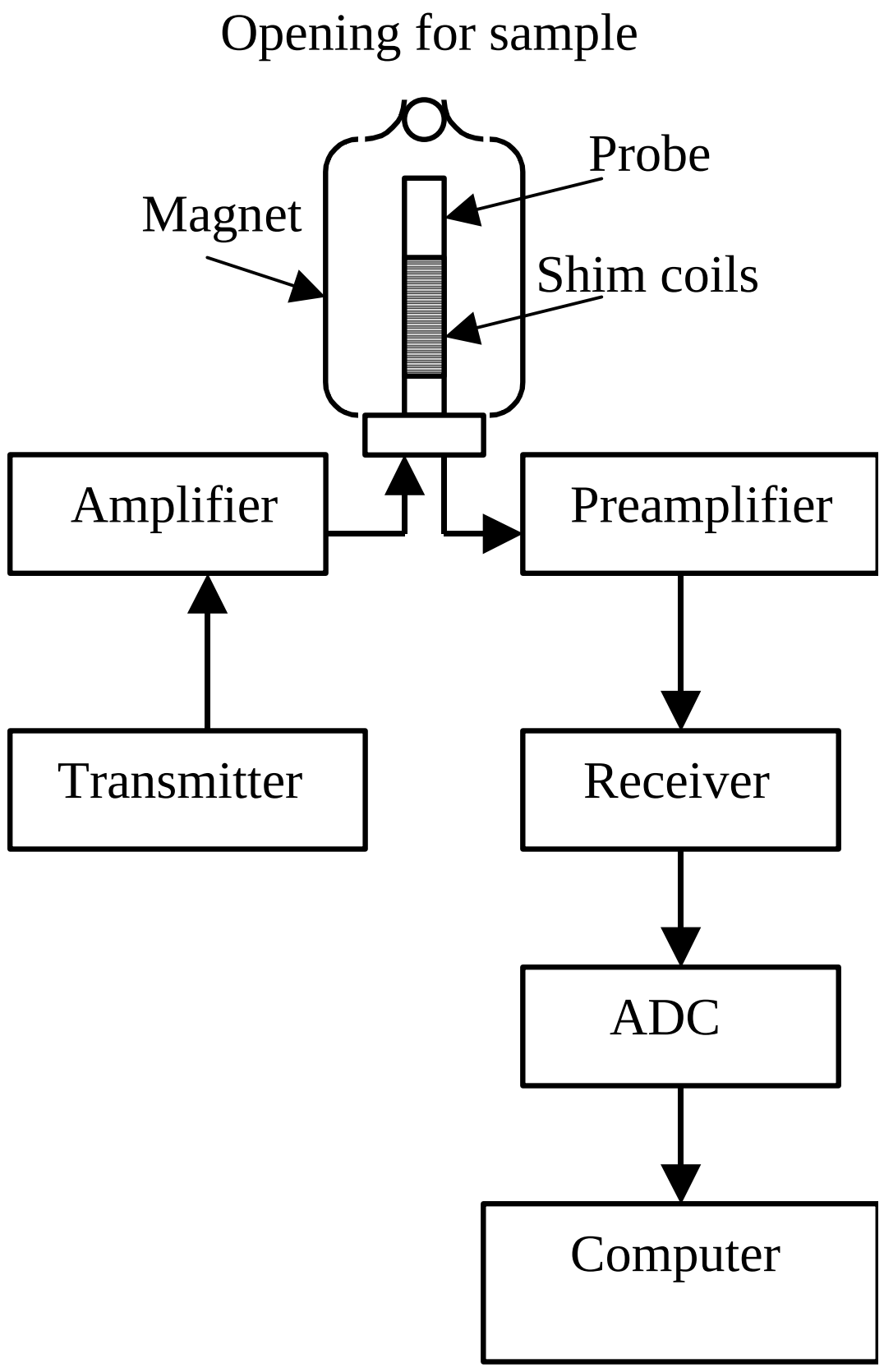

**Fig.4.1** Simplified block diagram of a pulse NMR spectrometer.

- The magnet produces a strong static magnetic field, which should be homogeneous and time-independent. It also houses the magnet bore and shim coil assembly, whose use allows improvement of the field homogeneity.
- The transmitter generates pulsed rf irradiation in several stages, which include the generation of a high-precision continuous radio-frequency signal in a rf synthesizer and manipulation of the signal phase and amplitude modulation of the continuous rf signal by the pulse programmer, allowing rf pulses to be delivered to the amplifier.
- The amplifier increases the power of the rf pulses produced by the transmitter to the level needed for broadband resonance excitation of the spins in the probe.
- The probe accomplishes several functions, which include, beyond the basic operations of irradiation of the sample by rf pulses delivered from the amplifier and detection of the resultant NMR signals, more specialised operations such as variable temperature control, sample spinning for averaging of transverse static magnetic field inhomogeneity, and any field gradient facility. As 3D imaging and automated shimming depend on the ability of either the probe or the shims to produce field gradient pulses, the field gradient facilities are described thoroughly in this chapter.

- The preamplifier magnifies the voltage of the NMR signal detected in the probe to a more convenient level, as the NMR signal induced in receiver coil of probe is intrinsically weak.
- The receiver for pulsed NMR normally uses a phase-sensitive detector (PSD)(4.2), whose output consists of $M_x$ and $M_y$ magnetizations.
- The analog-to-digital converter samples the NMR signal into digital form, required for processing by Fourier Transform and other digital signal processing algorithms.

The components, which are particularly important for 3D automated shimming are described in more detail in the rest of this chapter.

## 4.3 Superconducting NMR magnet

It is essential in high-resolution NMR that the static magnetic field generated by the magnet be strong, homogeneous and time-independent(4.3). Stronger fields allow better sensitivity and larger chemical shift dispersion, which simplifies analysis of complex spectra. High homogeneity of the magnetic field over the active volume of the sample is also essential, because spatial field inhomogeneity causes broadening of spectral lines, as was shown in Chapter 2. Normally, homogeneity of the order of one part per billion is required for high-resolution NMR. Here, high time-independence (or stability) of the field means that the field must not drift by more than about one part per billion during a routine experiment. Although in early NMR static magnetic fields were produced by permanent and electromagnets only, superconducting magnets are now used in high-resolution NMR due to the advantages explained in Chapter 1. Hence, the use of a superconducting magnet is assumed in this thesis. A superconducting magnet is illustrated in Fig.4.2.

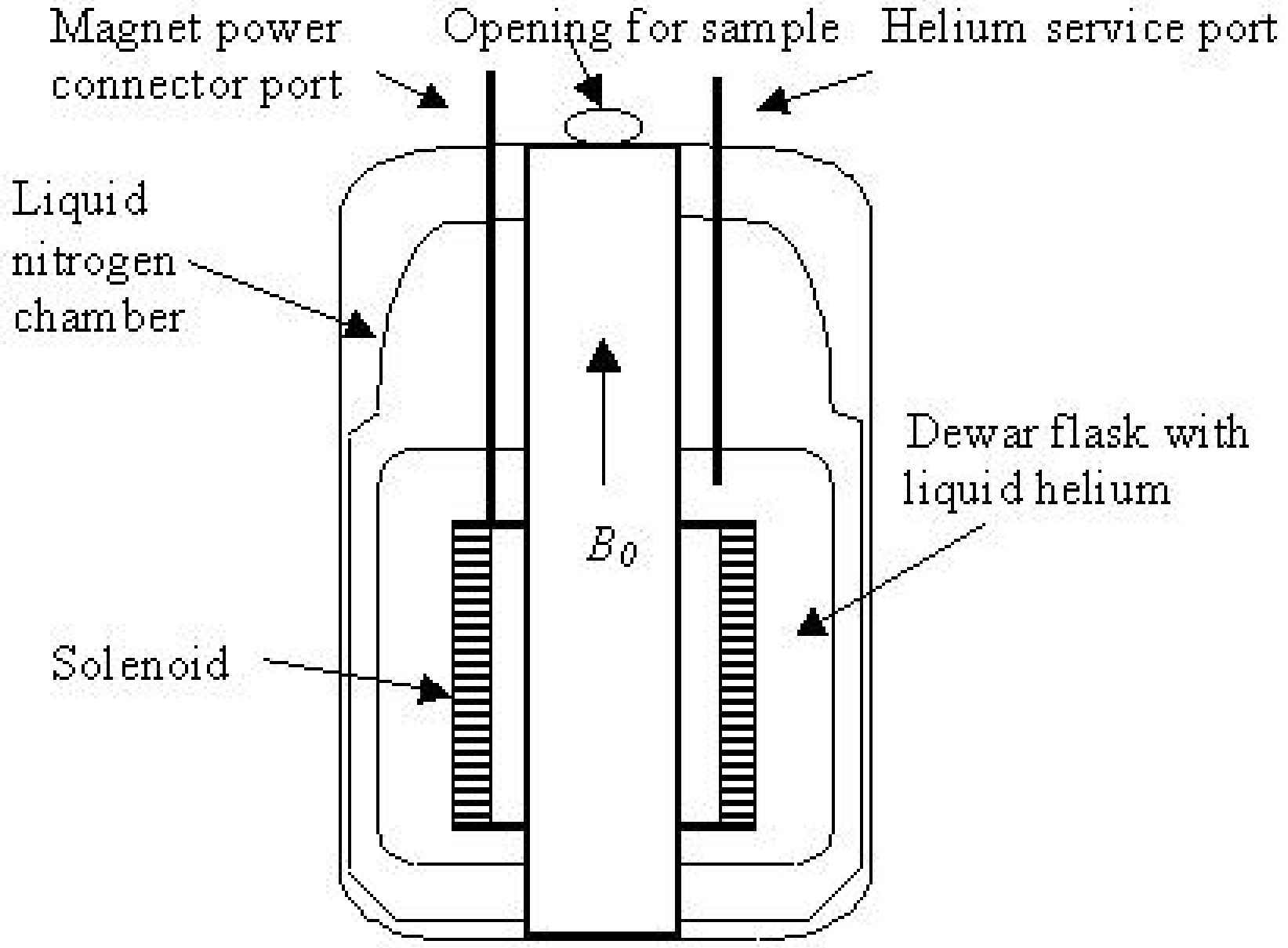

**Fig. 4.2.** Schematic representation of a superconducting magnet.

The solenoid is wound from a superconducting alloy, usually based on niobium, which is contained at the temperature of liquid helium in a Dewar flask, thermally isolated by an outer reservoir of liquid nitrogen from the outside environment.

## 4.4 Shim coils

The natural homogeneity of the field produced by a magnet is normally insufficient for high-resolution NMR; additional coils, called shims, are used to achieve the required field homogeneity of the order of one part per billion. As was mentioned in Part 1, shim coils of different geometries produce specific shapes of static magnetic field, which are used for compensation of the field inhomogeneity.

There are two sets of shims - the first is superconducting, and adjusted during the installation of the magnet; the second set, room temperature shims, are normally adjusted before each experiment by the user of the spectrometer. In routine operations only room temperature shims can be adjusted, and only their use is described here.

The shims are typically printed circuit coils, wrapped around a cylindrical former, which is placed inside of the magnet bore so that it embraces the part of probe that holds the sample. The set of shim coils with former is called the shim assembly. Generally, the number of shims required is larger for magnets with stronger field strengths and bore sizes. Normally an NMR magnet is equipped with a set of at least thirteen shims, as described in Section 3.2.1. Although in this thesis shimming using this minimum shim set is described, the 3D shimming procedure presented in the rest of the thesis is now being successfully applied elsewhere to much larger shim sets.

## 4.5 Pulsed field gradients

Although the effects of field gradients on NMR signals have been studied since the pioneering work of Gabillard(4.4) and the introduction of static field gradient coils(4.5), only since the introduction of pulsed field gradients(4.6) have gradually become a practical tool for NMR spectroscopy and imaging(4.7).

Normally, linear pulsed field gradients are used in imaging for spatial encoding of spins along $x, y$ and $z$ axes. The $x$ and $y$ linear gradients, referred to as transverse gradients, have strengths denoted $G_x$ and $G_y$ ; $G_z$ is the strength of the longitudinal gradient. The coils generating transverse gradients are normally Golay coils(4.8) or double saddle arrangements of four $120^0$ circular arcs arranged on a former(4.9) as shown in Fig.4.3. The design of a coil for a $x$ gradient is similar to that for a $y$ gradient, and can be obtained from latter by rotation through $90^0$.

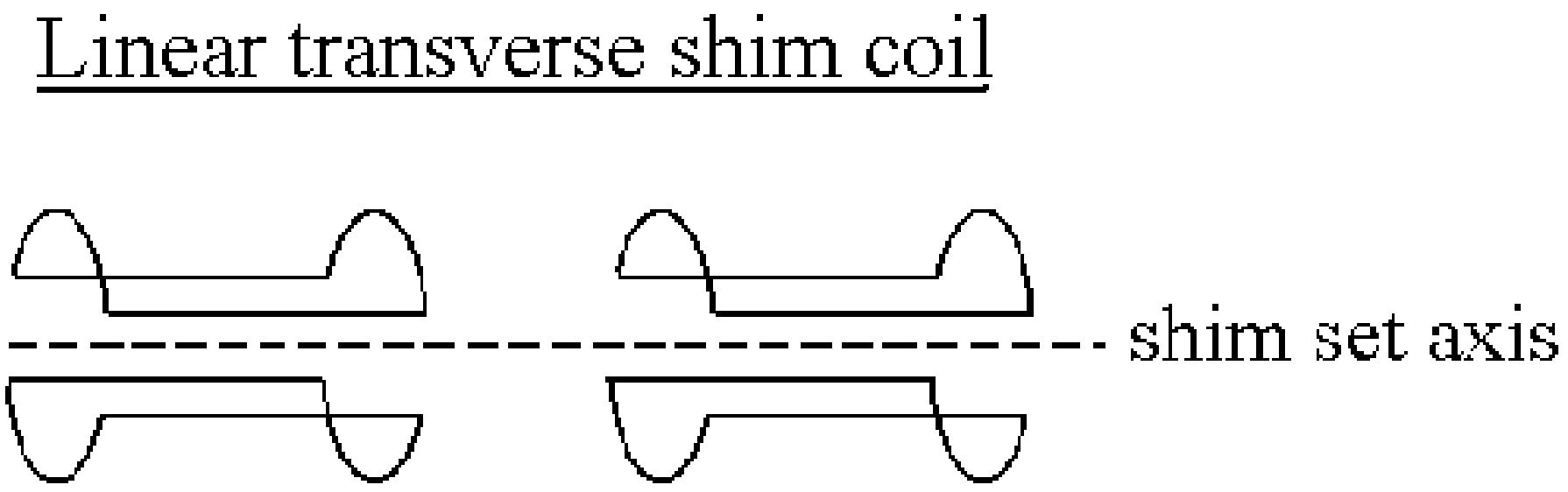

**Fig.4.3.** Double saddle arrangement of four $120^0$ circular arcs.

Transverse gradient coils are thus of similar design, arranged perpendicular to each other in the transverse plane.

A basic linear *z* gradient coil is a Helmholtz pair (Fig.4.4), consisting of two equal coaxial circular coils spaced along their axis. (In practice several pairs carrying different currents are combined to optimise gradient linearity.)

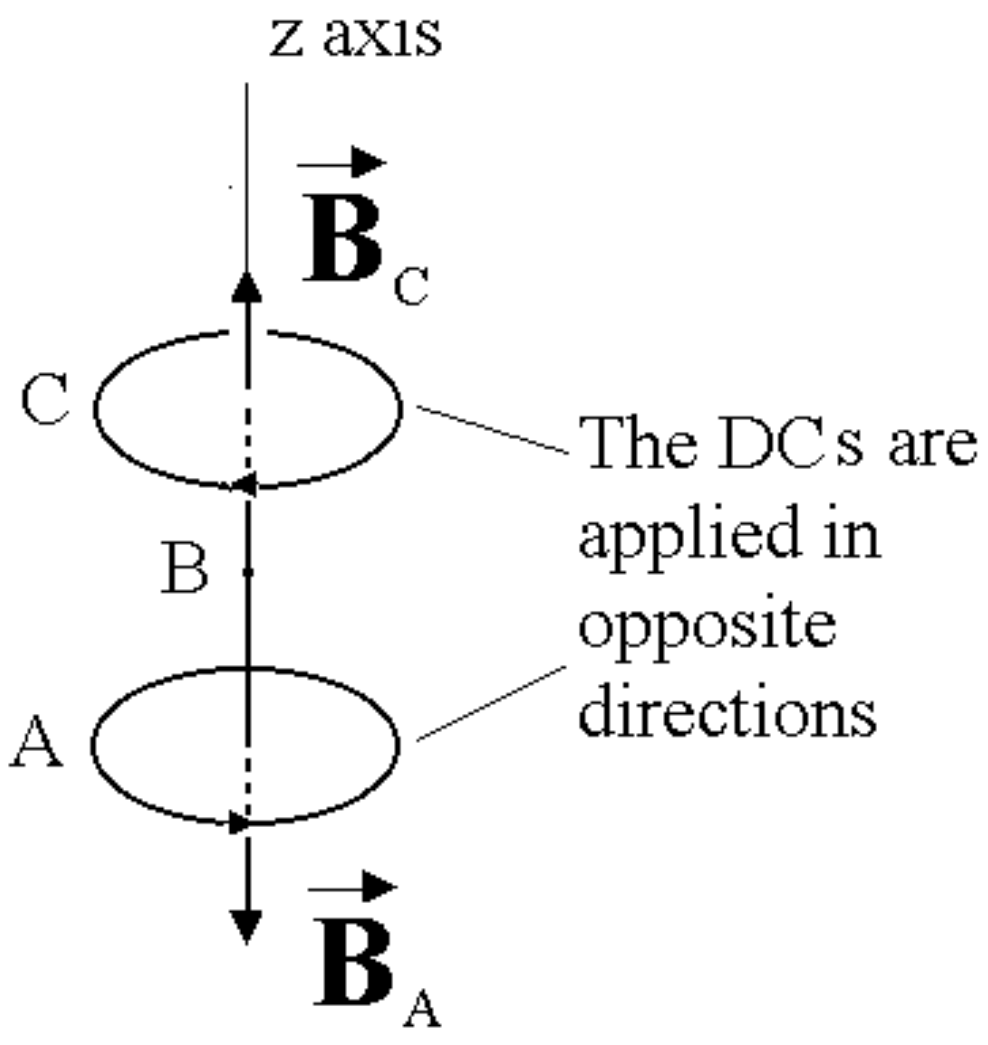

**Fig.4.4** Design of a *z* gradient coil. The coil consists of two parts, A and C, supplied by the direct currents (DCs) applied in opposite directions. Coil A produces a magnetic field $B_A$ which subtracts from the field $B_0$, generated by the magnet, and coil C produces field $B_C$ adding to field $B_0$, while the field at the centre (B) is equal to $B_0$. The variation of field from A to C is a linear gradient of the magnetic field.

Gradient coils can be characterised by their:

- Current efficiency, which is the ratio of gradient strength generated to current drawn.

- Switching time, the interval required for switching the field gradient between the on and off states.
- Gradient linearity, the deviation from linear of the field shape produced over specified sample volume.
- Power consumption, the power drawn by coil in operating mode.
- Induction of eddy currents, the coupling of the field produced with electrically conducting surroundings.

Two of these characteristics, switching time and induction of eddy currents, are especially relevant to 3D shimming. The switching time of a gradient coil depends on its inductance, and is a critical parameter for the timing of experiment. The inductance determines how rapidly a gradient coil may be switched by a given power supply. The ideal switching time can be defined as:

$$\tau = L \frac{I_{dc}}{V_{dc}} \tag{4.1}$$

where $L$ is the inductance, and $I_{dc}$ and $V_{dc}$ are the peak current and voltage delivered to the gradient coil by the power supply[(4.9)]. Though generally pulse field gradients can be of various shapes, in this thesis the use of rectangular gradient pulses is assumed.

### *4.5.1 Production of pulsed field gradients*

Generally, pulsed field gradients can be produced using:

- homospoil facility
- shim coils
- PFG modules

The homospoil facility and shim coils are available on almost all commercially produced NMR spectrometers, while PFG modules are a relatively recent development whose use is less common. In this thesis, spectrometers equipped with homospoil facility and shim gradients are referred to as NMR spectrometers with normal, as opposed to PFG, hardware.

PFG modules offer better control over field gradients than a homospoil facility or shim coils, and therefore are less restrictive on the timing of 3D shimming pulse sequences (although not all spectrometers allow control of shim coil currents in real time). However, as will be shown in the rest of the thesis, the modified PFGSTE pulse sequence used allows 3D automated shimming without the need for use of a PFG module, which can be an expensive addition to normal hardware. This sequence does not require rapid switching of gradient pulses and can use homospoil and shim gradients.

*4.5.1.1 Homospoil facility*

Homospoil (homogeneity spoiling) pulses are created by application of current pulses to the $z1$ shim coils in order to create field strength variation along the $z$ axis. Typically, the strength of a homospoil pulse is about $0.2 - 0.5\,G\,cm^{-1}$. The simplest application of a homospoil pulse is to dephase transverse magnetization when only $z$ component is of interest, for example in Hahn's early stimulated echo experiments[(4.10)].

Although a homospoil facility is available on almost all spectrometers and its applications are very useful, it also has limitations, which include long switching times, gradient shape distortions, phase instability, and disturbance of the field-lock system in high-resolution applications[(4.11)]. Nevertheless, homospoil pulses can be successfully used for spatial encoding of spins along $z$ axis in 1D automated shimming[(4.12)] and in 3D automated shimming with the PFGSTE pulse sequence.

*4.5.1.2 Shim coils*

Change of shim coil settings is the simplest way for user to create a field gradient over a sample. Not all instruments have as a standard the capability to control shim currents in real time during the application of a pulse sequence. Even without real-time control, shim gradients can be useful when a field gradient of a particular shape is required throughout an entire experiment. This is used in the shim mapping experiment, where the field shapes generated by shim coils are mapped. Besides lack of real-time control, shim coils are also not designed for rapid switching and are not actively shielded to reduce eddy currents. Hence the disturbances to the main field produced by changes in shim coil gradients take a considerable time to die down. In spite of these shortcomings, shim gradients can be used when real-time control is not crucial and the pulse sequence allows leisure for eddy currents[4.12] to decay. The INOVA series of spectrometers used in this work allow partial real-time control of $x1$ and $y1$ shims, used in this thesis for transverse phase encoding.

*4.5.1.3 PFG modules*

The Pulsed Field Gradient (PFG) module is a big improvement on other facilities for production of pulsed field gradients. It offers real-time programming of the field gradients during pulse sequences, and the strengths of the PFG pulses are typically larger than for homospoil pulses; common commercial systems have maximum gradient strengths between 10 and 100 $G\,cm^{-1}$. The switching times of PFG pulses are typically tens of microseconds, much shorter than for homospoil and shim gradient pulses. Thus PFG modules are more suitable for applications where the timing of a pulse sequence is critical. Another advantage of PFG modules is better reproducibility of field gradients for some modules, and a capability to produce shaped field gradient pulses. As this thesis mainly concerns the development of 3D automated shimming for spectrometers with normal hardware, PFG modules are not considered further here unless specially stated.

*4.5.2 Imperfections of the linear transverse field gradients*

In order to get an image of a sample with minimum, or ideally - without any distortions, the spatial encoding linear gradients must be produced with minimum errors. However, because of imperfections in the design and manufacture of gradient coils, common errors for linear gradients are[4.13]:

1. Imbalance in the gradient amplitude
2. Non-orthogonality of $x1$ and $y1$ gradients

These errors result in distortions of image shape and are undesirable. It will be shown in Chapter 6 that the errors can be compensated for appropriate calibration of $x1$ and $y1$ gradient strengths. In this experiment the linear gradient imbalance and non-orthogonality are measured in order to calibrate the linear transverse gradients used in the described 3D automated shimming technique.

## 4.6 Use of probe lock channel for experiments with deuterium

Since almost all samples used in high-resolution NMR have deuterated solvents, suitable deuterium signals are more commonly available for shimming than proton. Experiments on deuterium can be performed by observation of its NMR signal through either the probe lock channel, or through an observe coil tuned to deuterium. In this work both methods were used. Originally, the deuterium lock channel was introduced by Anderson as a solution of the field stability problem[4.14]. It uses as a feedback curcuit which compensates time variations of the static magnetic field strength by changing the $z0$ shim current. It normally works in background, without interference with main experiment on another nucleus or nuclei, and uses the solvent deuterium resonance. The lock channel is also routinely used for manual shimming, where field homogeneity is adjusted either by maximising the strength of the deuterium lock signal or by optimising the shape of its FID signal. The magnetogyric ratio of deuterium is about 6.5 times smaller than for protons, and although it facilitates simultaneous observation of their resonances without interference (as their Larmor frequencies are well spaced), this also results in lower sensitivity[(4.15)], which is approximately proportional to $\gamma^{5/2}$. In most high-resolution NMR spectrometers, the lock coil is actually also the proton coil, which is double tuned to the Larmor frequencies of proton and deuterium; thus sensitivity is reduced for deuterium since the probe is optimised for protons.

In summary, though the use of the lock channel allows 3D gradient shimming on deuterium, its important drawbacks are:

- Intrincically low signal-to-noise ratio of deuterium
- Reduced sensitivity of lock coil

On high-field instruments the resulting sensitivity is still adequate, but on low-field instruments (as used here) it is preferable to use broadband probes designed for observation of deuterium (as well as proton) resonance, as described in the following section. These can be partly resolved by use of broadband probes designed for observation of deuterium (as well as proton) resonance and described in the following section.

## 4.7 Use of broadband probe for X channel experiments with $^2$H

Since its discovery, NMR of the proton has been studied far more than that of any other nucleus, because of its high sensitivity, high natural abundance, and presence in the majority of the molecules of interest[(4.16)]. Thus, commercial NMR instruments were designed mainly for proton studies until the early 1970s, when the possibility of broadband pulse FT NMR observation was demonstrated[(4.17)]. The subsequent progress in design of commercially available broadband NMR probes made multinuclear NMR, including deuterium resonance, routine.

In this thesis, 3D automated shimming experiments on deuterium were tested using both the lock and deuterium channels of a dual $^1$H/BB probe. The results of these experiments, presented in Chapter 6, show considerable improvement in the signal- to-noise ratio of profiles, field and shim maps acquired through the deuterium channel of the dual probe, which results after shimming in a better homogeneity of the static magnetic field.

## 4.8 Experimental limitations

In spite of big advances in design, NMR instruments still suffer from the intrinsic weakness of NMR signals and other unfavourable experimental conditions. Some of these, which may result in difficulties with applications of 3D automated shimming in practice, are described in more detail here.

*4.8.1 Low signal-to-noise ratio*

The sensitivity of a spectrometer can be described as its ability to detect weak signals. This is limited by random noise present in NMR instruments, and can be described by the signal-to-noise ratio (SNR). There are many factors which influence SNR in the NMR experiment. They include the magnetogyric ratio $\gamma$ and strength of the static magnetic field $B_0$ (4.18):

$$SNR \propto |\gamma|^{5/2} B_0^{3/2} \qquad (4.2)$$

Hence, the SNR of deuterium is more than one hundred times smaller than for protons, observed with a similar receiver and the same magnetic field strength(4.15). At low fields, this SNR penalty makes the use of a broadband probe channel desirable. Although 3D phase shimming on deuterium through the probe lock channel is possible, its use in high-resolution applications at low fields require long time-averaging for optimal results.

### *4.8.2 Thermal convection*

Thermal convection in liquid state NMR is caused by temperature gradients, which make parts of the liquid within the sample move upwards and the remainder downwards[(4.19)]. Thermal convection occurs most readily in liquid samples with low viscosity; in a cylindrical sample, convection occurs when the temperature gradient satisfies the condition[(4.19)]:

$$\frac{dT}{dz} > \frac{k\nu}{g\alpha}\frac{R}{r^4} \tag{4.3}$$

where:

$\alpha$ - coefficient of thermal expansion,

$k$ - thermal diffusivity,

$\nu$ - kinematic viscosity,

$g$ - acceleration due to gravity,

$r$ - diameter of tube and $R$ is Rayleigh number.

The effect of thermal convection is deleterious (except for the thermal convection studies!) in experiments where gradients are used for spatial encoding of spins, since it introduces an extra phase variation in signals, and prevents the complete refocusing during the read-out gradient, described in Chapter 3. In this work, a strong effect of thermal convection was observed during some tests of 3D shimming with standard 5 mm line shape test samples. These results, presented in Chapter 6, show that field and shim maps acquired with these samples are heavily distorted and the subsequent 3D shimming does not converge to optimum shim values.

*4.8.3 Spin relaxation*

Spin relaxation drives spins towards the equilibrium state present before application of rf pulses. There are two different types of relaxation, spin-spin (transverse) and spin-lattice (longitudinal):

1. Transverse relaxation is a loss of microscopic phase coherence between spins, for example due to spin-spin interactions, characterized by a time constant $T_2$. The net dephasing of spins in the transverse plane due to the effect of relaxation and static magnetic field inhomogeneity is sometimes approximated by a time constant $T_2^*$.
2. Longitudinal relaxation describes exchange of energy between spins and their surroundings (the lattice), and is characterized by the spin-lattice relaxation time $T_1$.

The rate of spin dephasing due to field inhomogeneity can be represented by[(4.20)]:

$$\frac{1}{T_2^*} \cong \frac{1}{T_2} + \gamma \Delta B \qquad (4.4),$$

where $\Delta B$ is described in Chapter 2, Section 2.3.

It is assumed in the theoretical analysis carried out in Chapter 3 that the effects of relaxation can be neglected. However, this is correct only when the relaxation times of the sample are long enough that phases of spins are governed mainly by field gradients. Thus samples with very short relaxation times are unsuitable for mapping with the PFGSTE pulse sequence, which requires long time delays for decay of the field gradients and eddy currents produced with normal hardware. At the other extreme, the use of samples with very long relaxation times increases the duration of experiment, making it less practical. Hence, it is desirable for 3D mapping to use samples with relaxation times, which are:

1. Larger than or comparable to the duration of the PFGSTE pulse sequence, to minimise loss of magnetization due to relaxation processes during the pulse sequence.
2. Not sufficiently long that 3D shimming becomes too time consuming and impractical.

Hence, making a choice of sample for 3D shimming is a matter of compromise on relaxation times. The 5% PEO (poly-ethylene oxide) sample used for $^1H$ 3D shimming in this thesis was chosen for its strong singlet signal and its relaxation times about half second.

*4.8.4 Eddy currents*

Practically important drawbacks in the use of switched field gradients are the transient effects which follow rapid gradient switching. Sudden changes in field interact with the conductive materials of probe and magnet assembly to produce eddy currents[(4.21)]. The eddy currents, in turn, induce magnetic fields opposing those of the switched gradients, and disturbing the field. This results in prolonged field disturbance after switching, which can influence the NMR signal. For example, if the fields due to eddy currents from field gradient switching in the phase encoding period extend into the evolution period or acquisition time in mapping experiments, then the resultant phase error will distort the field map.

A highly effective way to reduce eddy currents, as proposed by Chapman and Mansfield[(4.22)], is to add shield coils to the gradient coils to cancel the magnetic fields induced outside the gradient coil pair and prevent formation of eddy currents. Although the current through the shielding coil disturbs to some extent the field produced inside the gradient coil, the advantage is that the error fields due to switched gradients can be reduced to about 1% of the unshielded values[(4.21)].

# 5

# Software for 3D Automated Shimming

## 5.1 Programming in VNMR Magical and C languages in Unix OS

Modern NMR spectrometers have diverse tools for the programming of pulse sequences and writing of signal processing software. The Varian spectrometer software allows(5.1)

- Macro programming by the use of the language MAGICAL
- Programming in the C language

MAGICAL (MAGnetics Instrument Control and Analysis Language) is a high level interpreter language designed for the programming of NMR experiments. It provides a full range of programming tools, which include statements, loop and conditional operators, and tools for access to NMR information. MAGICAL can also use the parameters of NMR experiments in data acquisition and processing software. When a macro written in MAGICAL is run, a command interpreter translates the high level MAGICAL programme into an intermediate form, which is then executed(5.2). It can lead to slow execution, which limits the use of MAGICAL in practice. Hence MAGICAL is usually used for writing small to medium size programmes for the control and processing of NMR experiments.

For the writing of the large and time-critical programmes, the use of the C language, originally designed by Ritchie(5.3) for the UNIX operating system, is advantageous. C is a high-level language whose instructions can be compiled into machine language, and therefore generally run faster than an interpreter. Other advantages of C include the availability of diverse libraries for mathematical calculations, and highly developed facilities for low-level programming, including a programme interface with Assembler.

## 5.2 Introduction to software for 3D automated shimming

Since $z$ inhomogeneity is usually the largest for the unshimmed static magnetic fields, automated $z$ shimming is normally performed before beginning 3D automated shimming. The use of $z$ shimming before 3D automated shimming software is particularly helpful when starting with shims set to zero ("cold" shims), for example for shimming of the static magnetic field of new magnets. *Z* shimming can be accomplished by the use of the **gmap** family of macros[5.4], which are part of the standard VNMR software installed on the Varian spectrometers. After $z$ shimming, initial 3D automated shimming begins with calibration of the *x1* and *y1* gradient strengths. The 3D automated shimming methods presented in this thesis are run in automation by the **gxyz** family of macros, which perform the operations after $z$ shimming as presented by the flowchart of Fig.5.1.

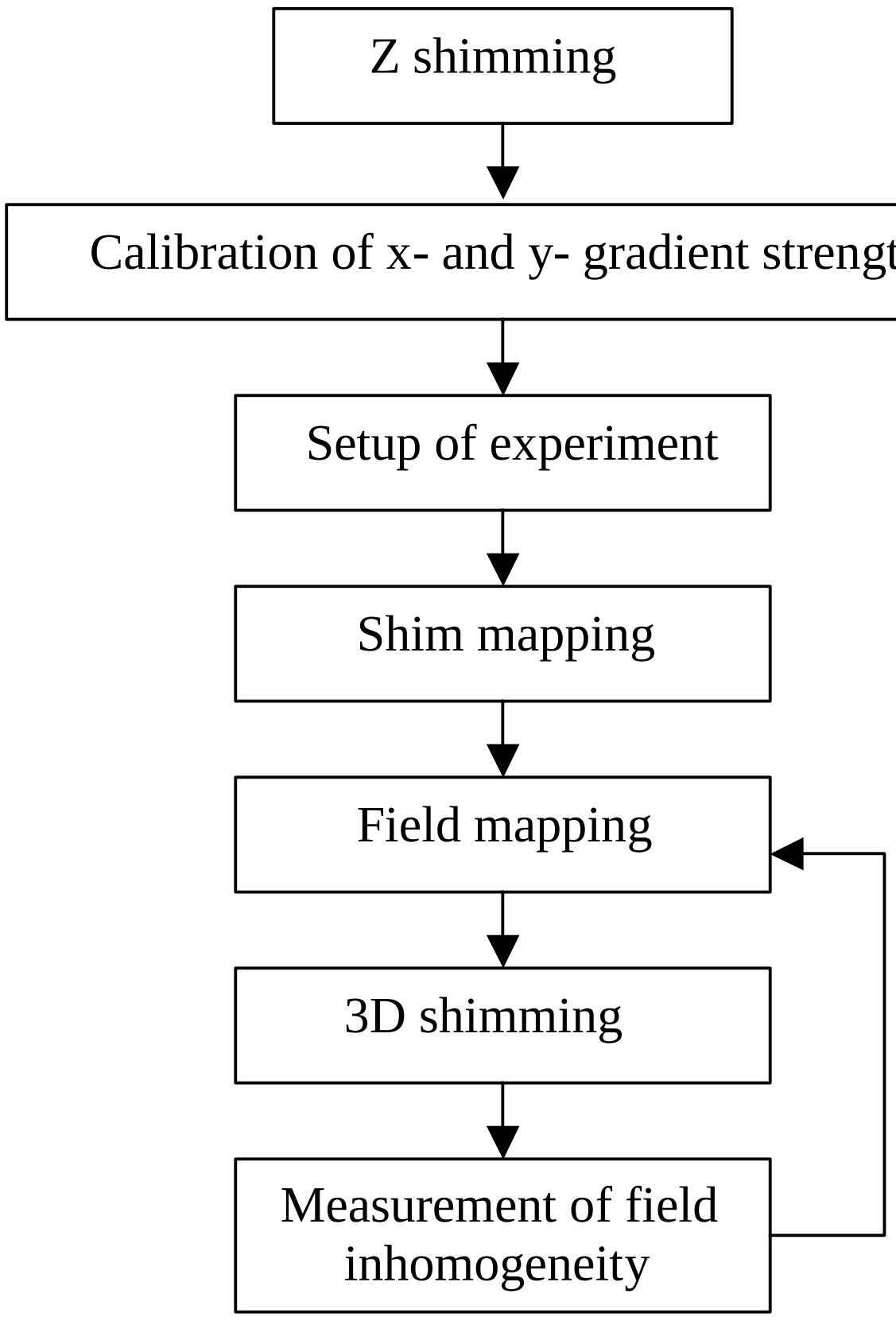

**Fig. 5.1** Simplified block diagram of the 3D automated shimming operations.

## 5.3 Calibration of $x1$ and $y1$ shim gradient strengths

The procedure presented here corrects imbalance of $x1$ and $y1$ shim gradient strengths as described in Chapter 3. It performs the following operations:

- Measurement of profile widths under a rotating gradient
- Calculation of the calibration parameters needed to produce orthogonal $x$ and $y$ gradients

The imbalance of $x1$ and $y1$ gradient strengths can be determined by measuring profile widths of spectra acquired with arrayed values of $x1$ and $y1$ shims, as described in Chapter 3. This experiment can be accomplished using either proton or deuterium. The values of $x1$ and $y1$ shims are arrayed by the macro **gxysweep**, as follows (for detail see Appendix C):

1. Reads the current values of $x1$ and $y1$ shims and assigns these to local variables:

   **$x1init=x1   $y1init=y1**

2. The gradient strength increment and number of steps for gradient rotation in the $xy$ plane are input arguments, assigned to the variables **gxystep** and **$n**, respectively. When arguments are not specified it uses default values equal, respectively, to 100 DAC points and 16 increments.
3. Macro checks that the values of $x1$ and $y1$ shims are within the allowed range (from –2048 to 2047 DAC points for the 12-bit shim module used here).
4. The $x1$ and $y1$ shim settings are arrayed in the loop:

```
$i=0   $th=0.0
repeat
      x1[$i]=$x1init+gxystep*cos($th)
      y1[$i]=$y1init+gxystep*sin($th)
      $th=$th+3.141592654*2.0/$n
      $i=$i+1
until $i>$n
```

When the experiment is completed, the set of spectra acquired with the arrayed values of $x1$ and $y1$ is processed by the macro **gxycal**, which reads the profile widths and then uses the C programme **calibxy.c** to fit the parameters for the model function (presented in Table 5.1) to the measured profile widths.

**Table 5.1 The parameters of $x1$ and $y1$ gradient calibration**

| Name of parameter | Description of parameter |
|---|---|
| **gcalx** | $x1$ gradient strength (represents α), $\left(G\,cm^{-1}\right)/$DAC points |
| **gcaly** | $y1$ gradient strength (represents β), $\left(G\,cm^{-1}\right)/$DAC points |
| **gcalang** | Angle between $x1$ and $y1$ gradients (represents θ), degrees |
| **xerr** | Estimated error in $x1$ shim setting (represents $x_0$), DAC points |
| **yerr** | Estimated error in $y1$ shim setting (represents $y_0$), DAC points |

The programme **calibxy.c** uses the Marquardt-Levenberg algorithm[(5.5)] for nonlinear least squares fitting[(5.6)]. This searches for the values of **gcalx**, **gcaly**, **gcalang**, **xerr** and **yerr**, which minimize the expression:

$$\sum_{i=0}^{N}\left[\nu_i^{\exp}-\nu_i^{calc}\left(gcalx, gcaly, gcalang, xerr, yerr\right)\right]^2 \qquad (5.1),$$

where $\nu_i^{calc}\left(gcalx, gcaly, gcalang, xerr, yerr\right)$ denotes model function and $\nu_i^{\exp}$ are the profile widths measured for the $N$ values of the arrayed shims. The model function for the fitting programme is described Chapter 3. The macros **gxysweep**, **gxycal** and programme **calibxy.c** are listed in Appendices C and D, respectively.

The calibration of $x1$ and $y1$ gradient strengths is a standalone procedure, normally performed before 3D shimming. Since the accuracy of calibration increases with improvement of the static magnetic field homogeneity, it can be useful to repeat the calibration after the first use of 3D shimming.

### 5.4 Parameter set up for field mapping experiments

Before beginning using 3D automated shimming, the parameters for experiments are set up for use by the pulse sequence, acquisition and processing macros. The C programme for the modified PFGSTE pulse sequence, **gmapxyz.c**, is presented in Appendix E. The pulse sequence parameters include types of gradients and nucleus, 90 degree pulse width and corresponding transmitter power level, spectral width, etc[5.7]. The values of $z0$ and transmitter offset are adjusted to bring the lock signal and nucleus of interest, respectively, to resonance. The parameters used by the acquisition and processing macros include the names of the shims to be mapped, the steps by which the shim values are incremented during mapping, field-of-view, arrayed delay $\tau$ (described in Chapter 3), the width of frequency domain window, and number of phase encoding increments. Shim mapping experiments require setting up of the shim arrays, which specify the shim changes for which fieldmap is required.

## 5.5 Shim and field mapping software

After the parameters of shim mapping experiment are set up, the PFG STE sequence is run with the arrayed shims, delay $\tau$ and the phase-encoding increments. The duration of the experiment is given by:

$$t_{\exp} = t_{ps} \times \mathbf{nt} \times \mathbf{arraydim} \qquad (5.2),$$

where $t_{ps}$ denotes the duration of the pulse sequence, **nt** is the number of transients and **arraydim** is the dimension of experiment, equal to product of the sizes of the arrays(5.8). For a shim mapping experiment, the value of **arraydim** is given by:

$$\mathbf{arraydim} = 2 \cdot (N_{shims} + 1)\mathbf{ni}^2 \qquad (5.3),$$

where **ni** is the number of phase-encoding increments in each of $x1$ and $y1$, and $N_{shims}$ is the number of shims. The operations performed by the field map processing software are summarised in Fig.5.2.

The result of a shim mapping experiment is a set of spectra, which encode the field distributions, for each value of the arrayed phase-encoding gradients and combination of shim settings and delay $\tau$. A field mapping experiment produces similar spectra for a single set of shim values. When acquisition is completed, the data, identified by parameter **mapname**, are used for processing by the Vnmr command **gxyzmap**, programmed in C, which first forms the data matrix data[**arraydim**][**np**], where **np** is the number of complex points in the acquisition domain.

A discrete 2D Fourier transform then is performed separately for the two subsets of the data acquired with different values of the $\tau$ delay. The results are used for calculation of:

- Amplitude maps
- Phase differences between the two images acquired with different values of $\tau$ delay.

Shim maps are calculated by subtracting a pair of field maps, measured with different shim values.

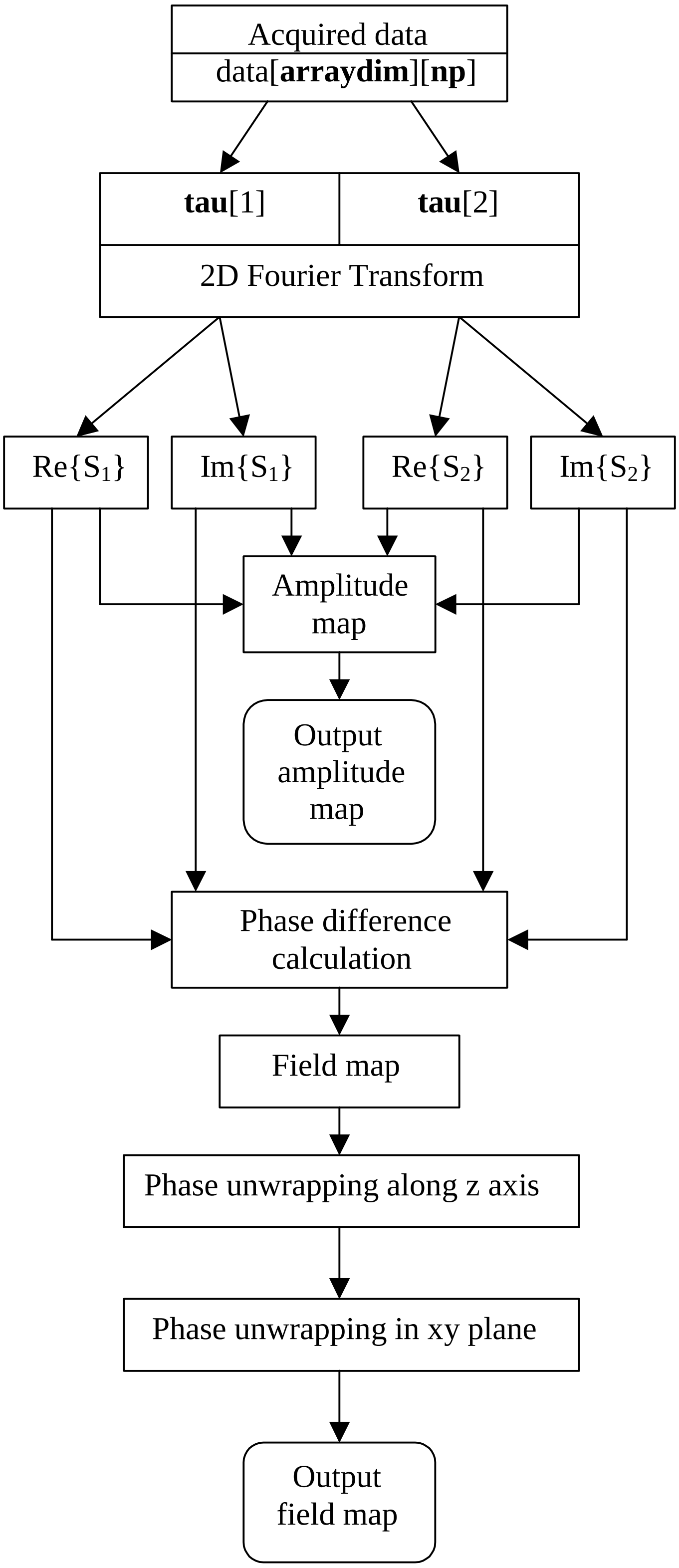

**Fig.5.2** Operations performed by the field map processing software.

If the calculated phase changes between adjacent points in the $x$, $y$ and $z$ directions exceed $\pi$ radians, this is corrected (as described in Chapter 3) by the use of a procedure called phase unwrapping. After this, the results representing amplitude, field and shim map data are saved in separate files.

## 5.6 Software for 3D shimming

Shimming is the optimisation of the shim settings to improve static magnetic field homogeneity. The flow chart of operations performed in an iteration of the 3D automated shimming software is presented in Fig.5.3

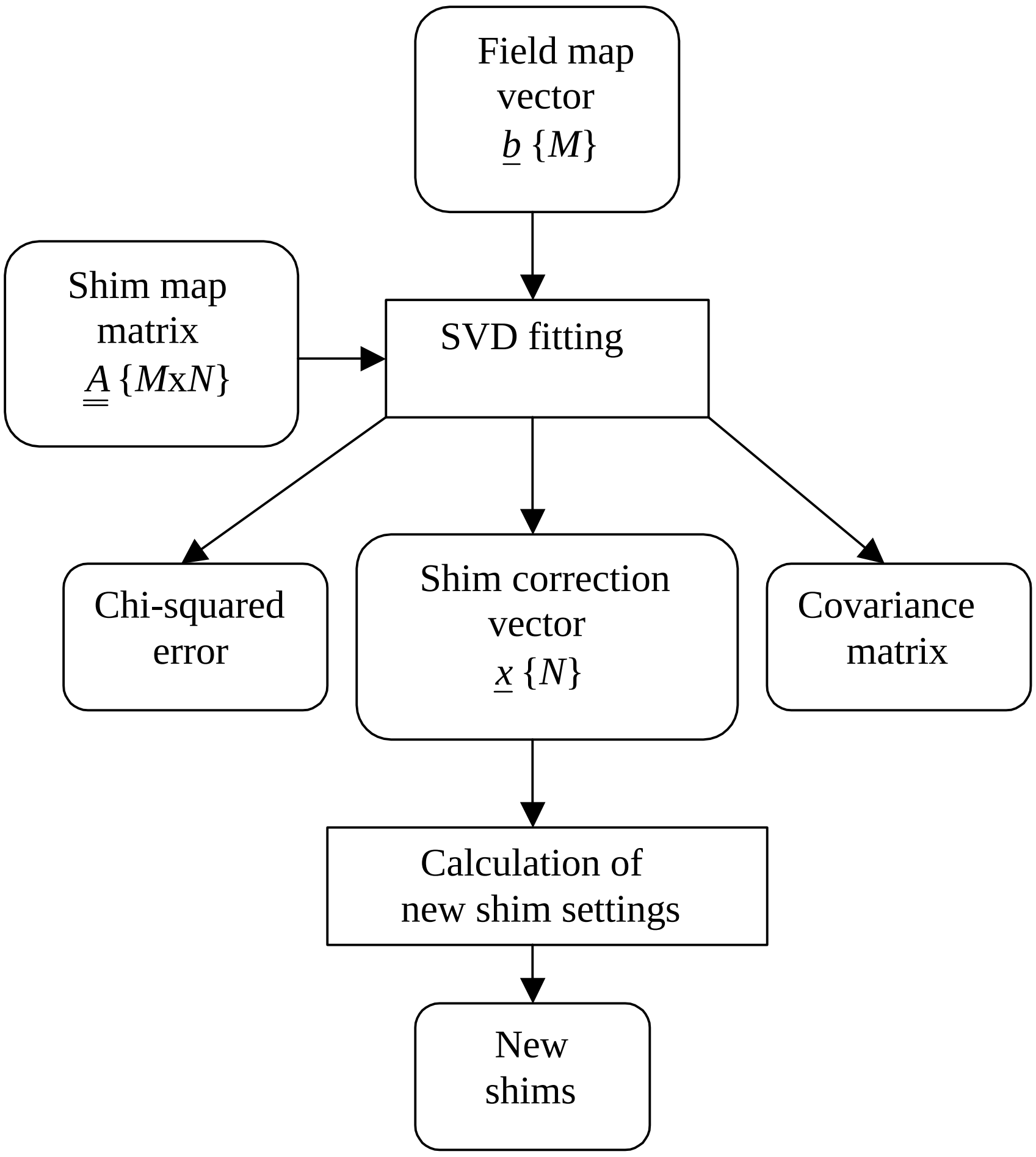

**Fig.5.3** Flow chart of operations performed by 3D shimming software. Rectangle and rounded blocks represent, operations and data, respectively.

The set of shim corrections that is sought can be represented as a vector $\underline{x}$ of length *N*, which is equal to the number of shims. The shim map data are assembled into a matrix $\underline{\underline{A}}$ of size *M*x*N*, where *M* is the number of data points in a fieldmap. The field map data are represented by vector $\underline{b}$ of length *M*.

The shim and field maps are used as input data to the Vnmr command **gxyzfit**, which calculates shim corrections by linear least squares fitting of the field map data with the stored shim maps. The SVD procedure (see Section 3.3.2 of Chapter 3 for detail) begins with the estimation of measurement error (standard deviation) from the input data. It assumes that all measurements have the same standard deviation $\sigma_i = \sigma$ and the model function fits to the experimental data well(5.9). The standard deviation is calculated in the following order:

1. Standard deviation is assumed uniform and initially assigned to 1.
2. Optimal model function parameters $x_0$ are found by least squares SVD fitting with the initial standard deviation, which minimises:

   $$\chi^2(\underline{x}) = \sum_{i=1}^{M} \left(\underline{\underline{A}}\underline{x} - \underline{b}\right)^2 \quad (5.4),$$

   where $\underline{\underline{A}}$ and $\underline{b}$ are, respectively, the shim and field maps and $\underline{x}$ is unknown vector whose optimum $x_0$ is searched.
3. Calculation of the squared standard deviation, which corresponds to the parameters $\underline{x_0}$ optimised in previous step by use of the formula:

   $$\sigma^2 = \frac{\sum_{i=1}^{M} \left(\underline{\underline{A}}\underline{x_0} - \underline{b}\right)^2}{M} \quad (5.5).$$

When the standard deviation is found, the shim corrections are calculated by the least squares SVD routine which minimises:

$$\chi^2(\underline{x}) = \sum_{i=1}^{M} \left( \frac{\underline{\underline{A}}\underline{x} - \underline{b}}{\sigma} \right)^2 \quad (5.6).$$

The results of the fitting include a vector of shim corrections $\underline{x}$, a value for chi-square $\chi^2$, and a covariance matrix. The new shims are calculated by subtracting the shim corrections found from the current shim settings.

## 5.7 Estimation of static magnetic field homogeneity from spectral line shape

The homogeneity achieved after shimming of the static magnetic field can be estimated from the line shape of spectrum acquired with the new shim settings. The widths of a singlet line with long $T_2$ corresponding to three different amplitude levels (at 50, 0.55 and 0.11%) are measured. As well as providing a criterion for acceptability of field homogeneity, the line width data are also used to set the change in delay $\tau$ to be used for the next shimming iteration. If the line broadening due to inhomogeneity is still larger than required, the field mapping and shimming procedure is repeated. In this thesis the value of $\tau$ was typically set to the inverse of the line width in Hz units, measured at 0.55% of the amplitude of a singlet line. It will be shown in Chapter 6 that several iterations of 3D field mapping and shimming are normally required before the desired field homogeneity can be achieved.

# 6

# Results

This chapter represents the results of 3D automated shimming for the nuclei $^{1}H$ and $^{2}H$, obtained on Unity INOVA 300 and 400 NMR spectrometers with normal hardware. The modified PFGSTE sequence described in Chapter 3 was used for 3D shimming. This sequence allows enough time for switching shim and homospoil gradients and therefore can be used with normal hardware.

Before starting 3D shimming, the transverse gradient calibration is normally performed as described in Chapter 3 and in Section 6.1 of this chapter. The results of experimental parameter optimisation in the 3D shimming procedure are presented in Section 6.2. Detailed, step-by-step descriptions of typical $^{1}H$ and $^{2}H$ 3D automated shimming experiments are presented in Sections 6.3 and 6.4, respectively. The effect of thermal convection on 3D shimming is described in Section 6.5, which completes this chapter.

## 6.1 Calibration of linear transverse shim gradients

The orthogonality of the gradients, produced by the $x1$ and $y1$ shims was investigated using the procedure described in Chapter 3. The $^{1}H$ standard line shape sample (1% $CHCl_3$ in 99% acetone-$d_6$) was used in this experiment. A sequence of 16 experiments (Fig.6.1) with arrayed values of $x1$ and $y1$ shims was performed, where a series of $^{1}H$ spectra was acquired using a $90^0$ pulse for a transverse linear gradient, rotated through different angles $\phi$. Fig.6.1 shows the spectra obtained for values of $\phi$ varied from 22.5 to 360 degrees with a constant increment of 22.5 degrees; a normal $^{1}H$ spectrum, acquired without application of $x1$ and $y1$ gradients, is shown at the right.

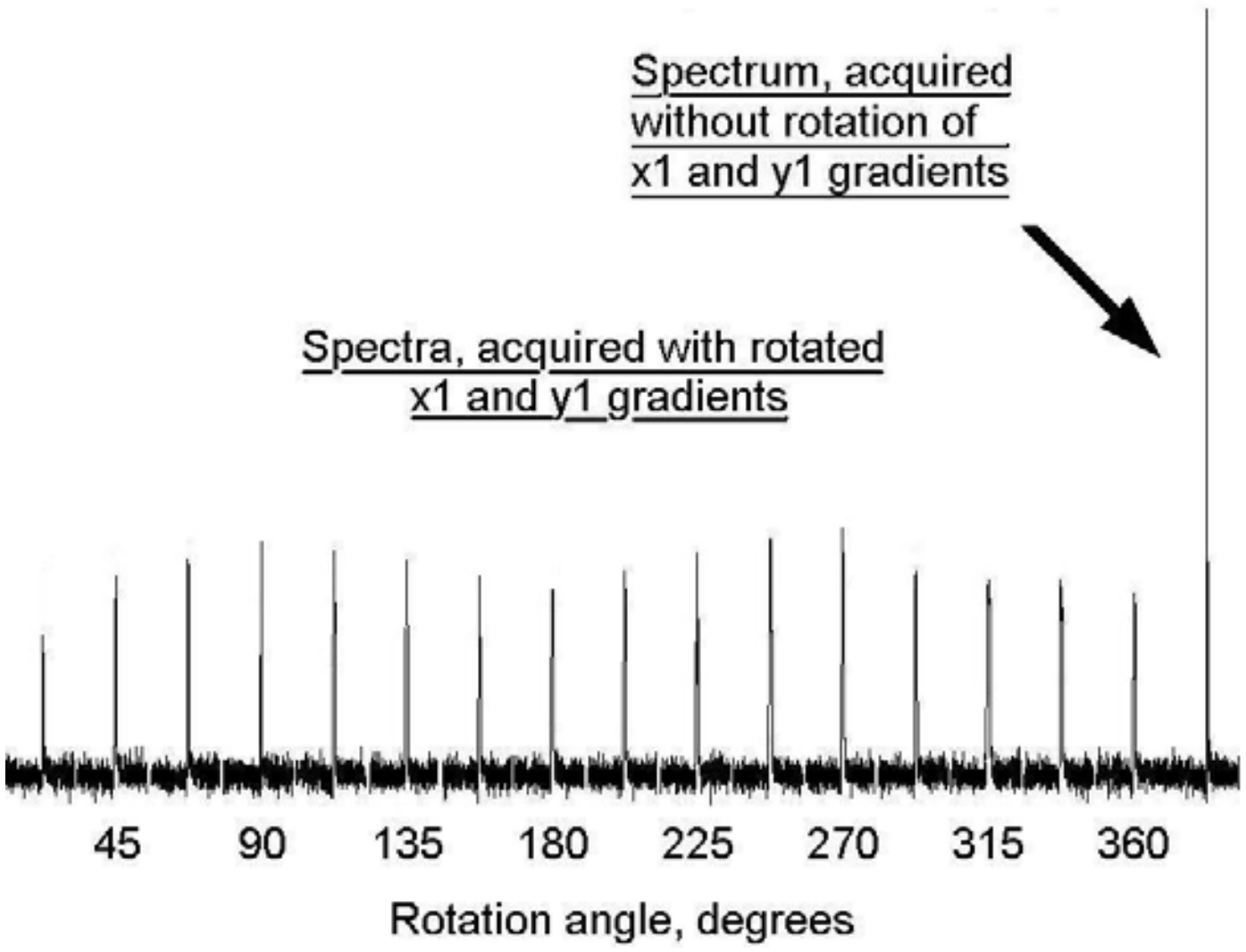

**Fig.6.1** $^{1}H$ profiles, acquired with linear transverse gradient rotated through angles from 0 to 360 degrees. When $x1$ and $y1$ gradients are imbalanced their amplitudes (and widths) change unevenly for different values of rotation angle, as shown.

The application of a field gradient broadens the NMR spectrum by a factor proportional to the gradient strength. When the $x1$ and $y1$ gradients are of different strength or are not orthogonal, the amplitudes and widths of profiles change with variation of the rotation angle, as described in Chapter 3 and shown in Figs.6.1 and 6.2, respectively. The relative strengths of linear transverse gradients and the angle between them were determined by fitting the experimental data for the width of the signal profile to the theoretical expression, with the result shown in Fig.6.3 After fitting, the parameters were used for calculating the $x1$ and $y1$ shim values.

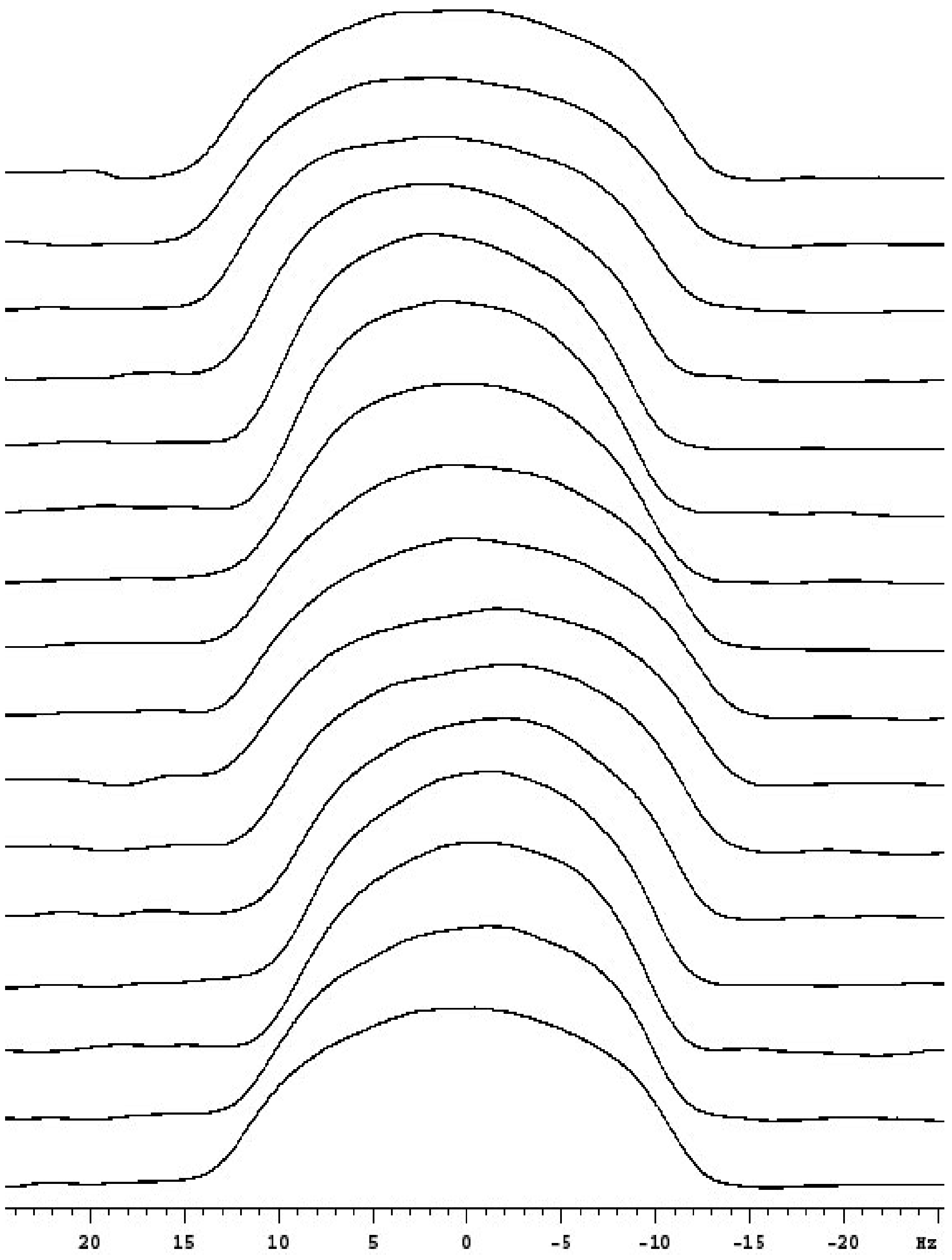

**Fig.6.2** The sample profiles obtained in the $x1$ and $y1$ gradient calibration experiment. The nominal rotation angle increases for profiles from bottom to top, as described in the text. These profiles were processed using line broadening.

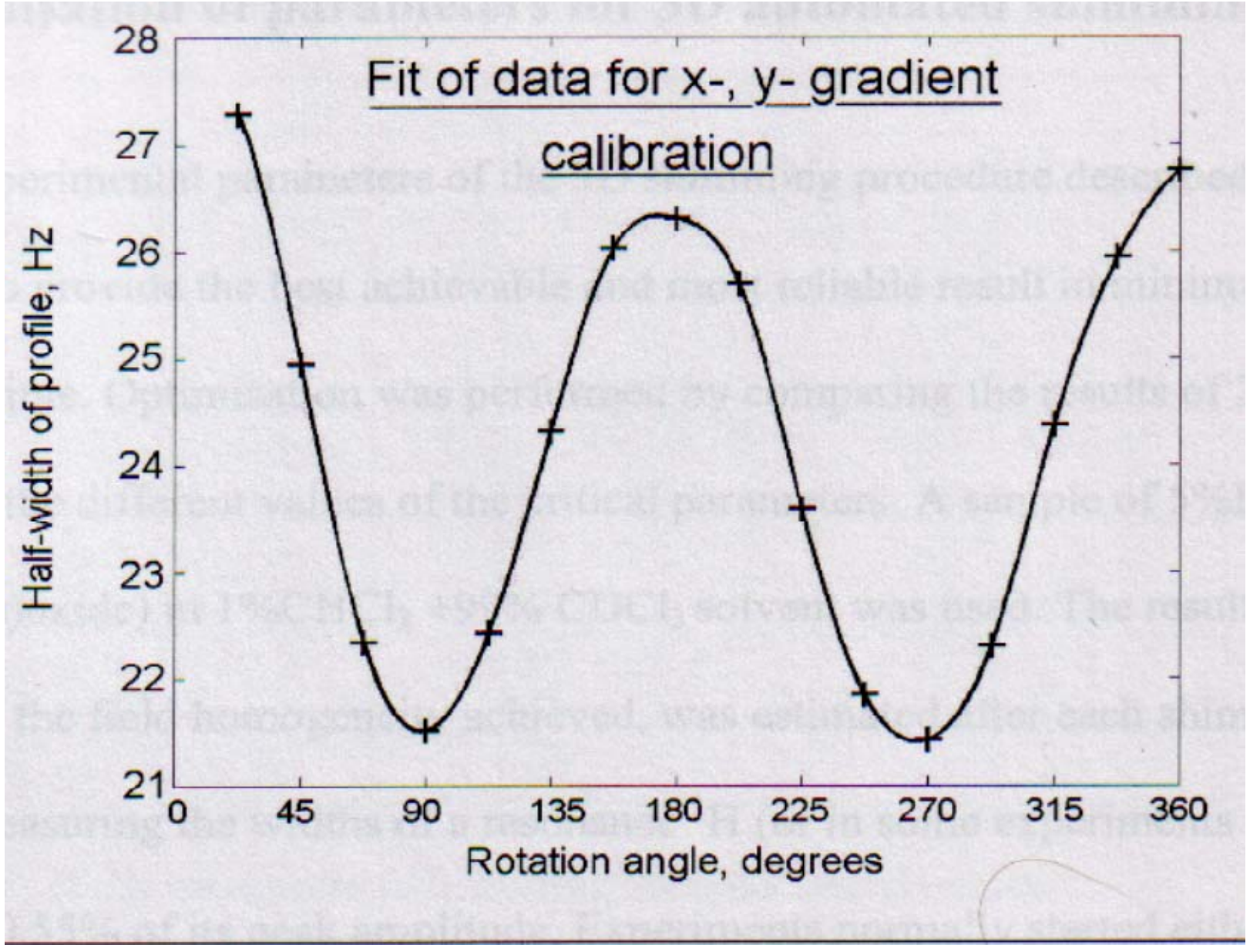

**Fig.6.3** Fitting of parameters for $x1$ and $y1$ gradient calibration. The continuous line represents the least squares best fit to the experimental data marked with crosses.

The reproducibility of the gradient calibration results was also investigated by repeating the calibration ten times on the 400 MHz spectrometer. The results of this testing, presented in Table 6.1, show good reproducibility, with standard deviations 0.0053, 0.0104, and 0.0805 for the parameters **gcalx**, **gcaly** and **gcalang** respectively.

**Table 6.1 Results of the calibration reproducibility testing**

| N | **gcalx**, $\times 10^{-5}$ $G\,cm^{-1} DAC^{-1}$ | **gcaly**, $\times 10^{-5}$ $G\,cm^{-1} DAC^{-1}$ | **gcalang**, degree | chi-square, $Hz^2$ |
|---|---|---|---|---|
| 1 | 9.273957 | 8.315921 | -11.754 | 0.145 |
| 2 | 9.277983 | 8.328627 | -11.659 | 0.041 |
| 3 | 9.276750 | 8.311021 | -11.586 | 0.020 |
| 4 | 9.277950 | 8.336126 | -11.789 | 0.125 |
| 5 | 9.265630 | 8.336437 | -11.618 | 0.049 |
| 6 | 9.270152 | 8.341528 | -11.783 | 0.020 |
| 7 | 9.281782 | 8.321534 | -11.626 | 0.101 |
| 8 | 9.279520 | 8.330687 | -11.647 | 0.063 |
| 9 | 9.280675 | 8.329960 | -11.605 | 0.054 |
| 10 | 9.270116 | 8.341195 | -11.768 | 0.020 |

## 6.2 Optimisation of parameters for 3D automated shimming

The experimental parameters of the 3D shimming procedure described needed optimisation to provide the best achievable and most reliable result in minimum experimental time. Optimisation was performed by comparing the results of 3D shimming obtained with the different values of the critical parameters. A sample of 5%PEO (poly-ethelene-oxide) in 1%$CHCl_3$ +99% $CDCl_3$ solvent was used. The result of shimming, i.e. the field homogeneity achieved, was estimated after each shimming iteration by measuring the widths of a resonance $^1$H (or in some experiments $^{13}$C or $^2$H ) line at 50/1.1/0.55% of its peak amplitude. Experiments normally started either from shims set to zero ('cold shims') or roughly adjusted ('warm shims'). All line shape data are reported for nonspinning sample unless otherwise stated.

To find optimum transverse digitization for field mapping (i.e. the number of phase-encoding increments required) experiments were performed with transverse data matrices from 3x3 to 8x8 increments. The number of the linear gradient strength increments used is called matrix size as field and shim maps are represented in *xy* plane by square matrices. After each iteration a new set of shim values was calculated and the shim currents set, and field homogeneity achieved was estimated from the line widths as described. The results of $^1$H 3D shimming with even and odd matrix sizes are presented in Tables 6.2 and 6.3 respectively. These show that the field homogeneity achieved with the use of even matrix sizes was generally better than with odd sizes under similar experimental conditions; the half-widths of $^1$H lines in spectra obtained after 3D shimming with matrix size 5x5 and 7x7 were greater than after shimming with a matrix size of 4x4.

**Table 6.2 The results of $^{1}$H 3D automated shimming for even matrix sizes**

| Number of iteration | Matrix size | | |
|---|---|---|---|
| | 4x4 | 6x6 | 8x8 |
| | $^{1}$H line widths, Hz measured at 50/1.1/0.55% of amplitude | | |
| 1 | 0.94/8.51/10.1 | 0.77/7.59/9.66 | 0.93/8.44/10.11 |
| 2 | 0.74/8.32/10.9 | 0.78/7.75/9.76 | 0.69/8.10/10.27 |
| 3 | 0.72/8.22/10.8 | 0.73/7.39/9.07 | 0.67/8.01/8.88 |
| 4 | 0.76/8.01/10.4 | 1.08/8.74/10.73 | 0.64/7.62/8.60 |
| 5 | 0.62/8.08/10.2 | 0.64/7.16/8.84 | 0.59/7.43/8.44 |
| 6 | 0.69/8.01/11.8 | 0.64/7.30/8.77 | 0.65/7.55/9.29 |
| 7 | 0.70/8.36/11.1 | 0.64/7.12/8.87 | 0.62/7.50/8.53 |
| 8 | 0.67/8.10/10.2 | 0.59/7.12/8.65 | 0.62/6.74/9.12 |
| 9 | 0.69/8.14/10.5 | 0.59/7.04/9.28 | 0.61/7.30/10.22 |
| 10 | 0.65/7.80/8.90 | 0.70/7.36/9.35 | 0.61/7.34/8.57 |

**Table 6.3 The results of $^{1}$H 3D automated shimming for odd matrix sizes**

| Number of iteration | Matrix size | | |
|---|---|---|---|
| | $3\times3$ | $5\times5$ | $7\times7$ |
| | $^{1}$H line widths, Hz measured at 50/1.1/0.55% of amplitude | | |
| 1 | 1.66/12.4/13.7 | 0.72/9.33/11.73 | 1.05/8.98/11.54 |
| 2 | 2.43/15.58/17.2 | 0.62/9.02/14.94 | 1.03/9.10/11.93 |
| 3 | 3.65/20.15/21.5 | 0.67/8.78/12.89 | 0.97/9.23/11.43 |
| 4 | 4.12/22.28/23.5 | 0.65/8.66/12.05 | 1.20/9.86/13.24 |
| 5 | 3.55/24.18/25.5 | 0.67/7.96/11.95 | 0.94/9.68/12.20 |
| 6 | 3.72/33.79/35.1 | 0.73/8.74/12.75 | 0.94/9.42/12.30 |
| 7 | 3.77/31.05/32.2 | 0.73/9.14/13.15 | 0.96/9.75/12.60 |
| 8 | 4.19/41.97/43.1 | 0.74/9.10 /12.2 | 0.97/9.57/11.69 |
| 9 | 5.15/41.19/43.4 | 0.72/9.12 /12.8 | 0.94/9.68/12.46 |
| 10 | 5.40/46.80/47.5 | 0.70/9.11 /12.5 | 0.93/9.54/11.93 |

It was found that the half-widths of spectra acquired after 3D shimming with a 4x4 matrix size are very close to the best results with any of the matrix sizes tried. This result is significant as decrease in matrix size allows a saving in experimental time. For example, field mapping with matrix size of 4x4 takes about 2.25 and 4 times less than with matrix sizes 6x6 and 8x8 respectively. The results of 3D shimming with odd matrix sizes 5x5 and 7x7 are presented in Table 6.3.

The half-widths of $^{1}$H lines in spectra acquired after 10 iterations and the durations of experiments are shown for different matrix sizes in Figs.6.4 and 6.5. 3D shimming with 2x2 matrix size is very quick, normally taking only a few minutes but its results are unsatisfactory for high-resolution applications.

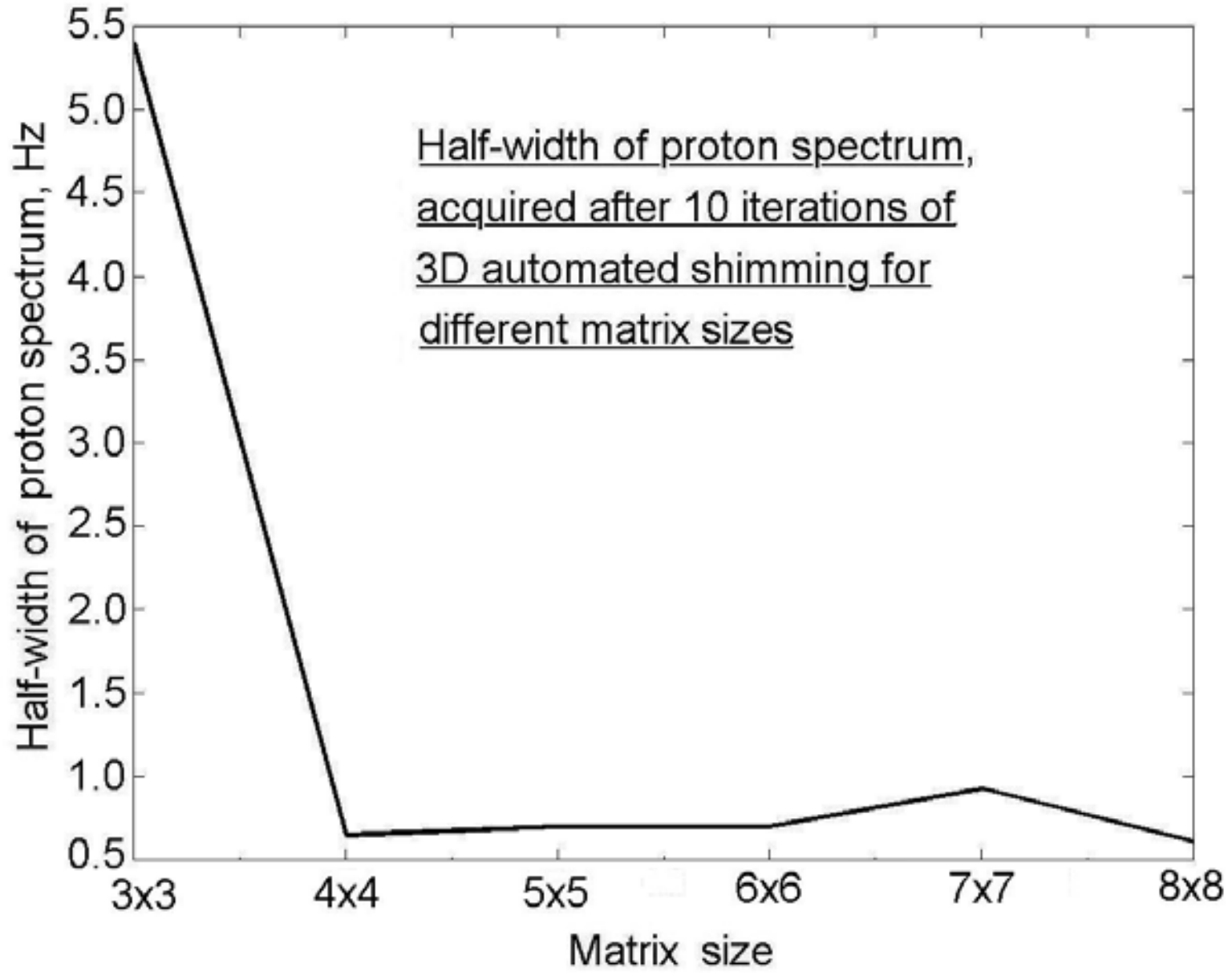

**Fig.6.4** $^1$H line half-width after 10 iterations of 3D automated shimming experiment versus matrix size.

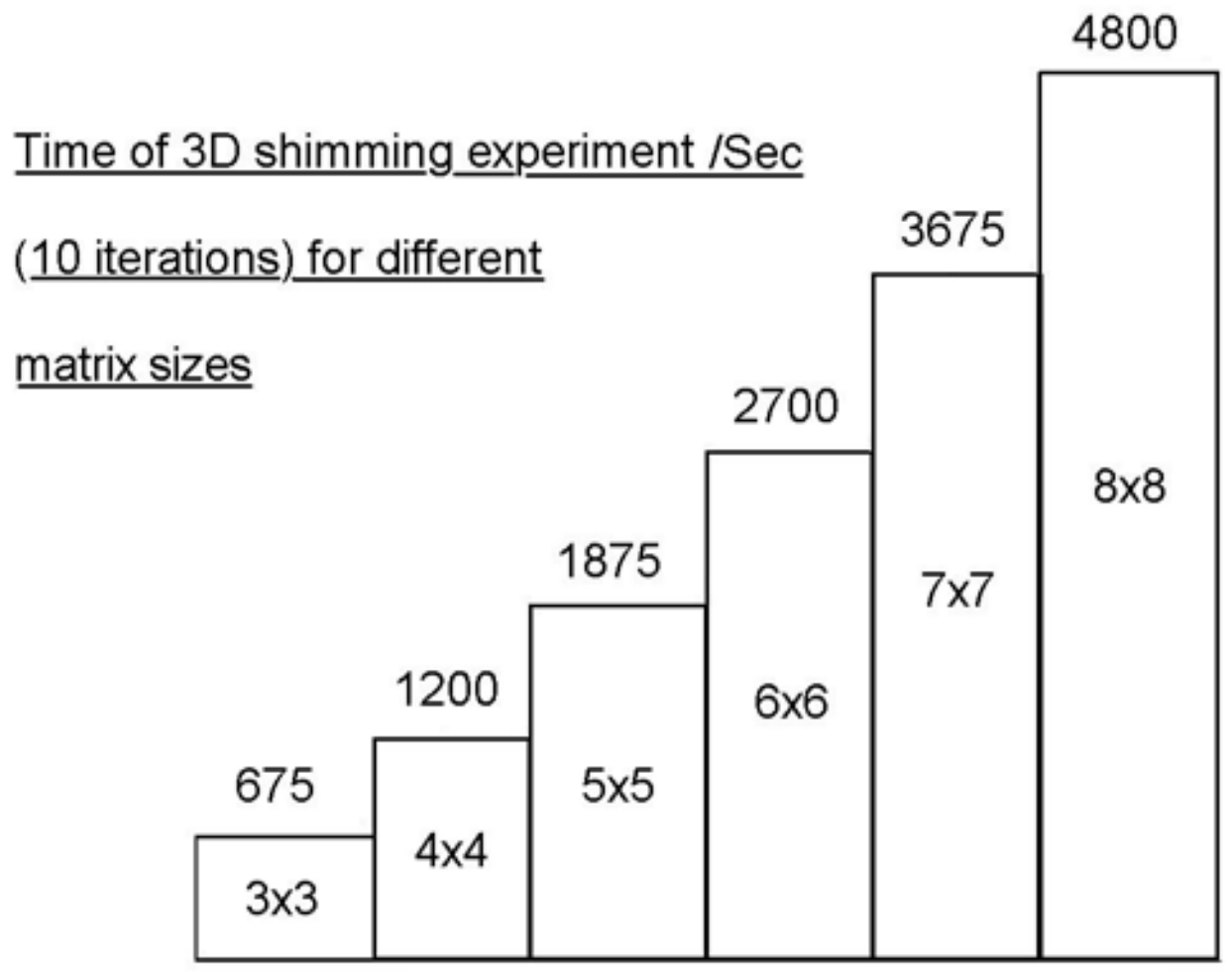

**Fig.6.5** Total experiment times of 10 iterations of 3D automated shimming for different matrix sizes as described in the text. These times are for shimming only, not including the measurements of shim maps.

Although the half-widths of spectra acquired after 10 iterations with the given matrix sizes differ very little, as already mentioned, the shapes of the lines become more symmetrical with increasing matrix size, especially at the foot of the line as illustrated in Figs.6.6 and 6.7.

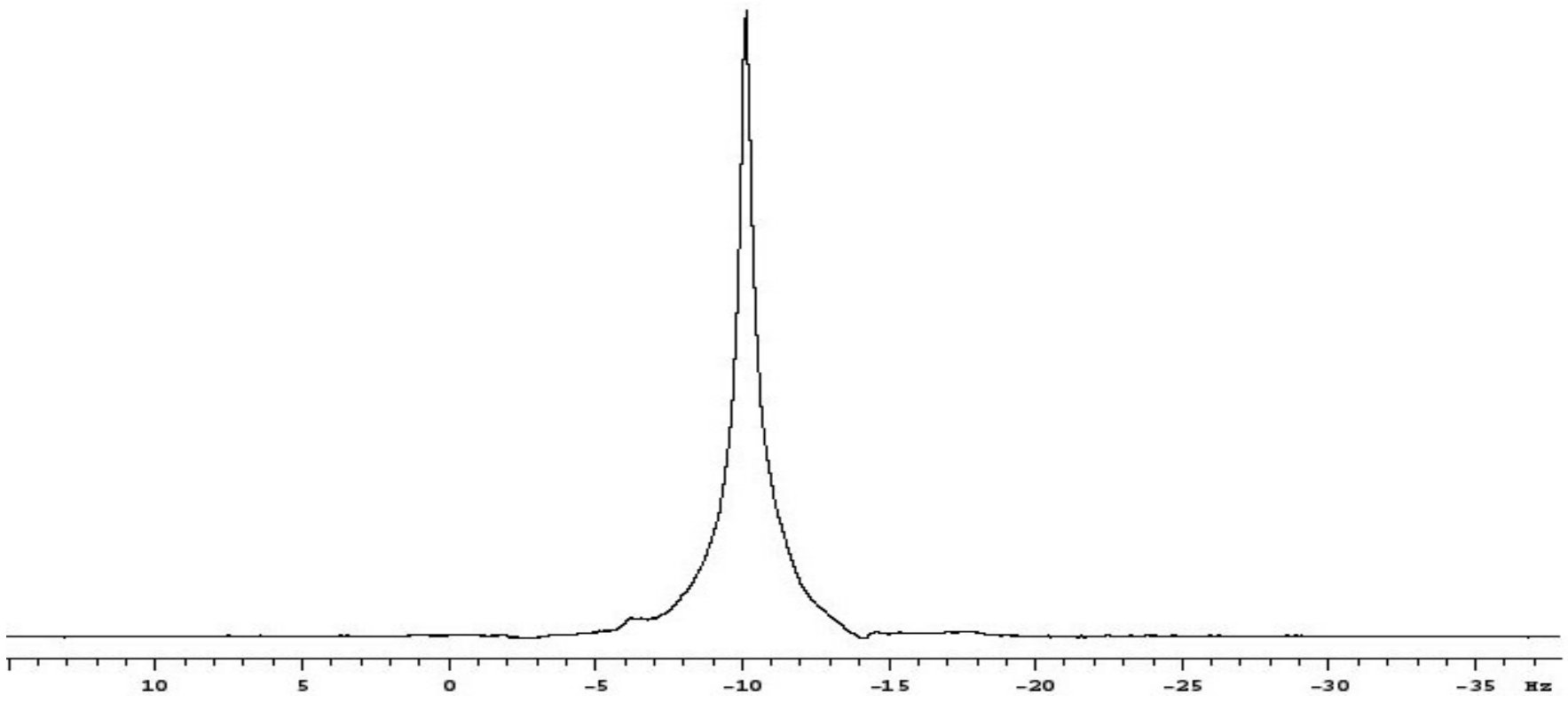

**a)**

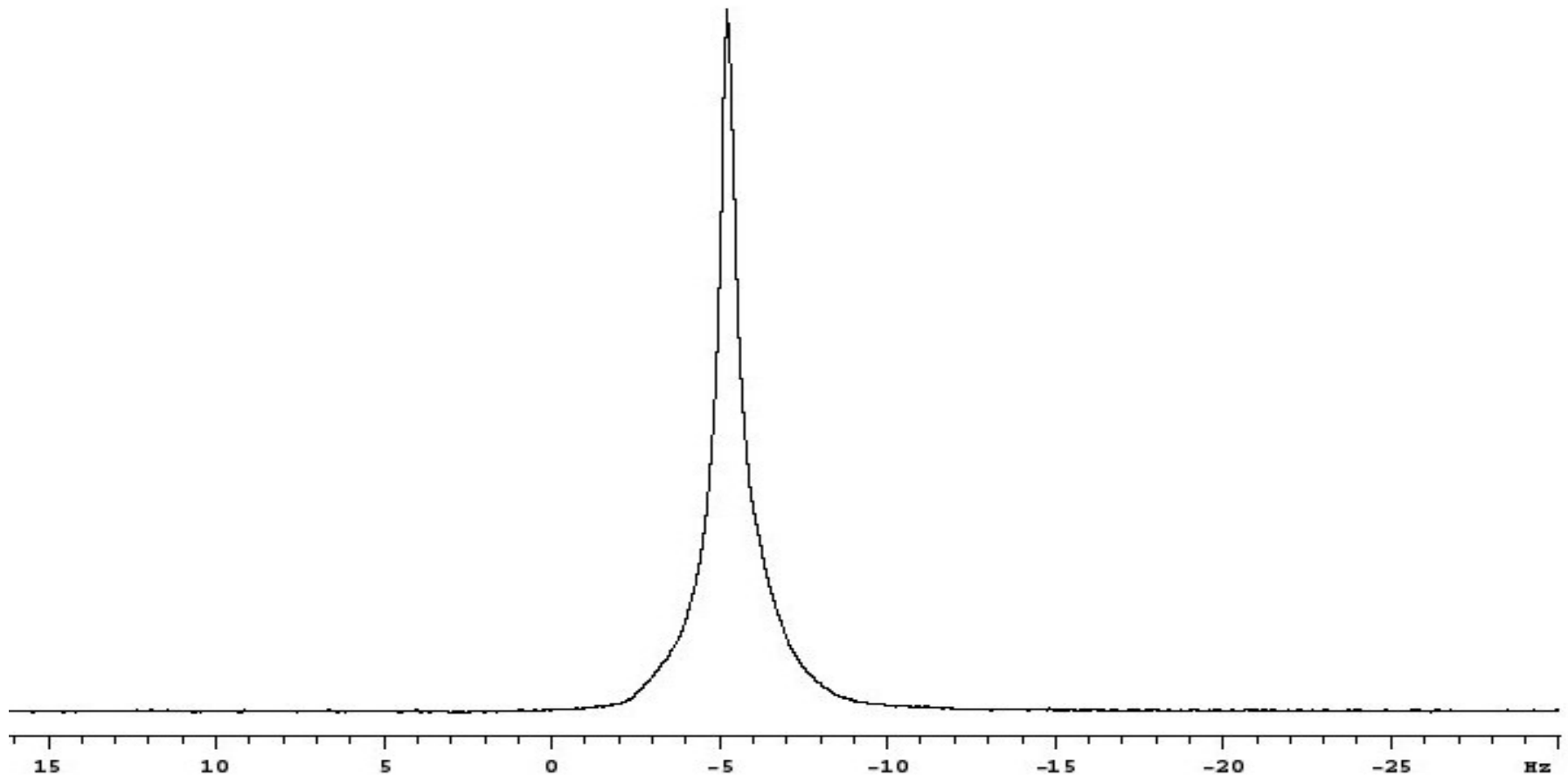

**b)**

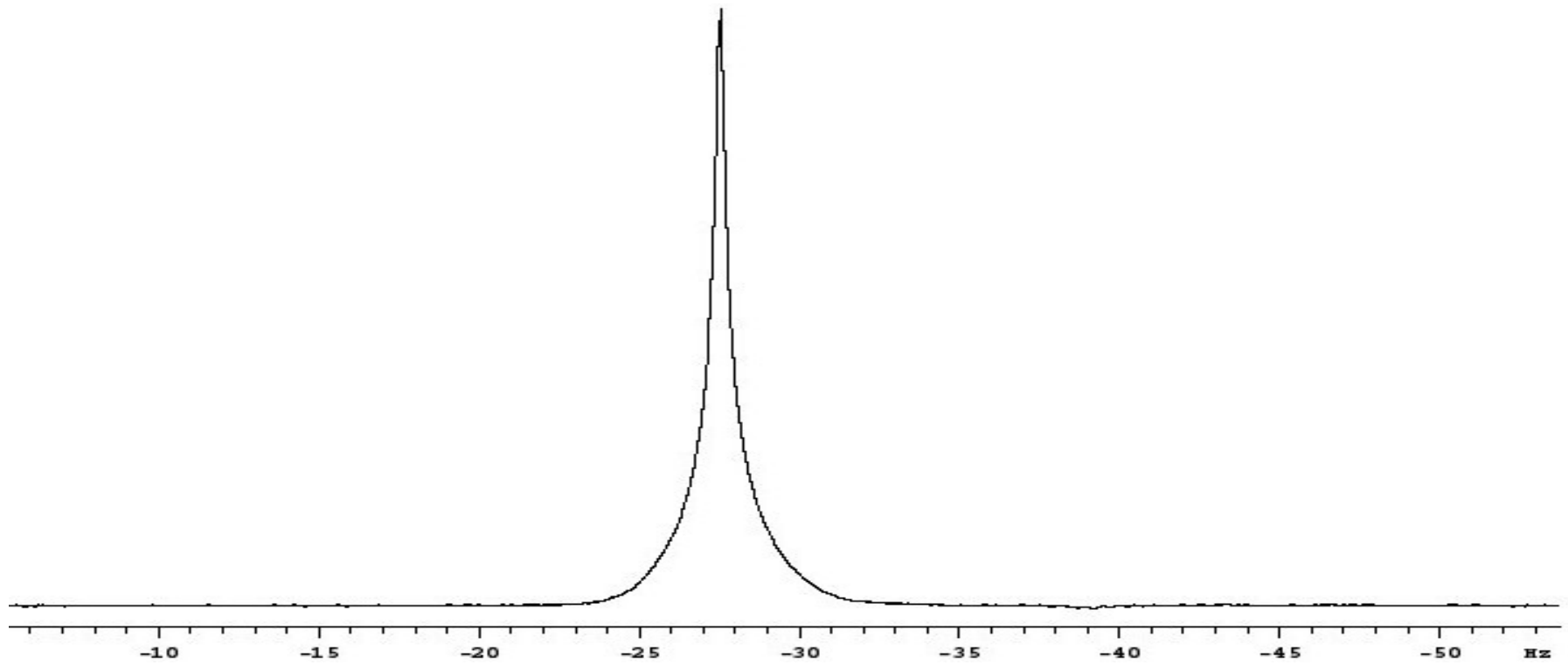

**c)**

**Fig.6.6** $^{1}H$ spectra, acquired after ten iterations of 3D automated shimming with even matrix sizes: **a)** 4x 4; **b)** 6x6; **c)** 8x8.

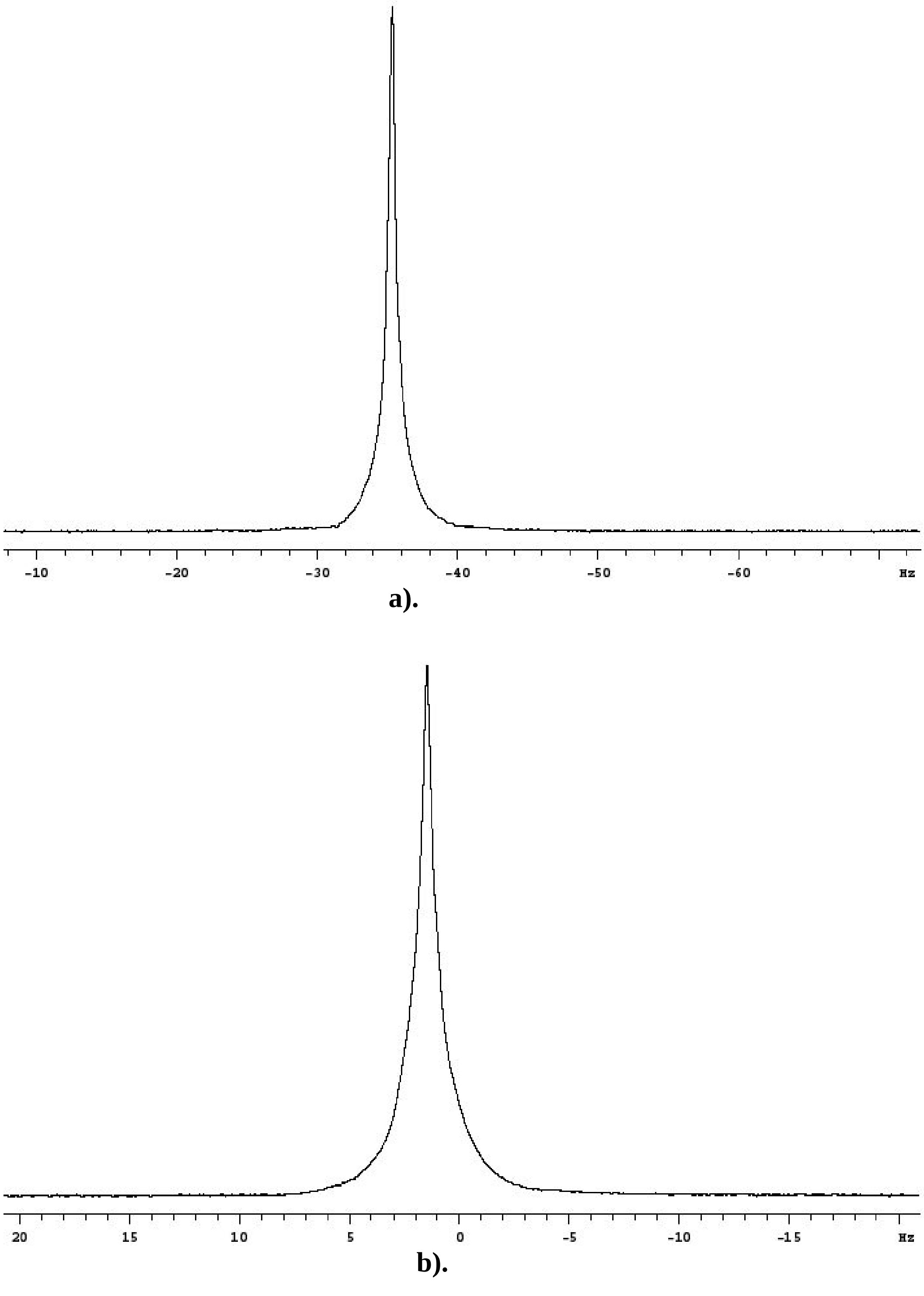

**Fig.6.7** $^{1}H$ spectra acquired after ten iterations of 3D automated shimming with odd matrix sizes: **a)** 5x5 and **b)** 7x7.

The data of Table 6.2 show that the half-widths of $^{1}H$ line gradually decrease during 3D shimming (indicating improvement in field homogeneity) and eventually reach a limit which corresponds to the optimum field homogeneity currently achievable with the use of this shimming technique. To confirm that the results of Tables 6.2 and 6.3 are representative, experiments from ‘cold’ shims were tried. These results (Table 6.4) show that the line widths of spectra converge to similar point irrespectively of the initial shim values.

**Table 6.4 The results of $^{1}$H 3D automated shimming from ‘cold’**

| Number of iteration | Matrix size | | |
|---|---|---|---|
| | $4\times4$ | $5\times5$ | $6\times6$ |
| | $^{1}$H line widths, Hz measured at 50/1.1/0.55% of amplitude | | |
| 1 | 5.61/31.26/35.33 | 3.10/31.28/38.81 | 5.71/29.42/37.12 |
| 2 | 2.48/16.63/19.36 | 3.92/24.01/38.33 | 0.92/8.48/11.61 |
| 3 | 1.88/11.29/12.09 | 3.10/22.77/32.80 | 0.67/8.24/8.63 |
| 4 | 1.5/10.28/13.59 | 2.57/17.56/20.05 | 0.89/9.86/10.41 |
| 5 | 1.55/7.51/7.93 | 2.60/15.32/18.18 | 0.79/6.82/7.38 |
| 6 | 0.80/9.26/10.75 | 2.02/13.25/17.61 | 0.62/8.50/9.33 |
| 7 | 0.71/8.50/9.30 | 1.74/10.61/15.99 | 0.66/8.39/8.70 |
| 8 | 0.61/9.48/11.55 | 1.58/10.32/12.88 | 0.71/7.67/11.28 |
| 9 | 0.59/7.90/8.79 | 1.41/10.95/14.68 | 0.62/7.45/9.79 |
| 10 | 0.62/8.05/9.03 | 1.36/10.35/13.49 | 0.60/7.55/8.71 |

The data presented in Tables 6.2-4 are typical for $^{1}$H 3D shimming under the described conditions.

However, it was observed that better field homogeneity for shimming from ‘cold’ shims was achieved when 1D $z$ shimming was performed before (and sometimes repeated after) 3D shimming. The $z$ inhomogeneity (especially $z1$) usually dominates the raw field, and can be compensated almost independently from other shims by using existing methods for automated $z$ gradient shimming. The use of such 1D $z$ shimming together with the 3D shimming procedure described in this thesis will be presented in the next Section of this chapter.

3D shimming for matrix sizes 4x4 and 6x6 with 30 iterations was also tested, with results presented in Table 6.5. These confirm that the $^{1}$H line rapidly reaches the limit, already observed for the data of Tables 6.2 and 6.4.

**Table 6.5 The results of $^1$H 3D automated shimming in 30 iterations from 'cold' for 6x6 matrix size and from 'warm' shims for matrix size of 4x4.**

| Number of iteration | Matrix size | |
|---|---|---|
| | $4\times4$ | $6\times6$ |
| | $^1$H line widths, Hz measured at 50/1.1/0.55% of amplitude | |
| 1 | 2.98/19.53/23.61 | 8.32/25.96/33.88 |
| 2 | 0.86/7.73/8.95 | 2.37/15.88/19.65 |
| 3 | 0.65/7.88/9.30 | 1.32/10.76/14.99 |
| 4 | 0.70/7.57/8.64 | 0.94/10.29/15.83 |
| 5 | 0.63/8.39/9.08 | 0.86/8.82/11.10 |
| 6 | 0.66/7.68/9.82 | 1.00/9.22/12.59 |
| 7 | 0.66/7.67/9.49 | 0.76/9.17/12.59 |
| 8 | 0.65/7.84/9.29 | 1.08/9.99/13.5 |
| 9 | 0.62/7.88/9.79 | 0.74/8.82/11.61 |
| 10 | 0.65/7.73/9.23 | 0.79/9.20/10.17 |
| 11 | 0.66/7.94/9.47 | 0.93/10.14/12.40 |
| 12 | 0.67/7.86/9.68 | 0.68/7.81/9.25 |
| 13 | 0.73/7.70/9.49 | 0.68/7.33/8.11 |
| 14 | 0.68/7.52/7.81 | 0.65/6.96/7.83 |
| 15 | 0.71/7.66/8.42 | 0.68/9.30/12.39 |
| 16 | 0.75/7.92/8.24 | 0.57/7.98/9.11 |
| 17 | 0.64/7.78/9.53 | 0.63/7.98/9.66 |
| 18 | 0.66/7.73/9.20 | 0.70/8.45/10.78 |
| 19 | 0.65/7.68/10.11 | 0.55/7.97/9.53 |
| 20 | 0.63/8.40/9.28 | 0.56/7.34/8.27 |
| 21 | 0.61/7.39/7.98 | 0.56/7.95/9.77 |
| 22 | 0.63/7.80/9.15 | 0.62/8.13/10.38 |
| 23 | 0.66/7.83/9.61 | 0.64/8.30/10.09 |
| 24 | 0.62/7.86/9.89 | 0.65/8.62/9.08 |
| 25 | 0.60/7.66/9.42 | 0.64/8.02/9.60 |
| 26 | 0.66/7.78/9.50 | 0.62/8.37/10.12 |
| 27 | 0.65/7.65/9.34 | 0.70/8.35/10.54 |
| 28 | 0.66/7.75/8.87 | 0.60/8.05/9.40 |
| 29 | 0.73/8.01/9.65 | 0.60/7.90/10.10 |
| 30 | 0.76/8.08/9.45 | 0.60/7.87/9.88 |

The data for 6x6 shimming in Table 6.5 show a slight improvement in half-width even after ten iterations of 3D shimming. The widths near the foot show that some field inhomogeneity of high orders remains uncompensated throughout the experiments. The most significant improvement in field homogeneity in these 3D shimming experiments (and in the experiments described above) was noted during the first 3-5 iterations.

It was interesting to find out how much time averaging could improve the results of 3D shimming near to the limit described above. This was investigated by performing ten iterations of 3D automated shimming in three experiments using the different numbers of transients (1, 32 and 64), with other parameters left identical. The results presented in Table 6.6 show that an increase in the number of transients only slightly improves field homogeneity. However, it was found that the use of 4 transients of phase cycling in order to eliminate unwanted NMR signals is desirable in 3D shimming.

**Table 6.6 Results of $^1$H 3D automated shimming with matrix size 4x4 for different numbers of transients**

| Number of iteration | Number of transients | | |
|---|---|---|---|
| | 1 | 32 | 64 |
| | $^1$H line widths, Hz measured at 50/1.1% of amplitude | | |
| 1 | 1.31/10.44 | 1.19/9.65 | 5.82/20.34 |
| 2 | 1.40/10.17 | 0.80/9.19 | 4.24/14.84 |
| 3 | 0.87/8.47 | 0.94/7.64 | 0.79/9.03 |
| 4 | 0.71/8.77 | 0.82/8.48 | 0.68/8.28 |
| 5 | 0.81/0.74 | 0.74/8.98 | 0.78/7.67 |
| 6 | 0.69/8.85 | 0.77/8.83 | 0.64/8.51 |
| 7 | 0.69/9.34 | 0.56/7.37 | 0.69/8.74 |
| 8 | 0.73/8.37 | 0.62/8.34 | 0.74/8.61 |
| 9 | 0.75/8.43 | 0.70/8.40 | 0.62/8.63 |
| 10 | 0.78/7.70 | 0.71/8.75 | 0.64/8.42 |

Matrix size and number of transients are parameters, which normally stay unchanged throughout a 3D automated shimming experiment. In contrast, the time interval $\tau$ (described in Chapter 3) represented by parameter **tau** is typically changed after each iteration. After iteration, this interval is automatically set by the software to the inverse of line width measured at 0.55% of line amplitude. This choice of **tau** is a compromise: if **tau**
is too short, the SNR of field map suffers, if **tau** is too long then the peak phase excursion exceeds $\pi$ radians and the field map ‘folds’, preventing proper shim convergence.

Since line widths depend on the magnetogyric ratio of the nucleus, this was taken into account in the calculation of **tau** when other than $^{1}H$ were used for shimming and acquisition of spectra. In this case the value of **tau**, defined as inverse of line-width, measured at 0.55% of its amplitude was multiplied by a ratio of magnetogyric ratio of the nucleus used for acquisition of spectra to that of nucleus used in shimming.

## 6.3 $^{1}$H 3D shimming

In this section a detailed, step-by-step description of the operations of a typical $^{1}H$ 3D shimming experiment, starting from 'cold' shims is presented. A 5mm sample of 5% PEO in [1% $CHCl_3$ + 99% $CDCl_3$] was used in these experiments, which were carried out on a Unity INOVA 300 spectrometer.

An ordinary $^{1}H$ spectrum acquired with 'cold' shims is presented in Fig.6.8.

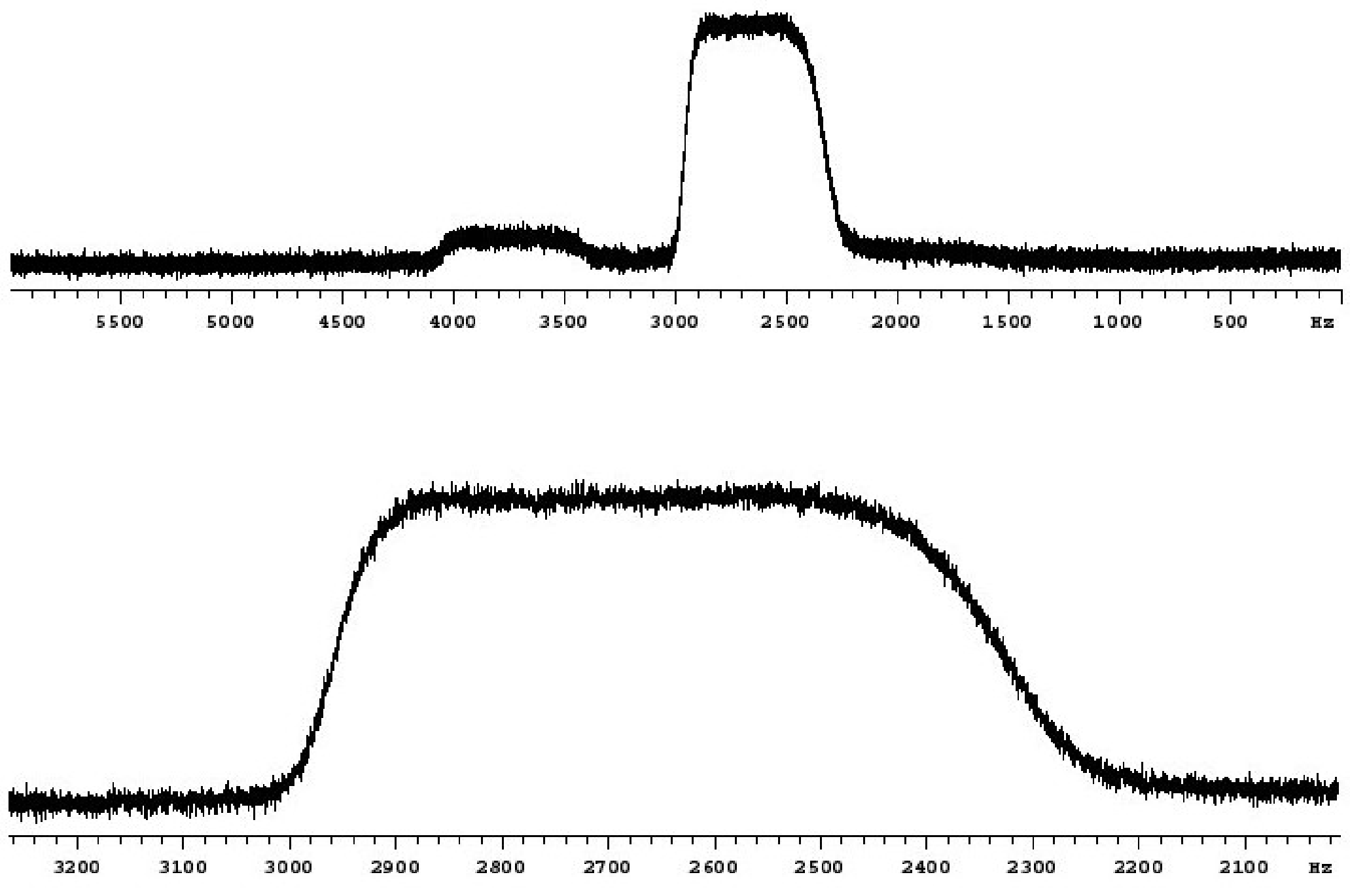

**Fig.6.8** The $^{1}H$ spectrum of 5% PEO in [1%$CHCl_3$ + 99%$CDCl_3$], acquired before 3D automated shimming, with all shims set to zero: (top) the $CHCl_3$ and PEO lines (from left to right, respectively), (bottom) enlarged PEO line; its half-width is 615 Hz which exceeds the natural line width by more than four thousand times.

It is heavily broaden and the task of the described 3D shimming procedure is to compensate field inhomogeneity by which this broadening is caused.

The procedure starts with 1D automated $z$ shimming. This uses homospoil pulses for $z$ gradients, and is controlled by the macro **gmapsys** provided with the standard Varian software. The first step is the mapping of the $z1-z5$ shims, followed by several iterations of automated $z$ shimming. The shim maps obtained are shown in Fig.6.9. The $z1-z4$ automated shimming converged in five iterations. After this, the field homogeneity achieved was estimated from the half-width of the PEO line in $^1$H spectrum acquired with the new shims; the half-width reduced from 615 to about 17 Hz. The value of line width measured at the foot, 62 Hz, was used for calculation of the parameter **tau** to be used in the first iteration of 3D shimming.

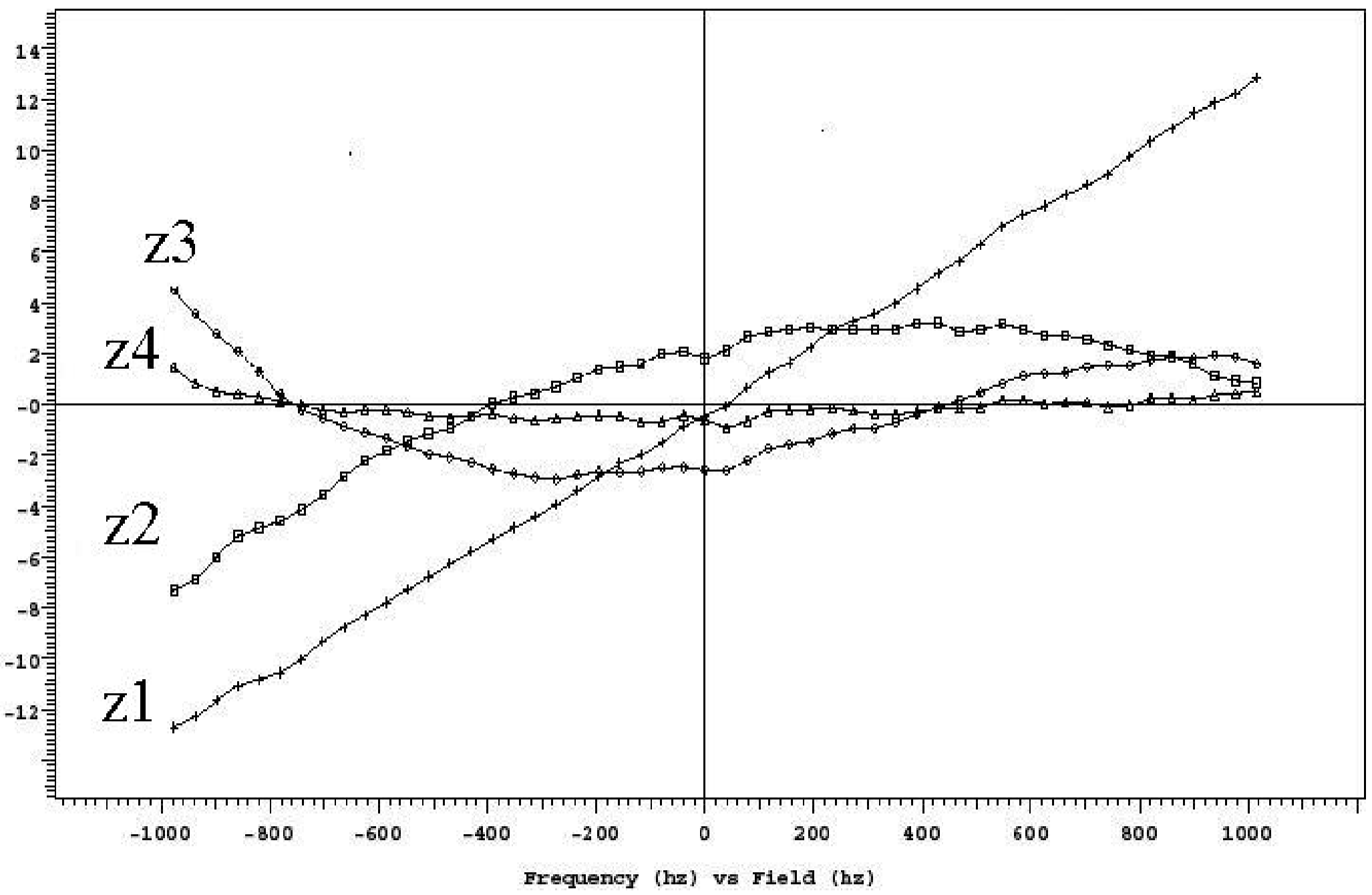

**Fig.6.9** The $z$ shim maps acquired with 'cold' shims before 1D $z$ shimming.

After the initial $z$ shimming, calibration of $x1$ and $y1$ gradients was performed as described in Section 6.1, with results presented in Table 6.7.

**Table 6.7. Results of transverse shim gradient calibration**

| **gcalx**, $\times 10^{-5}$ $G\,cm^{-1}DAC^{-1}$ | **gcaly**, $\times 10^{-5}$ $G\,cm^{-1}DAC^{-1}$ | **gcalang**, degree | x_error, *DAC* | y_error, *DAC* | chi-square, $Hz^2$ |
|---|---|---|---|---|---|
| 9.57 | 8.67 | -12.7 | -25 | 215 | 2.232 |

After this calibration, 3D shimming with matrix size of 4x4 was performed in automation. This started with a shim mapping experiment, in which the field shapes produced by the 13 shims of the shim set used were mapped. Then 3D automated shimming started. In the first iteration the set of profiles represented in Fig.6.10 was acquired.

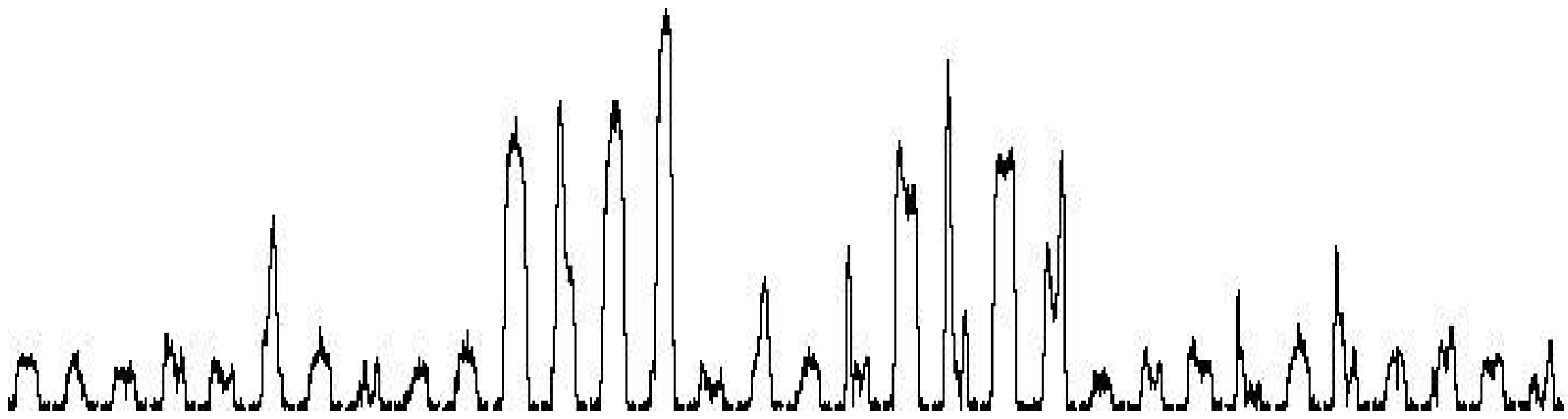

**Fig.6.10** The set of 16 profiles acquired in field mapping experiment with matrix size of 4x4 before first iteration of 3D shimming, as described in the text.

Next, the field map of Fig.6.11 was obtained by calculating phase difference between two 3D images as described in Chapter 3. This map is a transverse plane representation of the magnetic field variation in the sample after 1D $z$ shimming experiment. The frequency scale, shown at left of the map relates to maximum variation of the Larmor frequencies due to field inhomogeneity in the sample. Each square shows the Larmor frequencies as a function of $z$ for a given $(x, y)$ position.

This map shows stronger field variation in the outer areas of sample than at the middle; the $z^2$, $z^4$ field variations are especially strong at the top and bottom of the map, and some linear $z$ field variation is also present. The frequency scale is relatively coarse, so the other shapes of field variation of smaller magnitudes are masked by the described $z$ field variations.

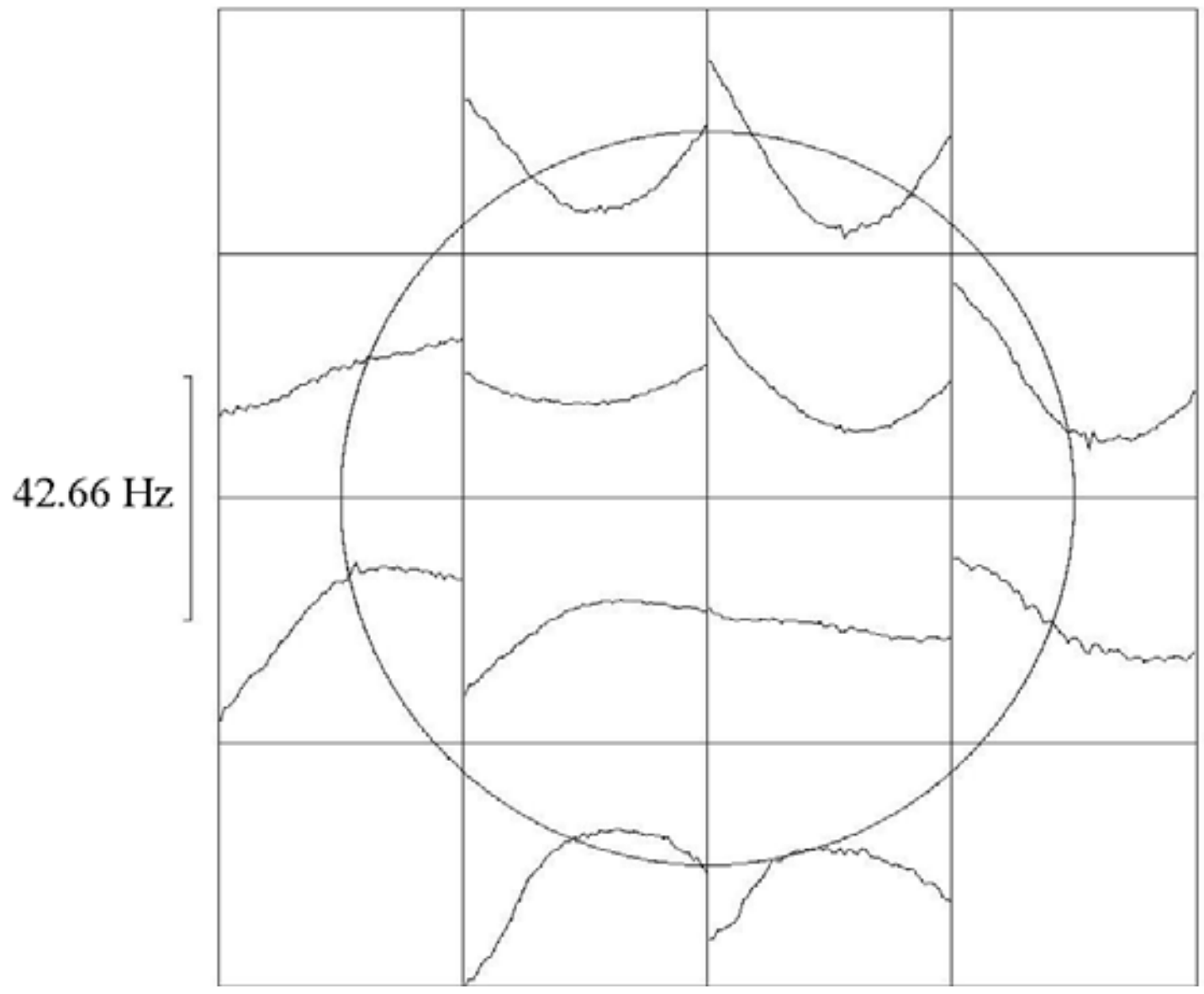

**Fig.6.11** Field map obtained in 1st iteration of 3D automated shimming.

The $^{1}H$ spectrum acquired after the first iteration is shown in Fig.6.12.

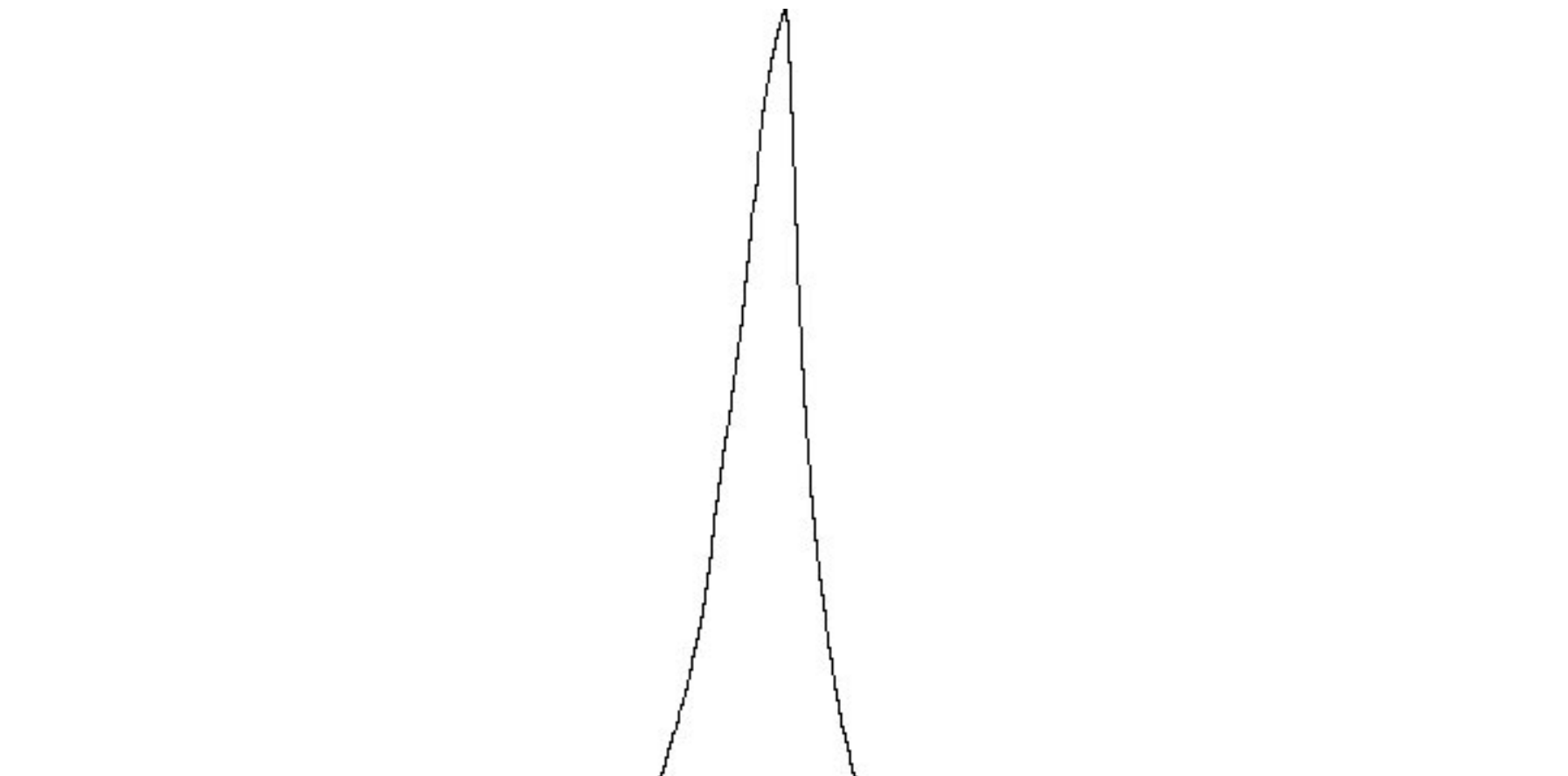

**Fig.6.12** $^{1}H$ spectrum acquired with the new shims found in the first iteration of 3D shimming. The widths measured at 50/1.1/0.55% of line amplitude are 2.61/13.5/14.19 Hz.

Next, 10 iterations of 3D automated shimming were performed. A set of the corresponding field maps is presented in Fig.6.13, in which each map shows the field variation after calculation and setup of the new shims from the previous iteration.

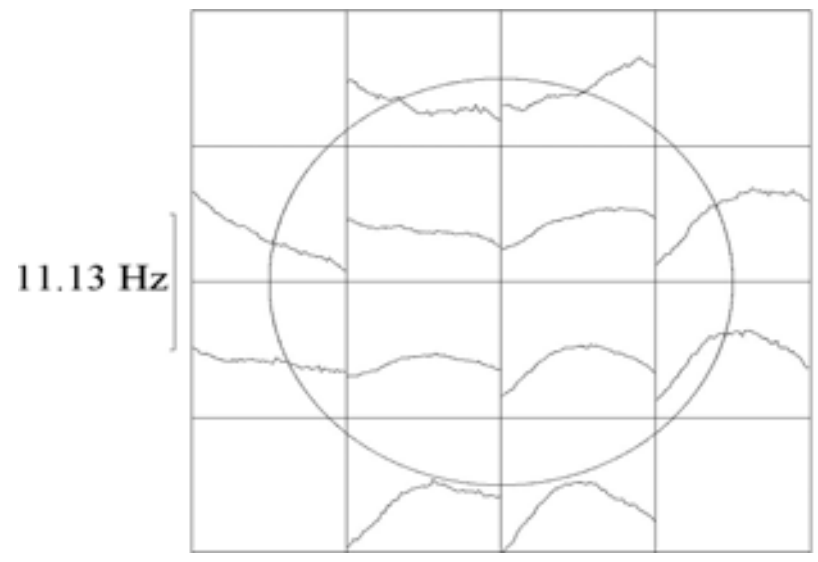

$2^{nd}$ iteration.

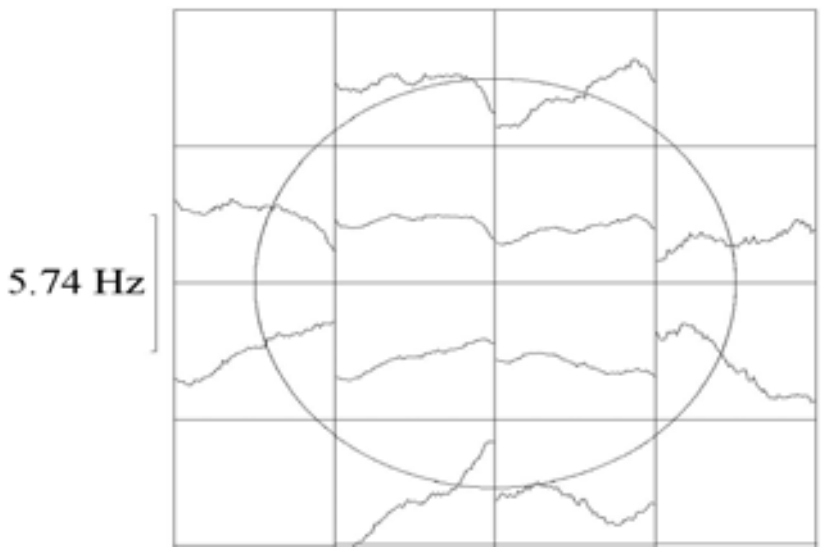

$3^{rd}$ iteration.

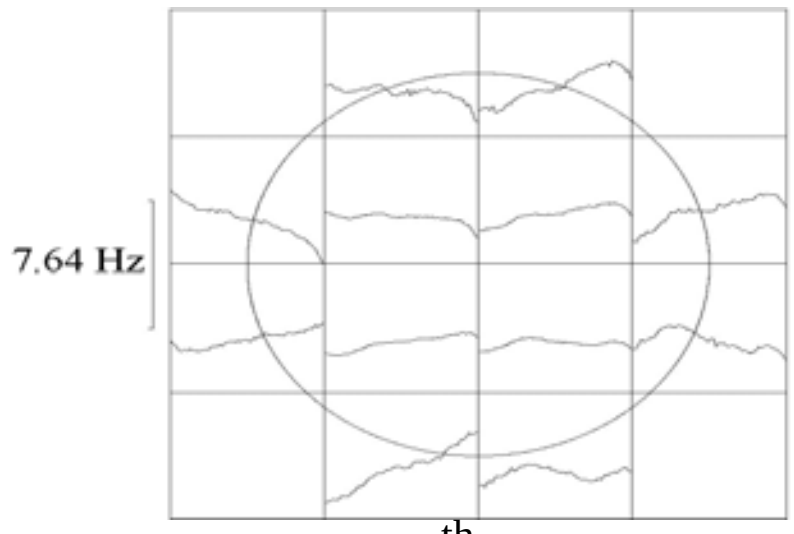

$4^{th}$ iteration.

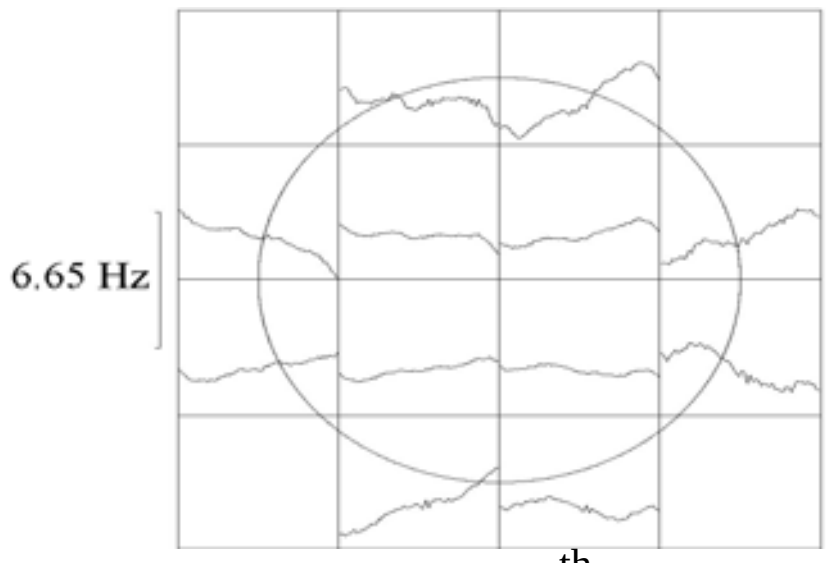

$5^{th}$ iteration.

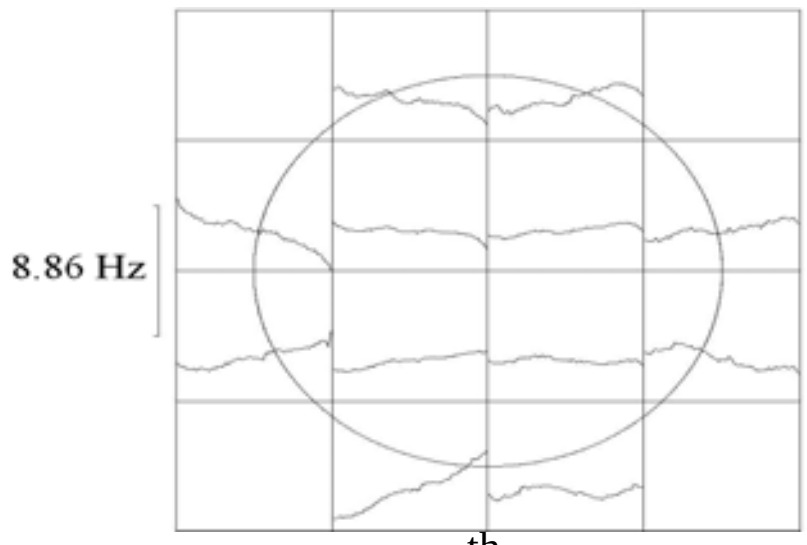

$6^{th}$ iteration.

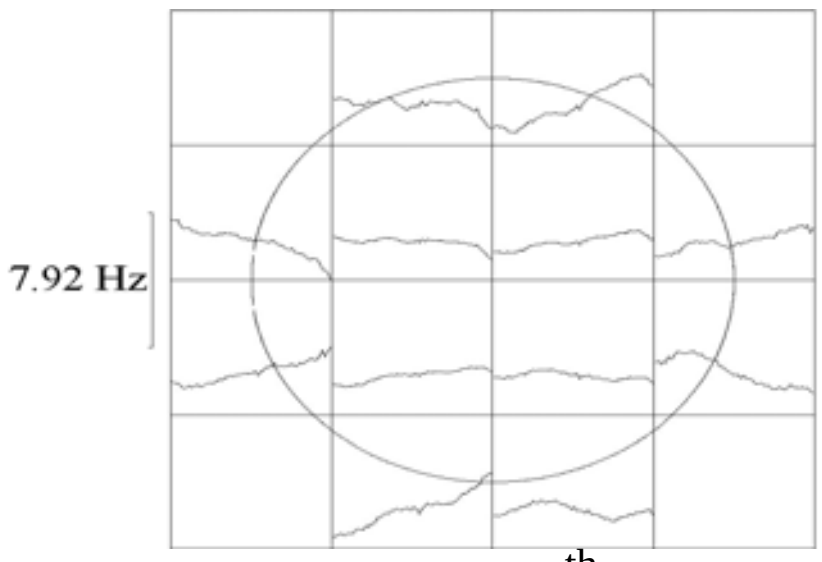

$7^{th}$ iteration.

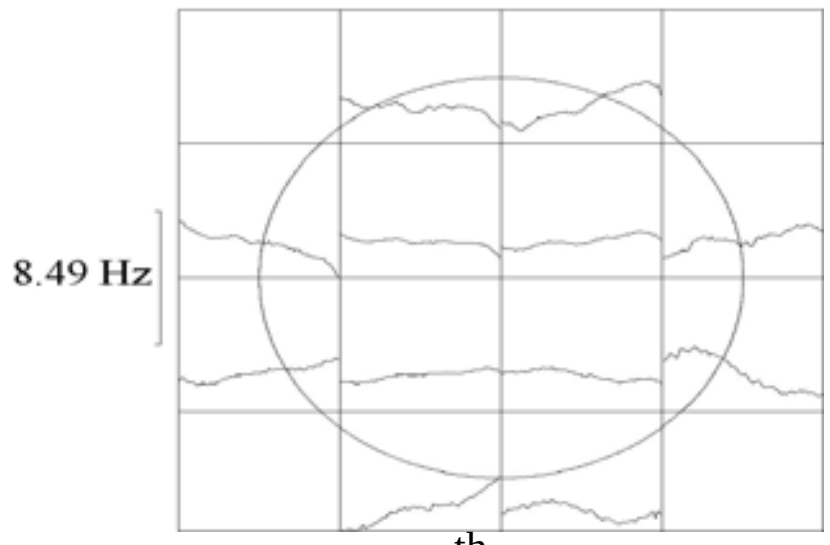

$8^{th}$ iteration.

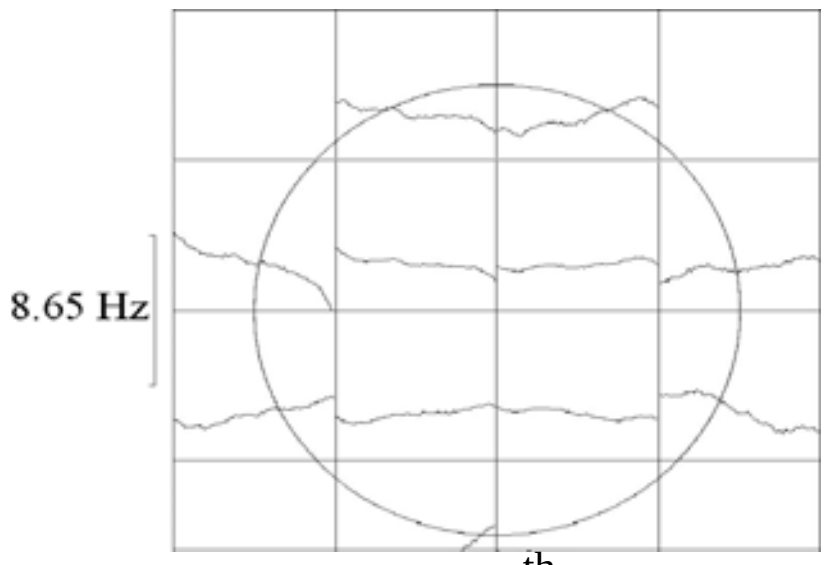

$9^{th}$ iteration.

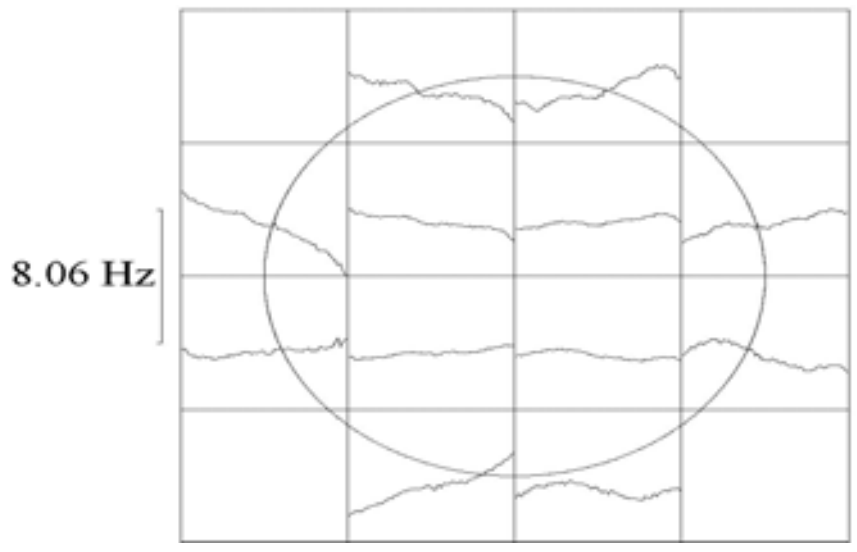

$10^{th}$ iteration.

**Fig.6.13** Field maps obtained in the $^{1}H$ 3D shimming experiment.

The shim values used are summarized in Fig.6.14.

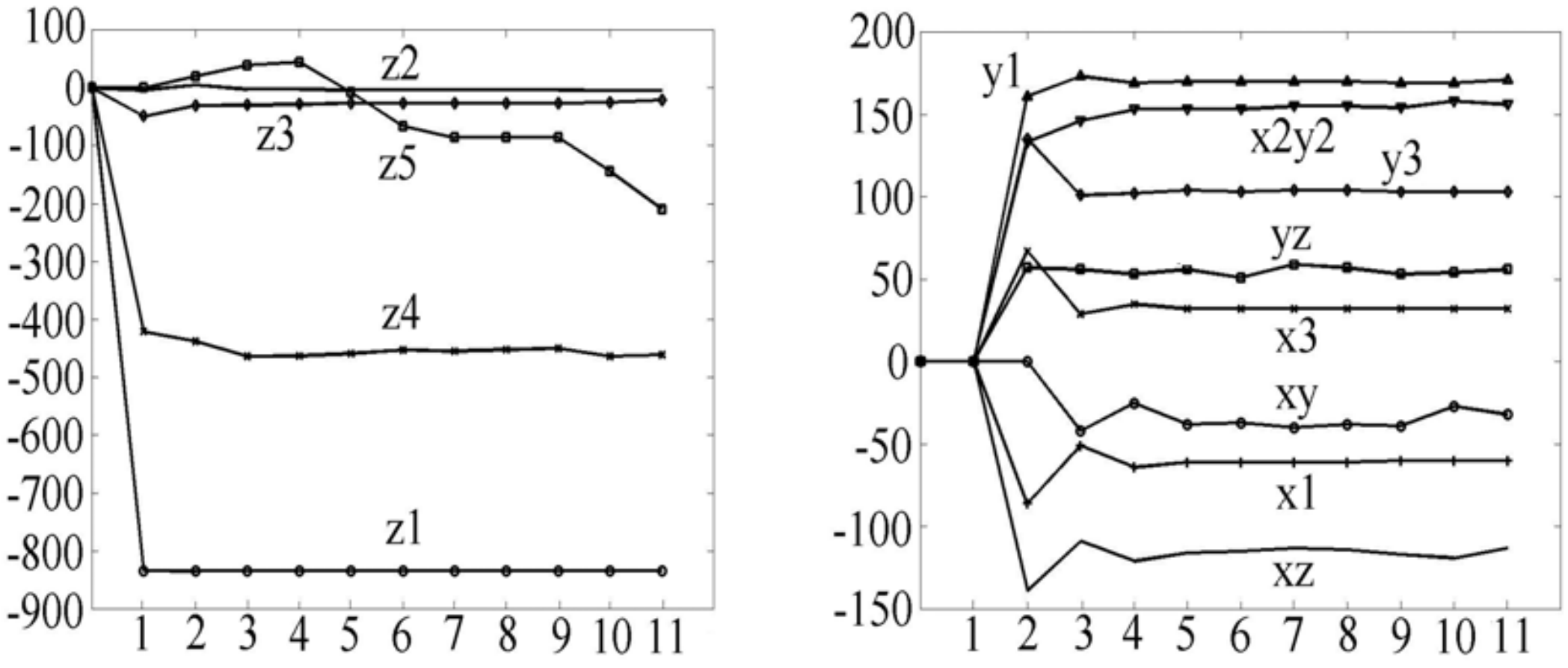

**a).** **b).**

**Fig.6.14** Changes in shim values versus iteration number in a $^1$H 3D automated shimming experiment for $z$ **(a)** and transverse **(b)** shims, as described in the text. The first iteration (1D $z$ shimming) is followed by 10 iterations of 3D shimming as described in the text.

The experimental field maps show how far the field homogeneity deviates from an ideally homogeneous field, whose map is flat. While field homogeneity improves the field maps become flatter and the frequency scale decreases; these changes represent the correction of the field variation over the sample volume, and can be seen in the maps in Fig.6.13. The low order $z$ shims $(z1 - z4)$ experienced only relatively small changes after the initial 1D $z$ shimming, while the transverse and, especially, higher order shims (including $z5$) were corrected throughout experiment.

The $^1$H spectrum shown in Fig.6.15 was acquired with sample spinning after 10 iterations of 3D shimming. It can be seen from Figs.6.8 and 6.15 that the half-widths of the spectra acquired, respectively, before and after the procedure described are in the ratio 2677 to one; the field homogeneity achieved meets comfortably the Varian specification for this spectrometer.

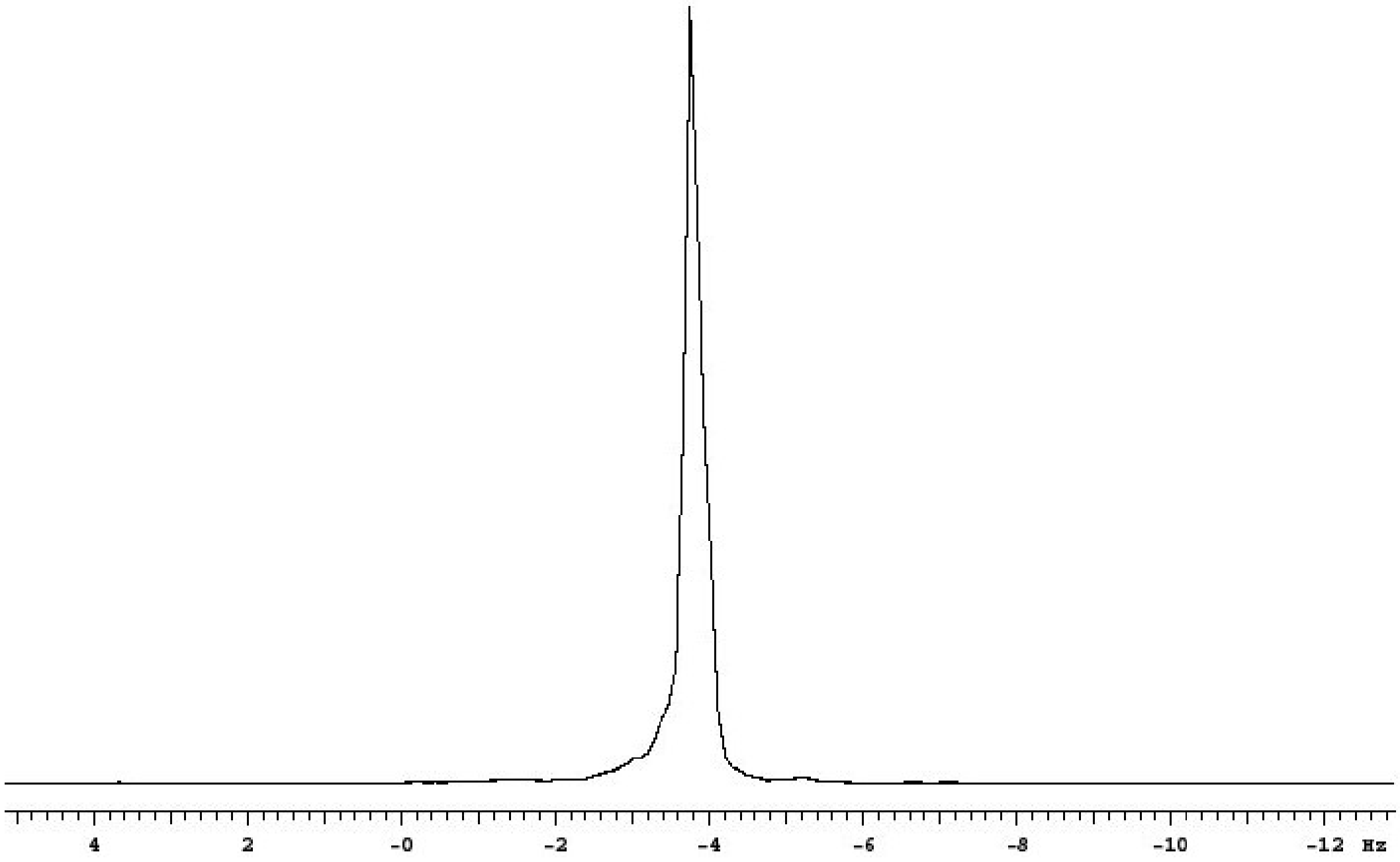

**Fig.6.15** $^{1}$H spectrum, acquired with sample spinning after 10 iterations of 3D automated shimming with matrix size of 4x4. The half-width of this line is 0.23 Hz; the widths at 0.55 and 0.11% of amplitude are 2.3 and 5.3 Hz respectively.

The final line shape is very good by the standards of manual shimming; the residual inhomogeneity at the foot reflects in part the limitations of the 13 channel shim set used and also those of the used matrix size 4x4, described in Section 6.2.

The total time required for 10 iterations of $^{1}$H 3D shimming was about 20 minutes, where the times taken for an iteration and for spectrum acquisition were 1 minute and 55 seconds and 11 seconds respectively.

## 6.4 $^{2}$H 3D shimming

It was found that $^{2}$H 3D shimming with the use of lock channel is less effective than with the X nucleus observation coil due to the low SNR of the acquired profiles. Hence, the X observation coil of the broadband probe was used, tuned to deuterium. In this section, a detailed description of a $^{2}$H 3D shimming experiment with matrix size of 4x4 and started from 'cold' shims is presented. A sample of 1% formic acid in 99% DMSO was used.

1D automated $z$ shimming was performed using the macro **gmapsys**, utilizing the homospoil facility for the production of $z$ gradients. This experiment started with mapping of $z1-z5$ shims and the field. The shim maps obtained are shown in Fig.6.16. The 1D automated $z$ shimming converged in ten iterations. The shim corrections found were used for calculation of new shim settings, which were set up. The $^1$H spectrum acquired with the new $z1-z5$ shims and all transverse shims set to zero is presented in Fig.6.17.

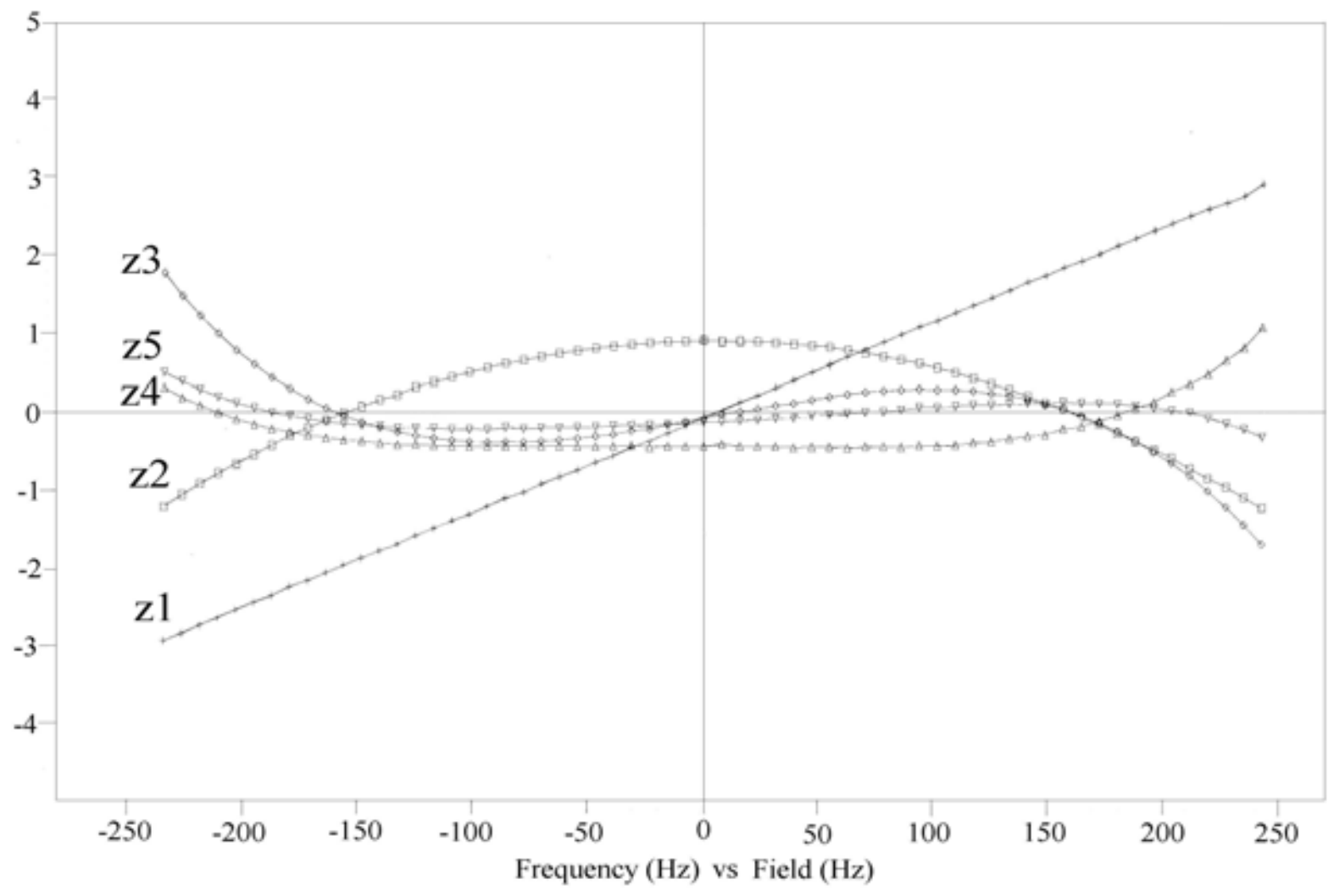

**Fig.6.16** $z1-z5$ shim maps obtained with 'cold' shims as described in the text

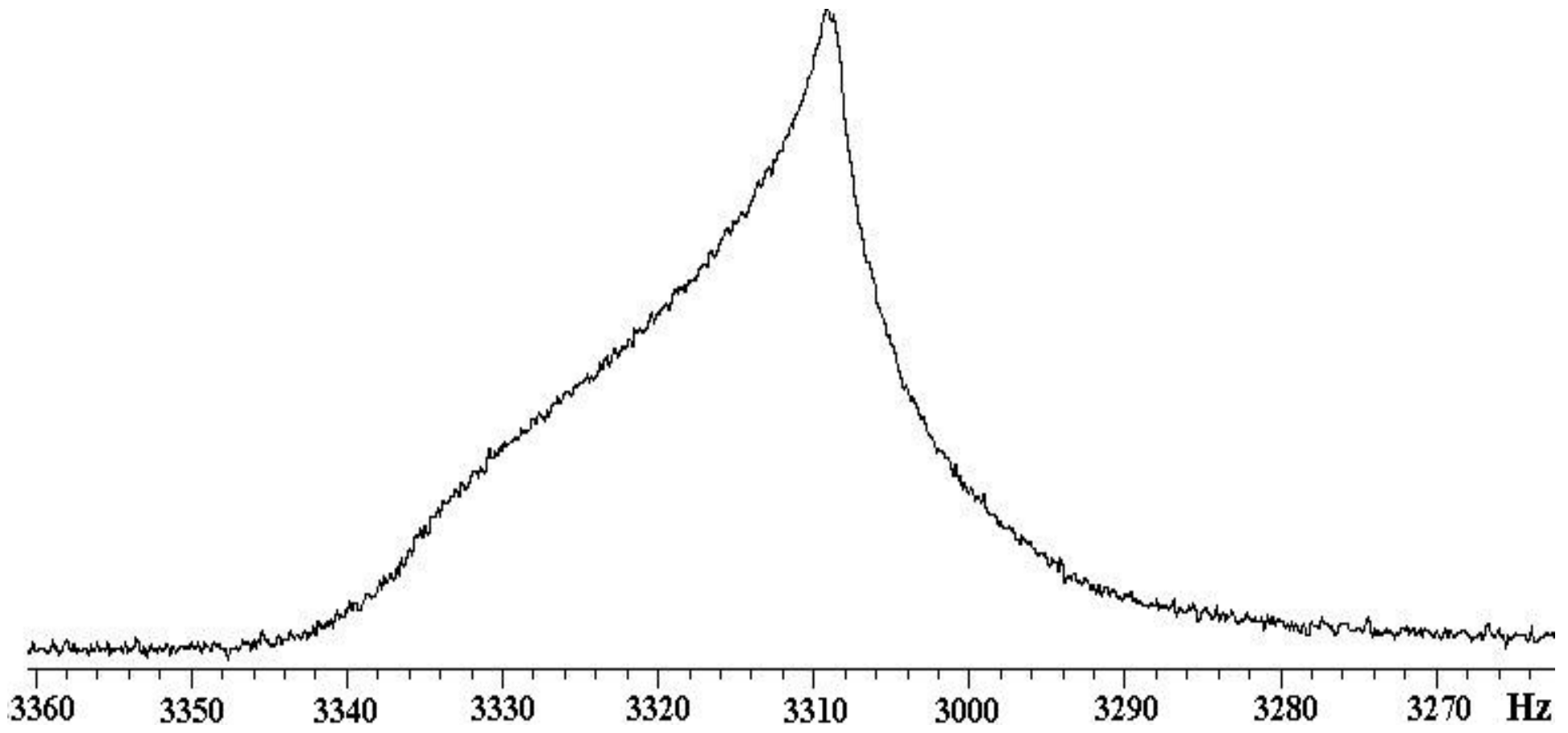

**Fig.6.17** $^1$H spectrum acquired after 10 iterations of 1D $z$ shimming as described in the text. The line widths, measured at 50/0.55/0.11% of amplitude level are, respectively, 15.19/63.7/69.8 Hz.

The spectrum broadening demonstrates strong field inhomogeneity over the sample volume, which can be corrected using 3D automated shimming. Before this, the $x1$ and $y1$ gradients were calibrated as described in Section 6.1. This was carried out on $^2$H nuclei and used single pulse spectrum experiments, performed in the presence of $x1$ and $y1$ gradients rotated through angles $\phi$ as described in Chapter 3.

The result is a set of profiles shown in Figs.6.18 and 6.19. These show an amplitude modulation pattern (Fig.6.18) similar with data in Fig.6.1, and the profile broadening (Fig.6.19) which depends on the angle of rotation and is governed by the imbalance of the gradients. The widths of the profiles were used for fitting as described in Chapter 3, and the results are summarised in Table 6.8.

**Table 6.8. The results of transverse shim gradient calibration in $^2$H shimming**

| **gcalx**, $\times 10^{-5}$ $G\,cm^{-1} DAC^{-1}$ | **gcaly**, $\times 10^{-5}$ $G\,cm^{-1} DAC^{-1}$ | **gcalang**, degree | x_error, *DAC* | y_error, *DAC* | chi-square, Hz$^2$ |
|---|---|---|---|---|---|
| 9.81 | 8.89 | -13.1 | -37 | 246 | 0.111 |

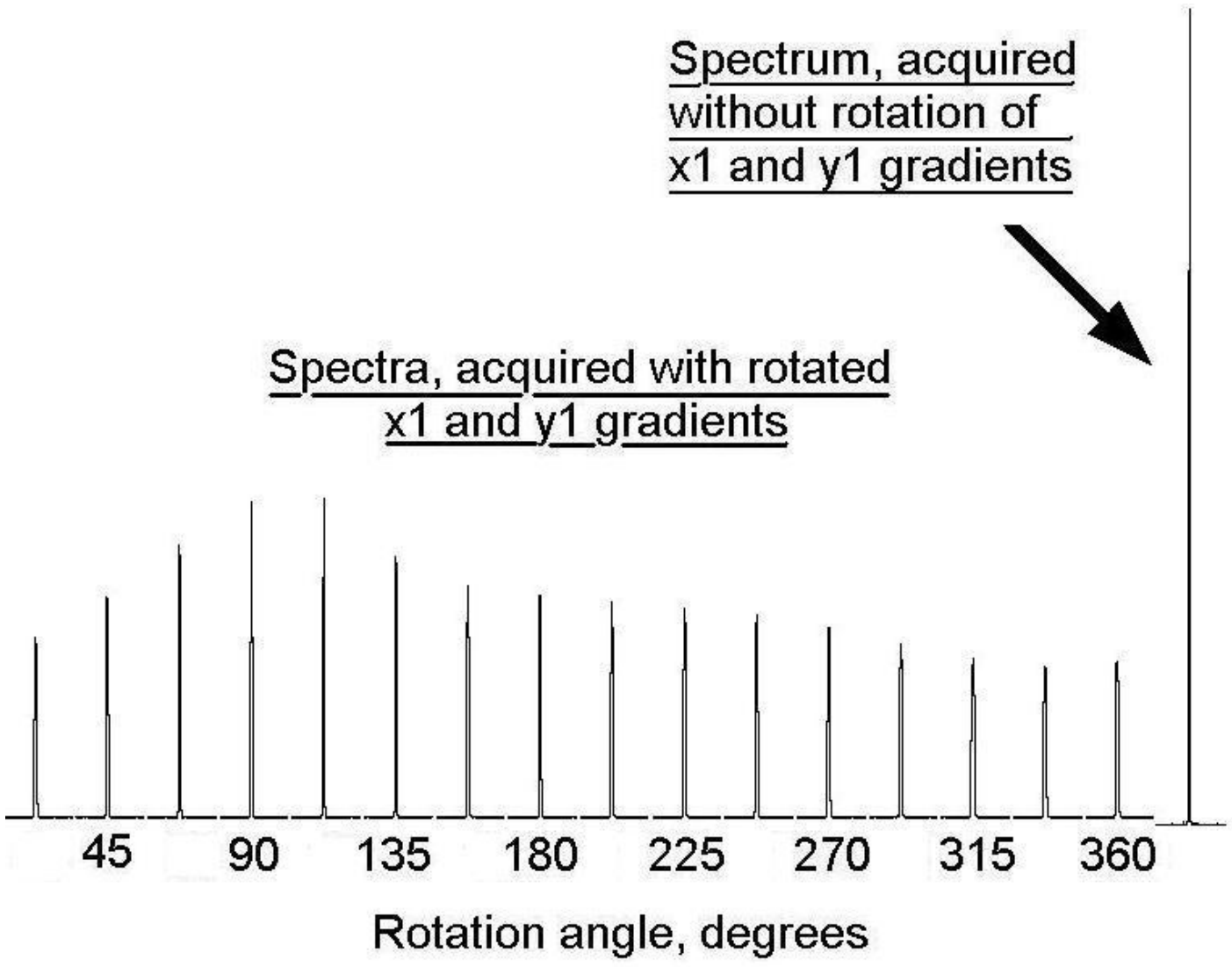

**Fig.6.18** The $^2$H profiles, acquired with the $x1$ and $y1$ gradients undergoing rotation through angle $\phi$ from 0 to 360 degrees, executed in 16 equal steps as described in the text. A spectrum acquired without the gradients is also shown.

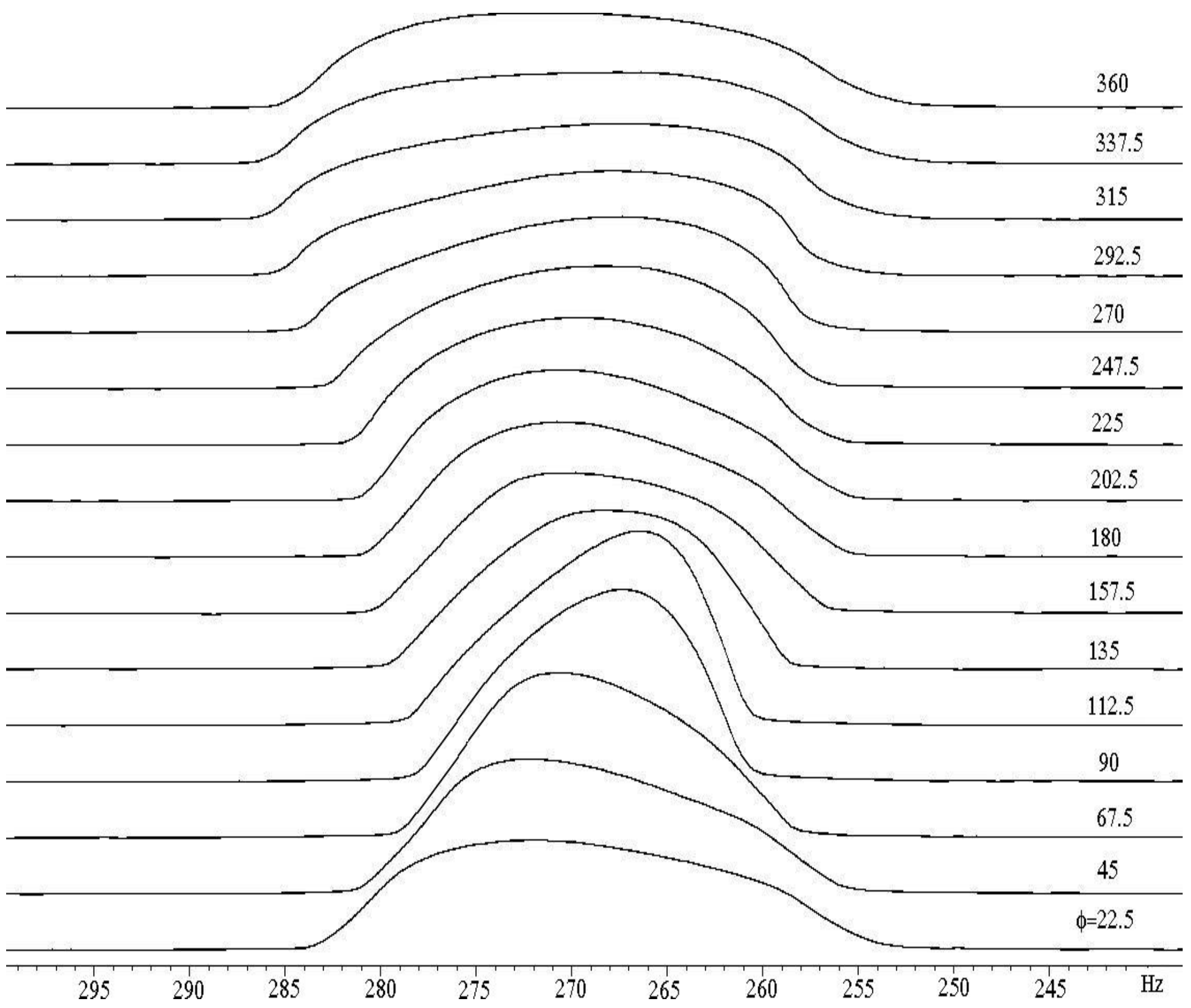

**Fig.6.19** The $^2$H profiles obtained in $x1$ and $y1$ gradient calibration experiment. The rotation angle $\phi$ increases for profiles from bottom to top as shown.

The calibrated $x1$ and $y1$ shim gradients were used for phase encoding of spins in the subsequent $^2$H 3D automated shimming experiment. This started from shim mapping carried out with the shim settings calculated in 1D $z$ shimming experiment.

The shim maps obtained (Fig.6.20) represent the field shapes generated by the shims. These were used in 3D shimming for calculation of the shim corrections by means of the linear least squares fitting programme described in Chapter 5.

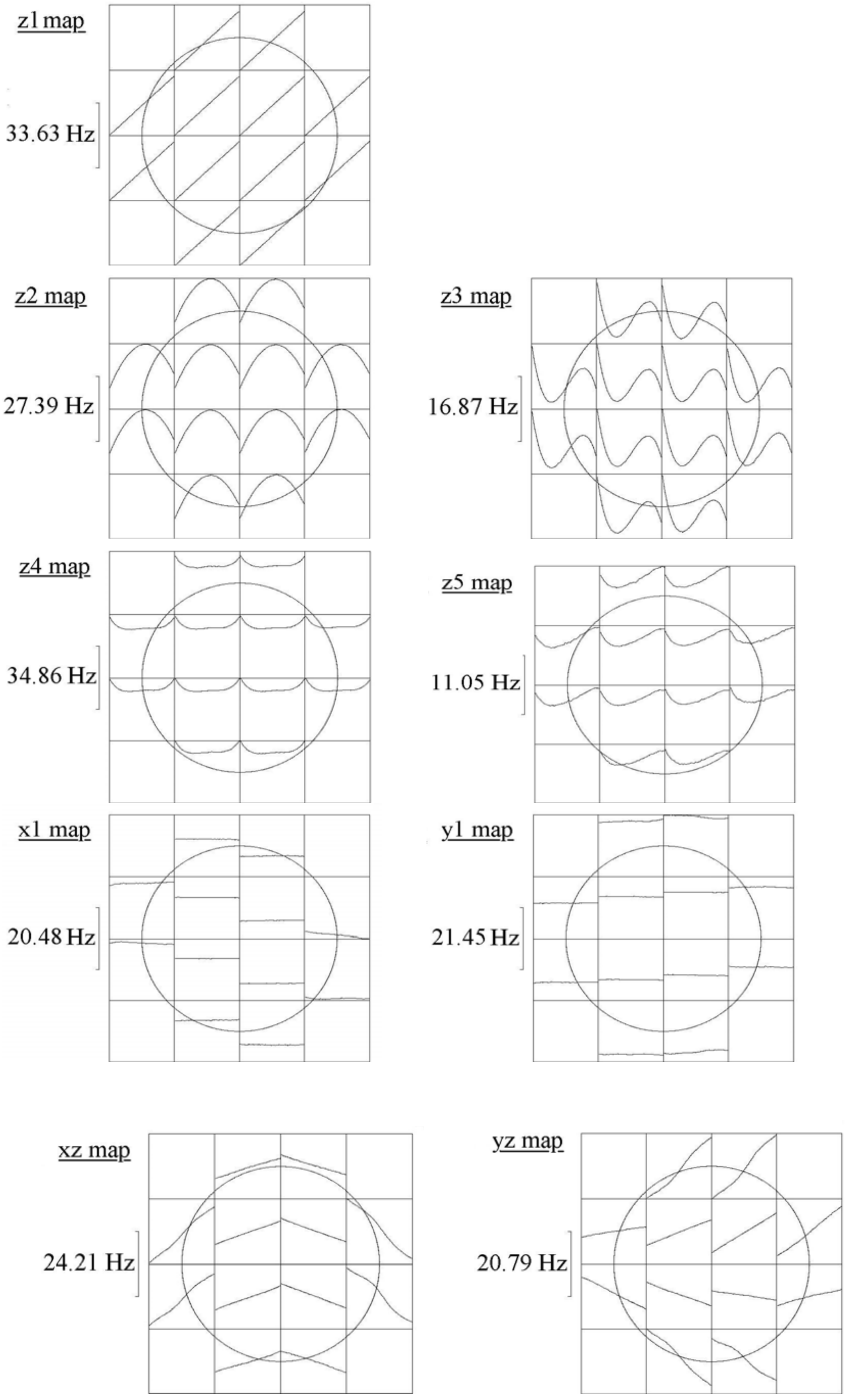

*(Fig.6.20 continued on following page)*

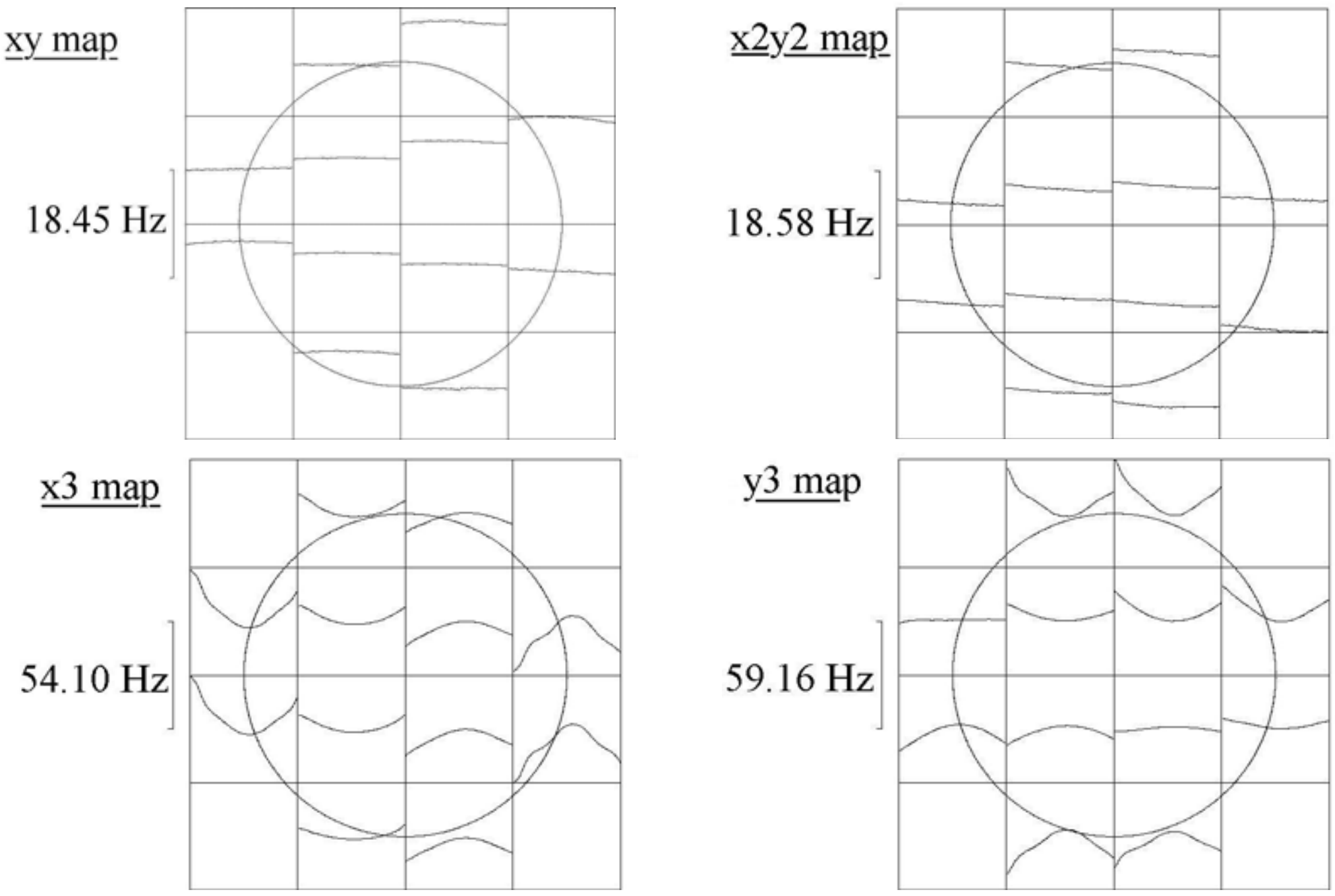

**Fig.6.20** The shim maps obtained in the $^{2}H$ 3D automated shimming experiment.

After shim mapping, six iterations of 3D shimming were performed. The field maps obtained during this experiment are shown in Fig.6.21. These show the field variations over the sample volume measured at each stage.

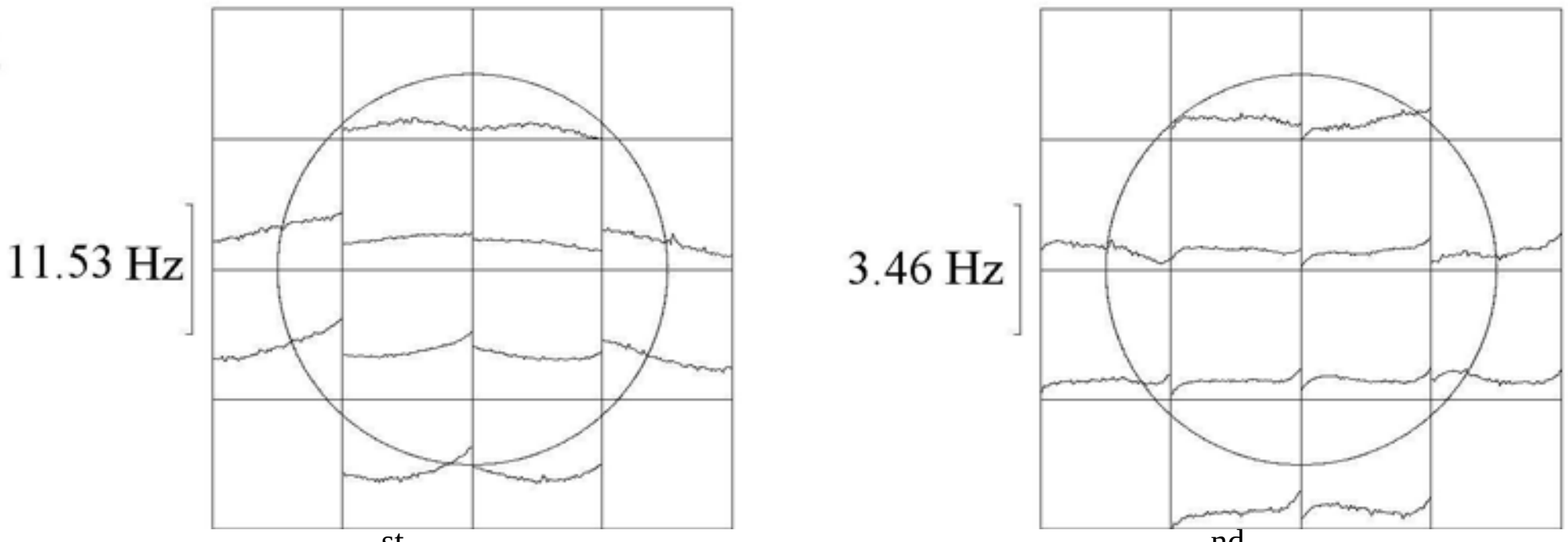

1st iteration 2nd iteration

*(Fig.6.21 continued on following page)*

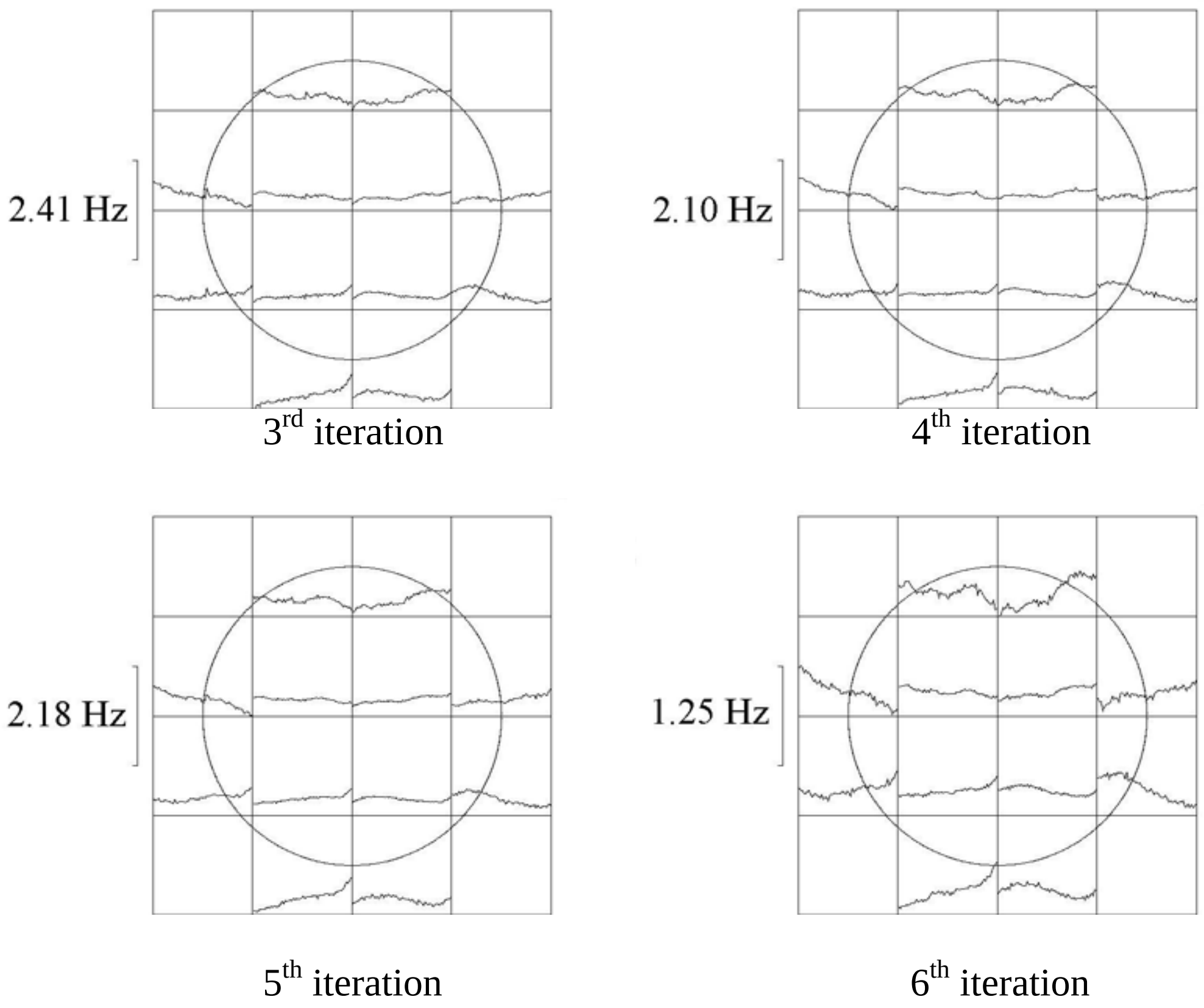

**Fig.6.21** Field maps obtained in the $^2$H 3D automated shimming experiment.

The changes in shim values are shown in Fig.6.22. It shows that after 1D $z$ shimming the main shim corrections were experienced by $z5$ and the transverse shims.

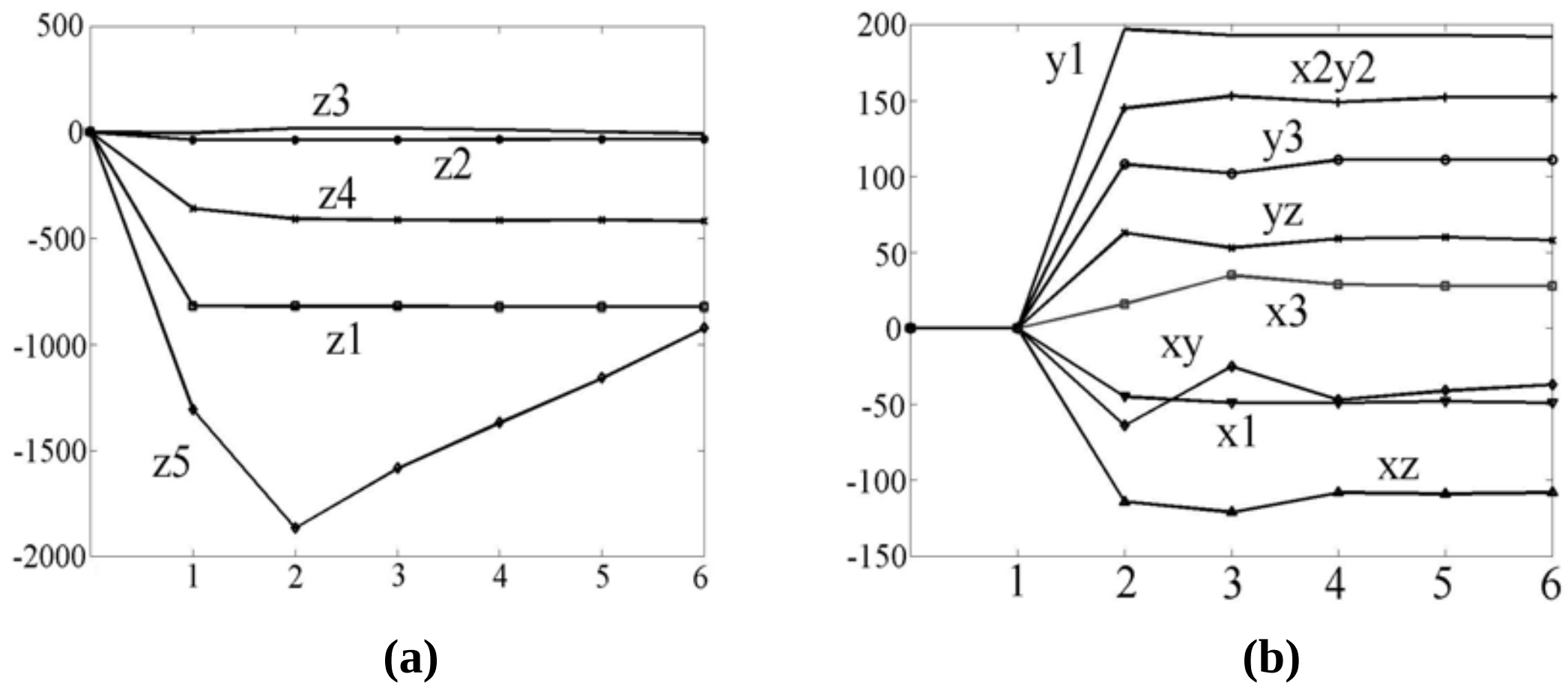

**Fig.6.22** The convergence of $z$ **(a)** and transverse **(b)** shims in the $^2$H 3D automated shimming experiment.

The $^{1}H$ spectrum acquired after the 5th iteration is shown in Fig.6.23. Its spinning line shape of 0.35/3.8/7.2 Hz is well within the manufacturer's specification for a 13-shim set (0.45/6.0/12.0 Hz). The line shape shows a slight asymmetry at the foot, which is typical, as explained in Section 6.2, of experiments using matrix size 4x4. The spinning $^{1}H$ line shape of 0.49/3.7/6.9 Hz measured after the 6th iteration is slightly broadened at half-width but narrower at the foot.

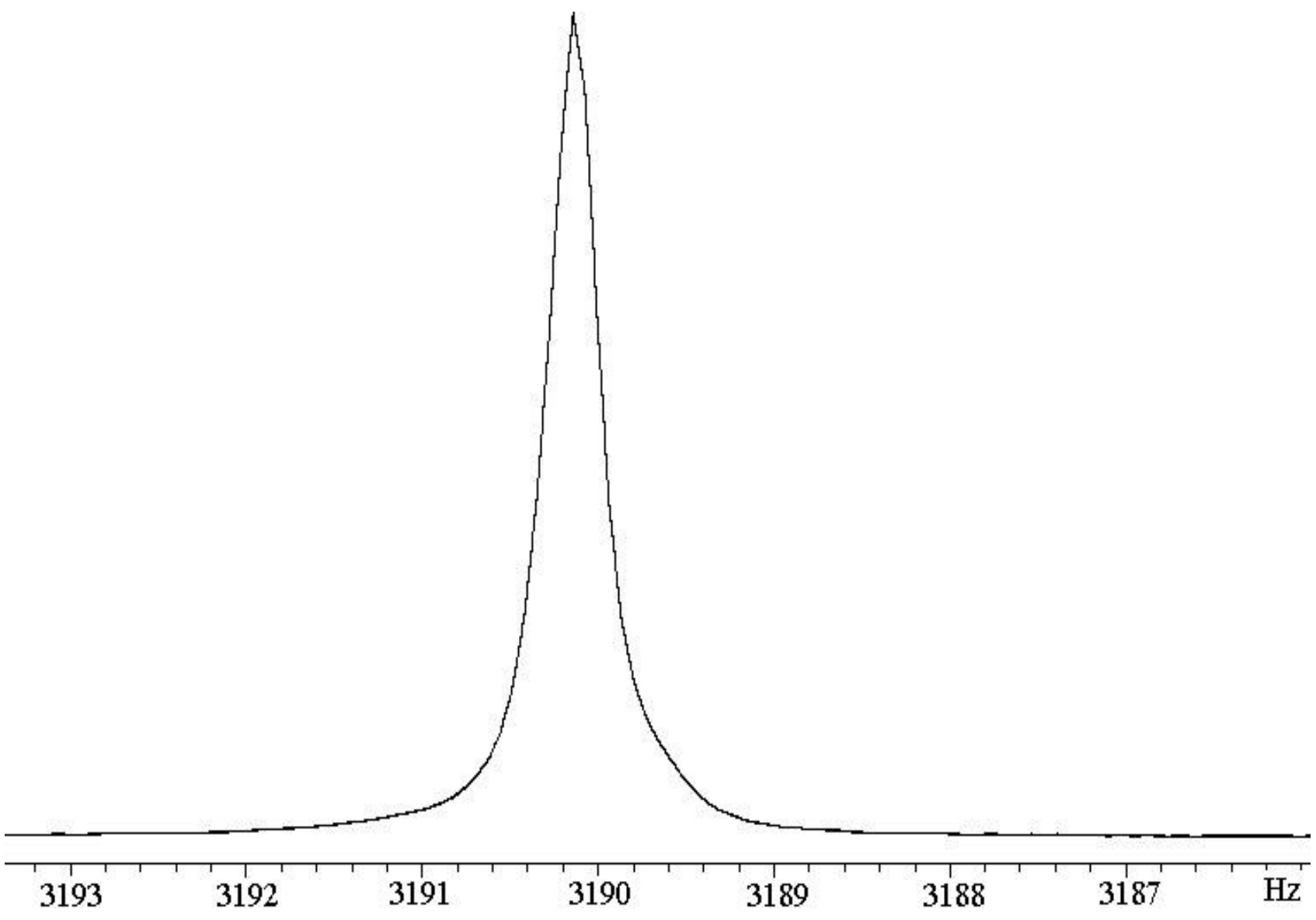

**Fig.6.23** $^{1}H$ spectrum acquired with sample spinning after 5th iteration of the $^{2}H$ 3D automated shimming experiment as described in the text. The half-width of is 0.35 Hz. The widths, measured at 0.55 and 0.11% of amplitude are, respectively 3.8 and 7.2 Hz.

## 6.5 Influence of thermal convection on 3D shimming

It was anticipated that the standard 5mm $^{1}H$ line shape sample (1% $CHCl_3$ +99% $CDCl_3$) could be used for routine shimming since this provides strong $^{1}H$ and $^{2}H$ signals and available in almost all NMR laboratories. However, 3D automated shimming experiments using this sample showed results that failed to converge, i.e. the shim corrections found spoiled field homogeneity instead of improving it. It was also found that the field mapping experiments suffered from artefacts, which were investigated and are described here.

The 3D shimming experiment was carried out using the transmitter/receiver coil tuned to $^2H$ and started with ‘cold’ shims with the sample at room temperature. The field homogeneity achieved was estimated from the line widths of the $CHCl_3$ peak measured at 50/0.55/0.11% of its amplitude. A slow improvement of the field homogeneity over 5 iterations was found as the data of Table 6.9 show.

**Table 6.9 The results of 1-5 iterations of $^1H$ 3D shimming** using the standard line shape sample, as described in the text.

| Number of iteration | Line widths of $CHCl_3$ line, Hz measured at | | |
|---|---|---|---|
| | 50% | 0.55% | 0.11% |
| **1** | 10.71 | 78.7 | 79.1 |
| **2** | 8.92 | 58.1 | 58.3 |
| **3** | 2.86 | 50.0 | 54.9 |
| **4** | 3.17 | 37.4 | 48.8 |
| **5** | 2.84 | 32.7 | 40.3 |

The field map, obtained in the 5th iteration and presented in Fig.6.24 shows an almost normal field map except for some additional curvature, which was not noticed before in similar experiments.

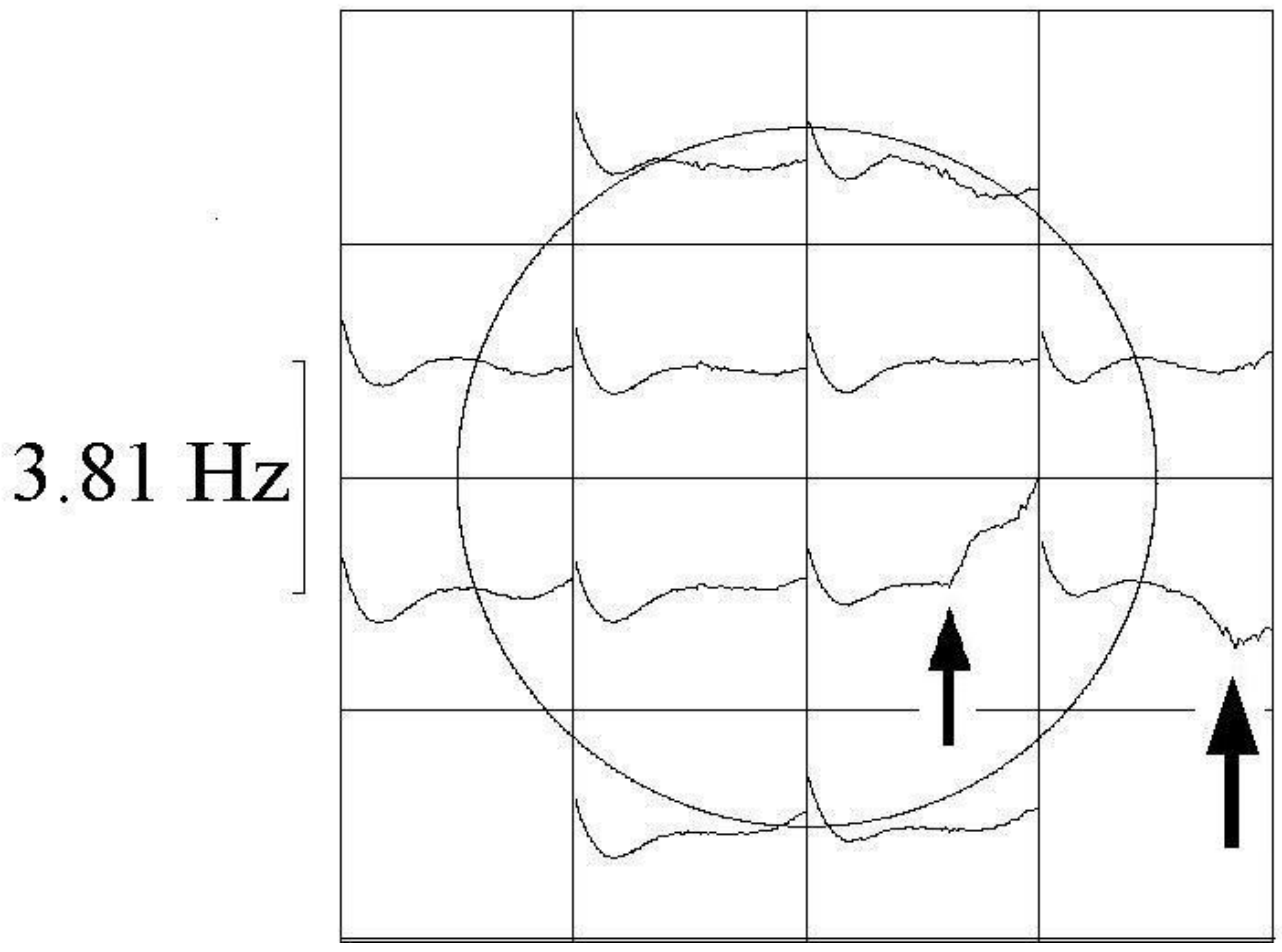

**Fig.6.24** Field map obtained in 5th iteration of 3D shimming, as described in the text.

It was found that the discontinuities in the field maps were associated with sudden drops in the amplitudes of the profiles acquired with delay $\tau$ in the field mapping experiment described in Chapter 3. As a test, three profiles were acquired with the same experimental parameters but with about 1 minute delay between their acquisitions. These (Fig.6.25) show that the drops in the profile amplitudes are different at different times.

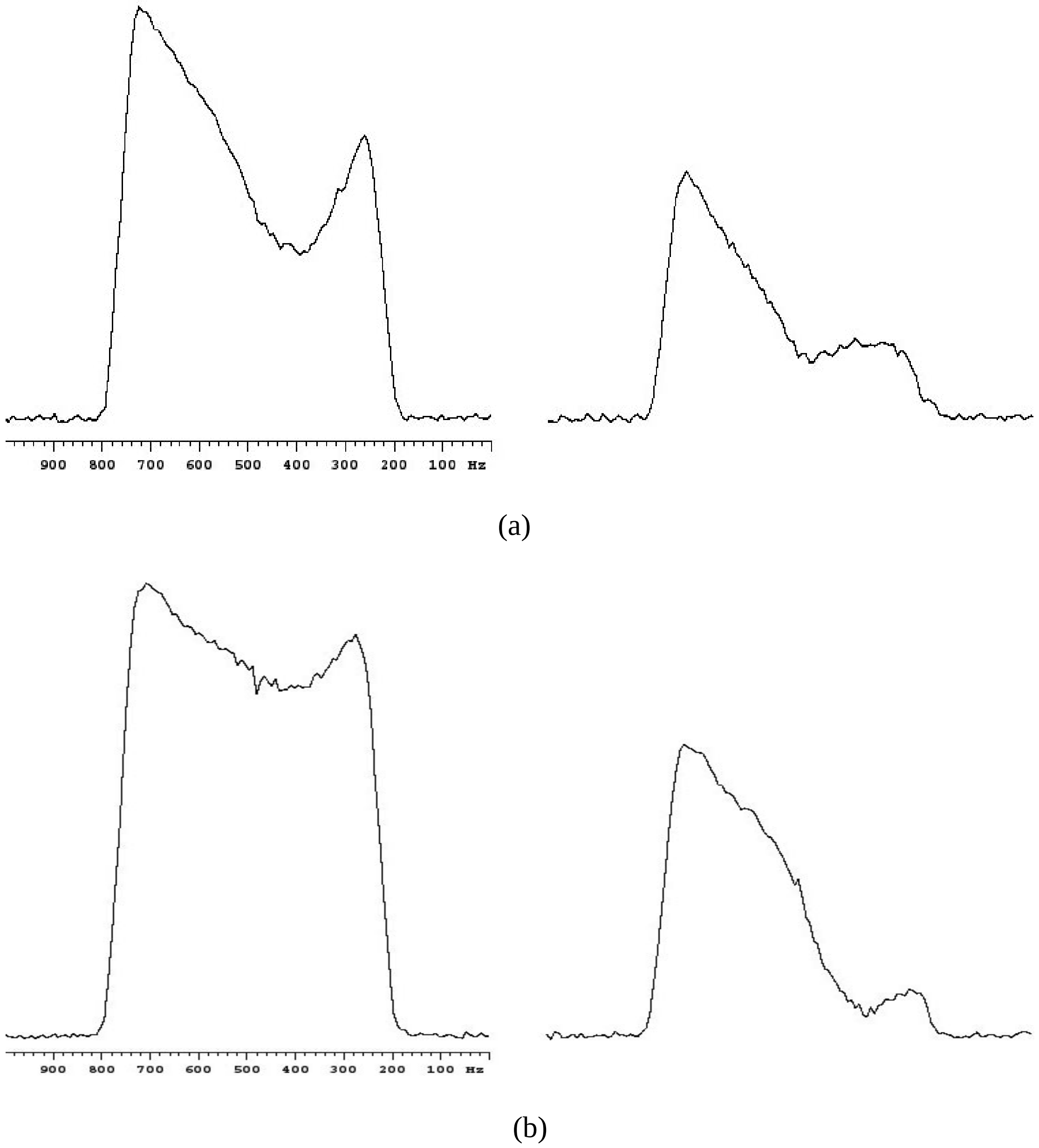

**Fig.6.25** Two pairs of sample profiles acquired without phase-encoding and with identical experimental parameters. Pair **(b)** was acquired a minute later than **(a)** as described in the text.

It was postulated that thermal convection could be a reason for the profile shape distortions resulting in the field map distortions described. Hence, field mapping experiments were tried at different temperatures in order to find out how temperature variation affects the results. The field and amplitude maps obtained at regulated sample temperatures ranging from 26 to 29 $^0$C are presented in Fig.6.26.

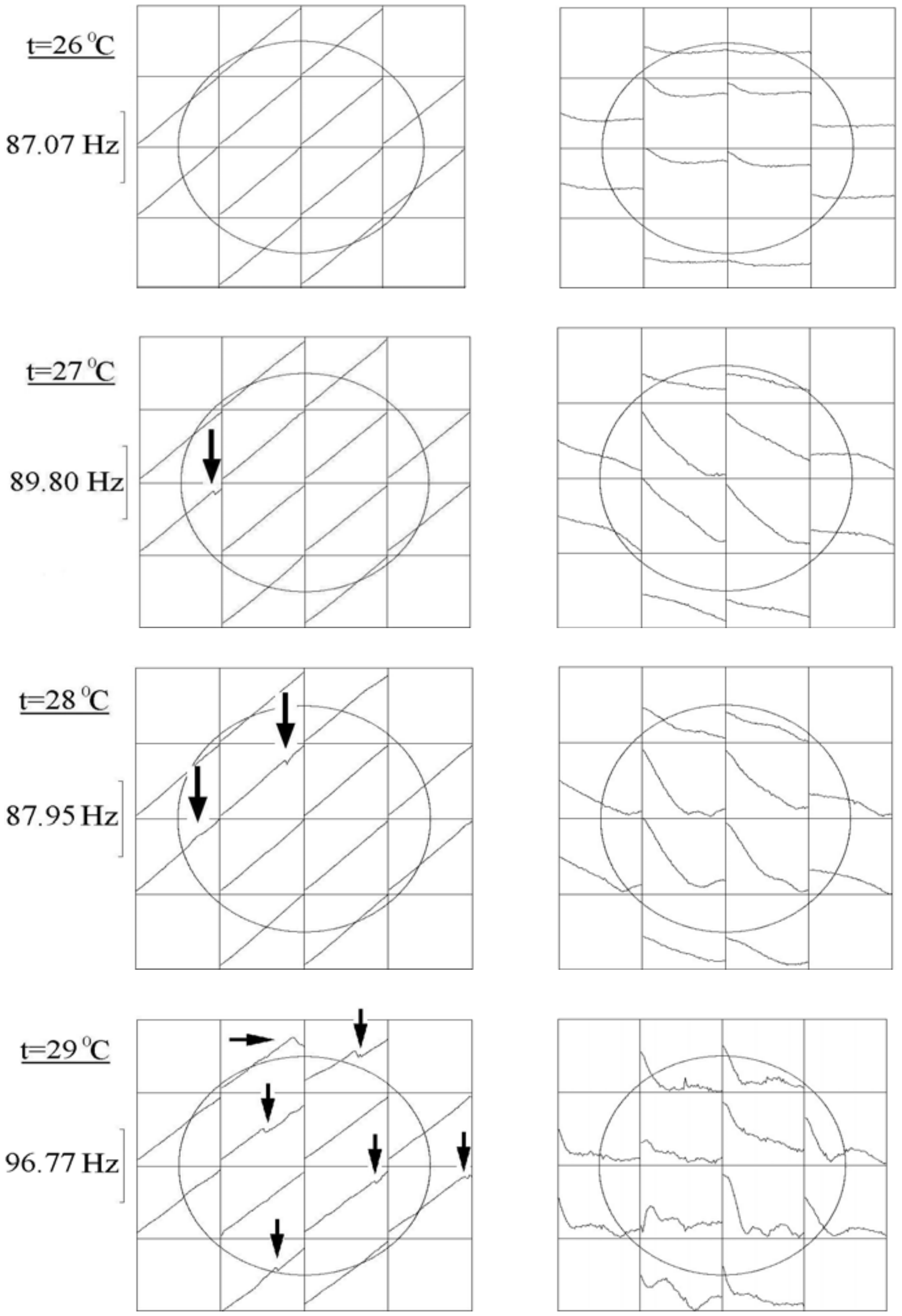

**Fig 6.26.** $^2$H field (left) and amplitude (right) maps obtained for temperatures in the range 26 to 29 $^0$C using the standard line shape test sample, as described in the text. The arrows indicate discontinuities in field maps caused by collapse of signals in amplitude maps.

The experiment was actually carried out up to 35 $^{0}$C, where distortions continuing to increase. It can be seen from the amplitude maps, shown in Fig.6.26 in the right side column, that signals in these maps start to collapse at temperatures near 27 $^{0}$C. This collapse can be attributed to thermal movements of spins caused by temperature gradients within the sample. This effect was observed exclusively using low viscosity samples, and can occur at room temperature as well as was demonstrated at the beginning of this Section (Fig.6.24). Since large changes in shim settings can lead to significant changes in the probe temperature, convection can cause particular problems when shimming from ‘cold’ shims. It is therefore preferable to use more viscous sample such as the PEO sample, described in Section 6.3.

## 6.6 Experiments with other solvents

The technique described was also tested for $^{2}$H 3D shimming with the following solvents: [1% $CHCl_3$+99% acetone-$d_6$] and [1% acetone+99% $D_2O$]. The results of these experiments had shown that field inhomogeneity within manufacturer specification for 5mm samples is achievable using these solvents just after few iterations of 3D shimming. However, some of these experiments suffered from random and systematic errors to larger extent than those described in Sections 6.3, 6.4 and their results were less reproducible.

# 7

# Conclusions

The work presented in this thesis illustrates some new developments in 3D automated shimming. The aim of the work was to implement and improve a previous technique for 3D shimming using normal hardware by optimising the experimental parameters and calibrating the strengths of the transverse linear shim gradients. The studies have taken place at different levels of the NMR experiment – from experimenting with different hardware, to programming of pulse sequences and signal processing software.

Investigations of the linear transverse shim gradients have shown that their non-orthogonality can be measured and corrected to a high precision, more than sufficient for the use of the shim gradients in field and shim mapping. The calibration technique can also be used for testing the shim gradient orthogonality and strengths on new instruments.

As the results have shown, the 3D automated shimming technique described can be used both with protonated and deuterated solvents, which together cover the vast majority of samples for high-resolution NMR spectroscopy. It was observed that the technique is very effective in correcting static magnetic field inhomogeneity, and when properly used allows field optimisation to meet the manufacturer's specification within a few iterations.

The performance of the described technique is limited both by systematic and by random errors. Temperature effects on the field generated by the magnet are a common source of changes in field inhomogeneity. External field perturbations can distort field mapping since it is not possible to maintain field-frequency lock during experiment. The SNR of profiles measured also limits the SNR of the field maps produced.

Some other factors may also cause phase changes of spins during the pulse sequence. One of these factors, thermal convection of spins, was identified during experiments and is described in Chapter 6. It was found that the result of 3D shimming does not improve much with increases in the number of transients, as presented in Section 6.2 of Chapter 6. It suggests that systematic errors outweight the random errors caused by limited SNR.

It was shown that the time taken for 3D shimming experiment can be minimized by optimising the number of phase-encoding gradient increments in the $xy$ plane. This determines the number of data points in the 3D Fourier Transform and is a critical parameter. The optimal number of increments, found here is 4; this was used both for $^1$H and $^2$H 3D shimming, as represented in Chapter 6. A comparison of the results using different numbers of increments was carried; it is found that normally the use of even numbers of increments gave better results that odd.

It has been shown that using 1D $z$ shimming before 3D shimming can be a very effective and rapid approach to shimming, since the $z$ inhomogeneity is normally larger than transverse inhomogeneities and can be approximately adjusted independently. This approach was found very effective in numerous experiments; some of those describing $^1$H and $^2$H 3D shimming from 'cold' shims are described in Chapter 6.

Choice of samples, for routine use of the 3D shimming was also investigated. It was found that stable and reproducible results can be obtained using 5% PEO in (1%$CHCl_3$ +99% $CDCl_3$) for $^1$H 3D shimming and a sample of 1% formic acid in 99% DMSO can be recommended for $^2$H 3D shimming. The use of the standard line shape (1% $CHCl_3$ in acetone-$d_6$) is limited by the effects of thermal convection on 3D field mapping, described in Section 6.5 of Chapter 6.

Another important limitation of the 3D shimming technique described stems from the design of the Golay coils themselves. While the field shapes generated by the coils represent a limited set of spherical harmonics, the actual field inhomogeneity have less regular patterns and would require a much larger set of harmonics (and shim coils) for full correction. Field maps sometimes include ‘spikes’ and discontinuities caused by magnetic susceptibility differences over the sample volume. These differences can often be attributed to the materials either within the sample or around the transmitter/receiver coil in the probe. For example, copper foil transmitter/ receiver coils can induce high-order field inhomogeneities, which are difficult to compensate by the use of normal Golay coils. One useful possible development that has yet to be fully explored practically is to find a better criterion for choosing the $\tau$ delay (described in Chapter 2), which would be less sensitive to noise in the measured spectra.

In summary, the results of described 3D automated shimming technique can more than meet manufacturer’s specification but are limited both by random and systematic errors and by the finite size of shim sets to an extent which still prevents observation of $^1$H natural line widths in many high-resolution applications. This method provides a sound basis of 3D homogeneity optimisation with normal hardware. The preliminary results show that this method can be successfully used with 10 mm samples as well. This method could be tested for use with Magic Angle Spinning (MAS) probes, which are not normally equipped with any PFG hardware.

In collaboration with Varian, the 3D shimming technique described was successfully used for shimming of higher field magnets (e.g. 700 MHz spectrometer with 39 channel shim set).

# REFERENCES

## Chapter 1

## Chapter 2

## Chapter 3

## Chapter 4

## Chapter 5

# Appendix A

## Calculation of NMR signal in the presence of static magnetic field inhomogeneity using density matrix formalism

Density operator of a spin ensemble in a static magnetic field may be written in the rotating frame of reference as:

$$\hat{\rho}_0(\vec{r}) = a + b(\vec{r})\hat{I}_z \qquad \text{(A.1)},$$

where $a$ and $b(\vec{r})$ are constants, described in Chapter 2. After 90 degree $x$ - pulse density operator becomes:

$$\hat{\rho}(\vec{r}, t_{90^0}) = e^{-\left(\frac{i}{\hbar}\right)\hat{H}_{xrf}{}^{rot} t_{90^0}} \hat{\rho}_0(\vec{r}) e^{\left(\frac{i}{\hbar}\right)\hat{H}_{xrf}{}^{rot} t_{90^0}} \qquad \text{(A.2)},$$

where

$$\hat{H}_{xrf}{}^{rot} = -\gamma\hbar B_1 \hat{I}_x \qquad \text{(A.3)}$$

is the Hamiltonian for interaction between the spins and the rf pulse.

For an ensemble of spins ½, equation (A.2) can be expressed in matrix form as follows:

$$\rho(\vec{r}, t_{90^0}) = a + b(\vec{r}) \begin{pmatrix} \cos\left(\frac{\alpha}{2}\right) & i\sin\left(\frac{\alpha}{2}\right) \\ i\sin\left(\frac{\alpha}{2}\right) & \cos\left(\frac{\alpha}{2}\right) \end{pmatrix} \begin{pmatrix} \frac{1}{2} & 0 \\ 0 & -\frac{1}{2} \end{pmatrix} \begin{pmatrix} \cos\left(\frac{\alpha}{2}\right) & -i\sin\left(\frac{\alpha}{2}\right) \\ -i\sin\left(\frac{\alpha}{2}\right) & \cos\left(\frac{\alpha}{2}\right) \end{pmatrix} \qquad \text{(A.4)}$$

where $\alpha = \gamma B_1 t_{90^0}$.

After multiplication, the density matrix becomes:

$$\rho(\vec{r}, t_{90^0}) = a + b(\vec{r})\begin{pmatrix} \frac{1}{2}\cos^2\left(\frac{\alpha}{2}\right) - \frac{1}{2}\sin^2\left(\frac{\alpha}{2}\right) & -i\cos\left(\frac{\alpha}{2}\right)\sin\left(\frac{\alpha}{2}\right) \\ i\cos\left(\frac{\alpha}{2}\right)\sin\left(\frac{\alpha}{2}\right) & \frac{1}{2}\sin^2\left(\frac{\alpha}{2}\right) - \frac{1}{2}\cos^2\left(\frac{\alpha}{2}\right) \end{pmatrix}$$

(A.5).

Since $\alpha = 90^0$ the density matrix simplifies to:

$$\rho(\vec{r}, t_{90^0}) = a + b(\vec{r})\begin{pmatrix} 0 & -\frac{i}{2} \\ \frac{i}{2} & 0 \end{pmatrix} \qquad \text{(A.6)},$$

what can be rewritten as:

$$\rho(\vec{r}, t_{90^0}) = a + b(\vec{r}) I_y \qquad \text{(A.7)}.$$

After the rf pulse is turned off, the spins evolve in the inhomogeneous part of the static magnetic field $\delta B(\vec{r})$. The Hamiltonian of this interaction is given by:

$$\hat{H}_0^{'} = -\gamma\hbar I_z \delta B(\vec{r}) \qquad \text{(A.8)}$$

Hence, density operator evolves after the rf pulse, according to:

$$\hat{\rho}(\vec{r}, t_{90} + t_1) = e^{-\frac{i}{\hbar}\hat{H}_0^{'} t_1}\, \hat{\rho}(\vec{r}, t_{90^0})\, e^{\frac{i}{\hbar}\hat{H}_0^{'} t_1} \qquad \text{(A.10)}.$$

Then density matrix is expressed by:

$$\rho(\vec{r}, t_{90^0} + t_1) = a + b(\vec{r})\begin{pmatrix} e^{i\frac{\varepsilon_1(\vec{r},t_1)}{2}} & 0 \\ 0 & e^{-i\frac{\varepsilon_1(\vec{r},t_1)}{2}} \end{pmatrix}\begin{pmatrix} 0 & -\frac{i}{2} \\ \frac{i}{2} & 0 \end{pmatrix}\begin{pmatrix} e^{-i\frac{\varepsilon_1(\vec{r},t_1)}{2}} & 0 \\ 0 & e^{i\frac{\varepsilon_1(\vec{r},t_1)}{2}} \end{pmatrix} =$$

$$a + b(\vec{r})\begin{pmatrix} 0 & -\frac{i}{2} e^{i\varepsilon_1(\vec{r},t_1)} \\ \frac{i}{2} e^{-i\varepsilon_1(\vec{r},t_1)} & 0 \end{pmatrix} \qquad \text{(A.10)}$$

where $\varepsilon_1(\vec{r}, t_1) = \gamma\delta B(\vec{r})t_1$ is the phase local precession angle proportional to field inhomogeneity at the position $\vec{r}$ . After decomposition of exponential:

$$e^{i\varepsilon_1} = \cos\varepsilon_1 + i\sin\varepsilon_1 \qquad \text{(A.11)},$$

the Eqn.(A.10) becomes:

$$\rho(\vec{r}, t_{90^0} + t_1) = a + b(\vec{r})\begin{pmatrix} 0 & -\frac{i}{2}(\cos\varepsilon_1(\vec{r}, t_1) + i\sin\varepsilon_1(\vec{r}, t_1)) \\ \frac{i}{2}(\cos\varepsilon(\vec{r}, t_1)_1 - i\sin\varepsilon_1(\vec{r}, t_1)) & 0 \end{pmatrix} \qquad \text{(A.12)}.$$

This can be rewritten in the form:

$$\rho(\vec{r}, t_{90^0} + t_1) = a + b(\vec{r})\left[I_y \cos\varepsilon_1(\vec{r}, t_1) + I_x \sin\varepsilon_1(\vec{r}, t_1)\right] \qquad \text{(A.13)}.$$

## Appendix B

# Density matrix evolution during field mapping using modified PFGSTE pulse sequence

The equilibrium density matrix of spin ensemble in presence of the static magnetic field can be given in the rotating frame of reference by

$$\rho_0(\vec{r}) = a + b(\vec{r}) I_z \qquad \text{(B.1).}$$

The sign convention for rotations here is consistent with used in[(2.64)]. In matrix form it can be written as:

$$\rho_0(\vec{r}) = a + b(\vec{r}) \begin{pmatrix} \frac{1}{2} & 0 \\ 0 & -\frac{1}{2} \end{pmatrix} \qquad \text{(B.2).}$$

After 90 degree pulse, it becomes:

$$\rho(\vec{r}, t_{90^0}) = a + b(\vec{r}) \begin{pmatrix} \cos\left(\frac{\alpha}{2}\right) & i\sin\left(\frac{\alpha}{2}\right) \\ i\sin\left(\frac{\alpha}{2}\right) & \cos\left(\frac{\alpha}{2}\right) \end{pmatrix} \begin{pmatrix} \frac{1}{2} & 0 \\ 0 & -\frac{1}{2} \end{pmatrix} \begin{pmatrix} \cos\left(\frac{\alpha}{2}\right) & -i\sin\left(\frac{\alpha}{2}\right) \\ -i\sin\left(\frac{\alpha}{2}\right) & \cos\left(\frac{\alpha}{2}\right) \end{pmatrix} \qquad \text{(B.3)}$$

where $\alpha = \gamma B_1 t_{90^0}$. After multiplication:

$$\rho(\vec{r}, t_{90^0}) = a + b(\vec{r}) \begin{pmatrix} \frac{1}{2}\cos^2\left(\frac{\alpha}{2}\right) - \frac{1}{2}\sin^2\left(\frac{\alpha}{2}\right) & -i\cos\left(\frac{\alpha}{2}\right)\sin\left(\frac{\alpha}{2}\right) \\ i\cos\left(\frac{\alpha}{2}\right)\sin\left(\frac{\alpha}{2}\right) & \frac{1}{2}\sin^2\left(\frac{\alpha}{2}\right) - \frac{1}{2}\cos^2\left(\frac{\alpha}{2}\right) \end{pmatrix} \qquad \text{(B.4).}$$

For a 90 degree pulse $\alpha = 90^0$, and the density matrix simplifies to:

$$\rho\left(\vec{r}, t_{90^0}\right) = a + b(\vec{r})\begin{pmatrix} 0 & -\frac{i}{2} \\ \frac{i}{2} & 0 \end{pmatrix} = a + b(\vec{r}) I_y$$

(B.5).

The Hamiltonian of time evolution during time $t_1$ is:

$$\hat{H}_G = -\gamma\hbar \hat{I}_z \vec{G} \cdot \vec{r} \qquad \text{(B.6)},$$

where $\vec{G}\left(G_x, G_y, G_z\right)$ is 3-dimensional phase encoding field gradient and $\vec{r}$ is a vector of position. The spin ensemble, represented by density matrix evolves due to gradient according to:

$$\rho\left(\vec{r}, t_{90^0} + t_1\right) = e^{-\frac{i}{\hbar} H_G t_1} \rho\left(\vec{r}, t_{90^0}\right) e^{\frac{i}{\hbar} H_G t_1} \qquad \text{(B.7)}.$$

In matrix form:

$$\rho\left(\vec{r}, t_{90^0} + t_1\right) = a + b(\vec{r})\begin{pmatrix} e^{i\frac{\varphi(\vec{r})}{2}} & 0 \\ 0 & e^{-i\frac{\varphi(\vec{r})}{2}} \end{pmatrix}\begin{pmatrix} 0 & -\frac{i}{2} \\ \frac{i}{2} & 0 \end{pmatrix}\begin{pmatrix} e^{-i\frac{\varphi(\vec{r})}{2}} & 0 \\ 0 & e^{i\frac{\varphi(\vec{r})}{2}} \end{pmatrix} \qquad \text{(B.8)}$$

where $\varphi(\vec{r}) = \gamma \vec{G}(\vec{r}) \cdot \vec{r} t_1$ is the local precession angle executed by spins at time $t_1$ and position $\vec{r}$ in presence of the linear gradient $\vec{G}(\vec{r})$.

After multiplication:

$$\rho\left(\vec{r}, t_{90^0} + t_1\right) = a + b(\vec{r})\begin{pmatrix} 0 & -\frac{i}{2} e^{i\frac{\varphi(\vec{r})}{2}} \\ \frac{i}{2} e^{-i\frac{\varphi(\vec{r})}{2}} & 0 \end{pmatrix}\begin{pmatrix} e^{-i\frac{\varphi(\vec{r})}{2}} & 0 \\ 0 & e^{i\frac{\varphi(\vec{r})}{2}} \end{pmatrix} =$$

$$a + b(\vec{r})\begin{pmatrix} 0 & -\frac{i}{2} e^{i\varphi(\vec{r})} \\ \frac{i}{2} e^{-i\varphi(\vec{r})} & 0 \end{pmatrix}$$

(B.9).

The exponential may be decomposed:

$$e^{i\varphi} = \cos\varphi + i \sin\varphi \qquad \text{(B.10)}.$$

Then (B.9) becomes:

$$\rho\left(\vec{r}, t_{90^0} + t_1\right) = a + b\left(\vec{r}\right)\begin{pmatrix} 0 & -\frac{i}{2}\left(\cos\varphi\left(\vec{r}\right) + i\sin\varphi\left(\vec{r}\right)\right) \\ \frac{i}{2}\left(\cos\varphi\left(\vec{r}\right) + i\sin\varphi\left(\vec{r}\right)\right) & 0 \end{pmatrix}$$

(B.11),

where the matrix on the right side may be rewritten as the sum of two matrices:

$$\rho\left(\vec{r}, t_{90^0} + t_1\right) = a + b\left(\vec{r}\right)\begin{pmatrix} 0 & -\frac{i}{2}\cos\varphi\left(\vec{r}\right) + \frac{1}{2}\sin\varphi\left(\vec{r}\right) \\ \frac{i}{2}\cos\varphi\left(\vec{r}\right) - \frac{1}{2}\sin\varphi\left(\vec{r}\right) & 0 \end{pmatrix} =$$

$$a + b\left(\vec{r}\right)\left\{\begin{pmatrix} 0 & -\frac{i}{2}\cos\varphi\left(\vec{r}\right) \\ \frac{i}{2}\cos\varphi\left(\vec{r}\right) & 0 \end{pmatrix} + \begin{pmatrix} 0 & \frac{1}{2}\sin\varphi\left(\vec{r}\right) \\ -\frac{1}{2}\sin\varphi\left(\vec{r}\right) & 0 \end{pmatrix}\right\} \quad \text{(B.12)}$$

It can be seen that the matrices in brackets represent $\hat{I}_y$ and $\hat{I}_x$ operators, respectively. Thus,

(B.12) can be finally rewritten in the form:

$$\rho\left(\vec{r}, t_{90^0} + t_1\right) = a + b\left(\vec{r}\right)\left\{I_y \cos\varphi\left(\vec{r}\right) + I_x \sin\varphi\left(\vec{r}\right)\right\} \quad \text{(B.13)}$$

After the second 90 degree pulse, the density matrix becomes:

$$\rho\left(\vec{r}, t_{90^0} + t_1 + t_{90^0}\right) =$$

$$a + b\left(\vec{r}\right)\begin{pmatrix} \cos\left(\frac{\alpha}{2}\right) & i\sin\left(\frac{\alpha}{2}\right) \\ i\sin\left(\frac{\alpha}{2}\right) & \cos\left(\frac{\alpha}{2}\right) \end{pmatrix}\begin{pmatrix} 0 & -\frac{i}{2}e^{i\varphi\left(\vec{r}\right)} \\ \frac{i}{2}e^{-i\varphi\left(\vec{r}\right)} & 0 \end{pmatrix}\begin{pmatrix} \cos\left(\frac{\alpha}{2}\right) & -i\sin\left(\frac{\alpha}{2}\right) \\ -i\sin\left(\frac{\alpha}{2}\right) & \cos\left(\frac{\alpha}{2}\right) \end{pmatrix} =$$

$$a + b\left(\vec{r}\right)\begin{pmatrix} -\frac{1}{2}e^{-i\varphi\left(\vec{r}\right)}\sin\left(\frac{\alpha}{2}\right) & -\frac{i}{2}e^{i\varphi\left(\vec{r}\right)}\cos\left(\frac{\alpha}{2}\right) \\ \frac{i}{2}e^{-i\varphi\left(\vec{r}\right)}\cos\left(\frac{\alpha}{2}\right) & \frac{1}{2}e^{i\varphi\left(\vec{r}\right)}\sin\left(\frac{\alpha}{2}\right) \end{pmatrix}\begin{pmatrix} \cos\left(\frac{\alpha}{2}\right) & -i\sin\left(\frac{\alpha}{2}\right) \\ -i\sin\left(\frac{\alpha}{2}\right) & \cos\left(\frac{\alpha}{2}\right) \end{pmatrix} \quad \text{(B.14)}$$

It results into:

$$\rho\left(\vec r, t_{90^0}+t_1+t_{90^0}\right)= a+b(\vec r)\begin{pmatrix} -\frac{1}{4}e^{-i\varphi(\vec r)}-\frac{1}{4}e^{i\varphi(\vec r)} & \frac{i}{2}e^{-i\varphi(\vec r)}-\frac{i}{4}e^{i\varphi(\vec r)} \\ \frac{i}{4}e^{-i\varphi(\vec r)}-\frac{i}{4}e^{i\varphi(\vec r)} & \frac{1}{4}e^{-i\varphi(\vec r)}+\frac{1}{4}e^{i\varphi(\vec r)} \end{pmatrix} \tag{B.15}$$

$$\rho\left(\vec r, t_{90^0}+t_1+t_{90^0}\right)= a+b(\vec r)\begin{pmatrix} -\frac{1}{2}\cos\varphi(\vec r) & \frac{1}{2}\sin\varphi(\vec r) \\ \frac{1}{2}\sin\varphi(\vec r) & \frac{1}{2}\cos\varphi(\vec r) \end{pmatrix}$$

$$= a+b(\vec r)\left\{\begin{pmatrix} -\frac{1}{2} & 0 \\ 0 & \frac{1}{2} \end{pmatrix}\cos\varphi(\vec r)+\begin{pmatrix} 0 & \frac{1}{2} \\ \frac{1}{2} & 0 \end{pmatrix}\sin\varphi(\vec r)\right\} \tag{B.16}$$

As the matrices in brackets represent $-\hat I_z$ and $\hat I_x$ operators, respectively, (B.16) can be presented as:

$$\rho\left(\vec r, t_{90^0}+t_1+t_{90^0}\right)= a+b(\vec r)\left\{-I_z\cos\varphi(\vec r)+I_x\sin\varphi(\vec r)\right\} \tag{B.17}$$

During $t_2$ time spins evolve mainly due to “purge” z gradient, assuming that spin relaxation and dephasing due to inhomogeneity are neglegible. The Hamiltonian of spin interaction with $z$ ”purge” gradient is given by:

$$\hat H_z^{\ purge} = -\gamma\hbar I_z G_z^{\ purge} z \tag{B.18}$$

The phase change due to “purge” gradient with duration $t_z^{\ purge}$ is:

$$\delta = \gamma G_z^{\ purge} z t_z^{\ purge} \tag{B.19}$$

Therefore, the density matrix at $t = t_{90^0}+t_1+t_{90^0}+t_2$ can be given by:

$$\rho\left(\vec r, t_{90^0}+t_1+t_{90^0}+t_2\right)= e^{-\frac{i}{\hbar}H_Z^{\ purge}}\rho\left(\vec r, t_{90^0}+t_1+t_{90^0}\right)e^{\frac{i}{\hbar}H_Z^{\ purge}}$$

(B.20)

It can be rewritten in matrix form:

$$\rho\left(\vec{r}, t_{90^0} + t_1 + t_{90^0} + t_2\right) =$$

$$a + b(\vec{r})\begin{pmatrix} e^{i\frac{\delta}{2}} & 0 \\ 0 & e^{-i\frac{\delta}{2}} \end{pmatrix}\begin{pmatrix} -\frac{1}{2}\cos\varphi(\vec{r}) & \frac{1}{2}\sin\varphi(\vec{r}) \\ \frac{1}{2}\sin\varphi(\vec{r}) & \frac{1}{2}\cos\varphi(\vec{r}) \end{pmatrix}\begin{pmatrix} e^{-i\frac{\delta}{2}} & 0 \\ 0 & e^{i\frac{\delta}{2}} \end{pmatrix} \tag{B.21}$$

After multiplication, the density matrix becomes:

$$\rho\left(\vec{r}, t_{90^0} + t_1 + t_{90^0} + t_2\right) =$$

$$a + b(\vec{r})\begin{pmatrix} -\frac{1}{2}e^{i\frac{\delta}{2}}\cos\varphi(\vec{r}) & \frac{1}{2}e^{i\frac{\delta}{2}}\sin\varphi(\vec{r}) \\ \frac{1}{2}e^{-i\frac{\delta}{2}}\sin\varphi(\vec{r}) & \frac{1}{2}e^{-i\delta}\cos\varphi(\vec{r}) \end{pmatrix}\begin{pmatrix} e^{-i\frac{\varphi(\vec{r})}{2}} & 0 \\ 0 & e^{i\frac{\varphi(\vec{r})}{2}} \end{pmatrix} =$$

$$a + b(\vec{r})\begin{pmatrix} -\frac{1}{2}\cos\varphi(\vec{r}) & \frac{1}{2}e^{i\delta}\sin\varphi(\vec{r}) \\ \frac{1}{2}e^{-i\delta}\sin\varphi(\vec{r}) & \frac{1}{2}\cos\varphi(\vec{r}) \end{pmatrix} \tag{B.22}$$

After decomposition of exponentials, the density matrix simplifies to:

$$\rho\left(\vec{r}, t_{90^0} + t_1 + t_{90^0} + t_2\right) =$$

$$a + b(\vec{r})\left\{\begin{pmatrix} -\frac{1}{2} & 0 \\ 0 & \frac{1}{2} \end{pmatrix}\cos\varphi(\vec{r}) + \begin{pmatrix} 0 & \frac{1}{2}e^{i\delta} \\ \frac{1}{2}e^{-i\delta} & 0 \end{pmatrix}\sin\varphi(\vec{r})\right\} \tag{B.23}$$

The second matrix in brackets includes transverse magnetization oscillating with frequency proportional to strength of “purge” gradient. This dephases in strong “purge” gradient becoming negligible and therefore, can be ignored. As a result, the transverse magnetization is "filtered out", leaving $-z$ component:

$$\rho\left(\vec{r}, t_{90^0} + t_1 + t_{90^0} + t_2\right) =$$

$$a + b(\vec{r})\begin{pmatrix} -\frac{1}{2} & 0 \\ 0 & \frac{1}{2} \end{pmatrix}\cos\varphi(\vec{r}) = a + b(\vec{r})\left(-I_z \cos\varphi(\vec{r})\right) \qquad \text{(B.24)}$$

After the third 90 degree pulse, the density matrix becomes:

$$\rho\left(\vec{r}, t_{90^0} + t_1 + t_{90^0} + t_2 + t_{90^0}\right) =$$

$$\begin{pmatrix} \cos\left(\frac{\alpha}{2}\right) & i\sin\left(\frac{\alpha}{2}\right) \\ i\sin\left(\frac{\alpha}{2}\right) & \cos\left(\frac{\alpha}{2}\right) \end{pmatrix}\begin{pmatrix} -\frac{1}{2}\cos\varphi(\vec{r}) & 0 \\ 0 & \frac{1}{2}\cos\varphi(\vec{r}) \end{pmatrix}\begin{pmatrix} \cos\left(\frac{\alpha}{2}\right) & -i\sin\left(\frac{\alpha}{2}\right) \\ -i\sin\left(\frac{\alpha}{2}\right) & \cos\left(\frac{\alpha}{2}\right) \end{pmatrix} \qquad \text{(B.25)}$$

As $\alpha = 90^0$, the density matrix after multiplication is:

$$\rho\left(\vec{r}, t_{90^0} + t_1 + t_{90^0} + t_2 + t_{90^0}\right) =$$

$$\begin{pmatrix} -\frac{1}{2}\cos\varphi(\vec{r})\cos\left(\frac{\alpha}{2}\right) & \frac{i}{2}\cos\varphi(\vec{r})\sin\left(\frac{\alpha}{2}\right) \\ -\frac{i}{2}\cos\varphi(\vec{r})\sin\left(\frac{\alpha}{2}\right) & \frac{1}{2}\cos\varphi(\vec{r})\cos\left(\frac{\alpha}{2}\right) \end{pmatrix}\begin{pmatrix} \cos\left(\frac{\alpha}{2}\right) & -i\sin\left(\frac{\alpha}{2}\right) \\ -i\sin\left(\frac{\alpha}{2}\right) & \cos\left(\frac{\alpha}{2}\right) \end{pmatrix} =$$

$$\begin{pmatrix} -\frac{1}{4}\cos\varphi(\vec{r}) + \frac{1}{4}\cos\varphi(\vec{r}) & \frac{i}{4}\cos\varphi(\vec{r}) + \frac{i}{4}\cos\varphi(\vec{r}) \\ -\frac{i}{4}\cos\varphi(\vec{r}) - \frac{i}{4}\cos\varphi(\vec{r}) & -\frac{i}{4}\cos\varphi(\vec{r}) + \frac{1}{4}\cos\varphi(\vec{r}) \end{pmatrix}$$

(B.26).

It simplifies to:

$$\rho\left(\vec{r}, t_{90^0} + t_1 + t_{90^0} + t_2 + t_{90^0}\right) = \begin{pmatrix} 0 & \frac{i}{2}\cos\varphi(\vec{r}) \\ -\frac{i}{2}\cos\varphi(\vec{r}) & 0 \end{pmatrix} = -I_y \cos\varphi(\vec{r}) \qquad \text{(B.27).}$$

After the third 90 pulse spin system evolves in inhomogeneous static magnetic field. The Hamiltonian of spin interaction with the static magnetic field inhomogeneity can be given by:

$$H_z^{inh} = -\gamma\hbar I_z \delta\vec{B}(\vec{r}) \cdot \vec{r} \qquad \text{(B.28)}$$

Evolution of the density matrix in inhomogeneous static magnetic field is allowed for time $\tau$ and can be described by:

$$\rho\left(\vec{r}, t_{90^0} + t_1 + t_{90^0} + t_2 + t_{90^0} + \tau\right) = e^{-\frac{i}{\hbar}H_Z^{inh}\tau} \rho\left(\vec{r}, t_{90^0} + t_1 + t_{90^0} + t_2 + t_{90^0}\right) e^{\frac{i}{\hbar}H_Z^{inh}\tau} \qquad \text{(B.29)}$$

It can be presented in matrix form as:

$$\rho\left(\vec{r}, t_{90^0} + t_1 + t_{90^0} + t_2 + t_{90^0} + \tau\right) =$$

$$a + b(\vec{r}) \begin{pmatrix} e^{i\frac{\Omega\tau}{2}} & 0 \\ 0 & e^{-i\frac{\Omega\tau}{2}} \end{pmatrix} \begin{pmatrix} 0 & \frac{i}{2}\cos\varphi(\vec{r}) \\ -\frac{i}{2}\cos\varphi(\vec{r}) & 0 \end{pmatrix} \begin{pmatrix} e^{-i\frac{\Omega\tau}{2}} & 0 \\ 0 & e^{i\frac{\Omega\tau}{2}} \end{pmatrix} \qquad \text{(B.30)}$$

After multiplication, density matrix becomes:

$$\rho\left(\vec{r}, t_{90^0}+t_1+t_{90^0}+t_2+t_{90^0}+\tau\right)=$$

$$a+b(\vec{r})\begin{pmatrix} 0 & \frac{i}{2}\cos\varphi(\vec{r})e^{i\frac{\Omega\tau}{2}} \\ -\frac{i}{2}\cos\varphi(\vec{r})e^{-i\frac{\Omega\tau}{2}} & 0 \end{pmatrix}\begin{pmatrix} e^{-i\frac{\Omega\tau}{2}} & 0 \\ 0 & e^{i\frac{\Omega\tau}{2}} \end{pmatrix}=$$

$$a+b(\vec{r})\begin{pmatrix} 0 & \frac{i}{2}\cos\varphi(\vec{r})e^{i\Omega\tau} \\ -\frac{i}{2}\cos\varphi(\vec{r})e^{-i\Omega\tau} & 0 \end{pmatrix} \tag{B.31}$$

Decomposition of exponentials yields:

$$\rho\left(\vec{r}, t_{90^0}+t_1+t_{90^0}+t_2+t_{90^0}+\tau\right)=$$

$$a+b(\vec{r})\begin{pmatrix} 0 & \begin{matrix}\frac{i}{2}\cos\varphi(\vec{r})\cos\Omega\tau- \\ \frac{1}{2}\cos\varphi(\vec{r})\sin\Omega\tau\end{matrix} \\ \begin{matrix}-\frac{i}{2}\cos\varphi(\vec{r})\cos\Omega\tau- \\ \frac{1}{2}\cos\varphi(\vec{r})\sin\Omega\tau\end{matrix} & 0 \end{pmatrix} \tag{B.32}$$

This can be rewritten as a sum of two matrices:

$$\rho\left(\vec{r}, t_{90^0}+t_1+t_{90^0}+t_2+t_{90^0}+\tau\right)=$$

$$=a+b(\vec{r})\left\{\begin{pmatrix} 0 & \frac{i}{2} \\ -\frac{i}{2} & 0 \end{pmatrix}\cos\varphi(\vec{r})\cos\Omega\tau+\begin{pmatrix} 0 & -\frac{1}{2} \\ -\frac{1}{2} & 0 \end{pmatrix}\cos\varphi(\vec{r})\sin\Omega\tau\right\} \tag{B.33}$$

As matrices in brackets respectively represent $-I_y$ and $-I_x$ operators, this can be finally written as:

$$\rho\left(\vec{r}, t_{90^0} + t_1 + t_{90^0} + t_2 + t_{90^0} + \tau\right) =$$

$$a + b(\vec{r})\left(-I_y \cos\varphi(\vec{r})\cos\Omega\tau - I_x \cos\varphi(\vec{r})\sin\Omega\tau\right) \qquad \text{(B.34)}$$

The interactions of spins with magnetic field, produced by read-out $z$ gradient is given by:

$$H_z^{\ rg} = -\gamma\hbar I_z G_z^{\ rg} z \qquad \text{(B.35)}$$

Application of the read-out gradient during time $t_z^{\ rg}$ leads to evolution of the density matrix, described by equation:

$$\rho\left(\vec{r}, t_{90^0} + t_1 + t_{90^0} + t_2 + t_{90^0} + \tau + t_z^{\ rg}\right) =$$

$$e^{-\frac{i}{\hbar}H_Z^{\ rg} t_Z^{\ rg}} \rho\left(\vec{r}, t_{90^0} + t_1 + t_{90^0} + t_2 + t_{90^0} + \tau\right) e^{\frac{i}{\hbar}H_Z^{\ rg} t_Z^{\ rg}} \qquad \text{(B.36)}$$

It can be rewritten in matrix form as:

$$\rho\left(\vec{r}, t_{90^0} + t_1 + t_{90^0} + t_2 + t_{90^0} + \tau + t_Z^{\ rg}\right) =$$

$$a + b(\vec{r})\begin{pmatrix} e^{i\frac{\varphi_Z}{2}} & 0 \\ 0 & e^{-i\frac{\varphi_Z}{2}} \end{pmatrix}\begin{pmatrix} 0 & \frac{i}{2}\cos\varphi(\vec{r})e^{i\Omega\tau} \\ -\frac{i}{2}\cos\varphi(\vec{r})e^{-i\Omega\tau} & 0 \end{pmatrix}\begin{pmatrix} e^{-i\frac{\varphi_Z}{2}} & 0 \\ 0 & e^{i\frac{\varphi_Z}{2}} \end{pmatrix} \qquad \text{(B.37)}$$

After multiplication, density matrix becomes:

$$\rho\left(\vec{r}, t_{90^0} + t_1 + t_{90^0} + t_2 + t_{90^0} + \tau + t_Z^{\ rg}\right) =$$

$$= a + b(\vec{r})\begin{pmatrix} 0 & \frac{i}{2}\cos\varphi(\vec{r})e^{i\left(\Omega\tau + \frac{\varphi_Z}{2}\right)} \\ -\frac{i}{2}\cos\varphi(\vec{r})e^{-i\left(\Omega\tau + \frac{\varphi_Z}{2}\right)} & 0 \end{pmatrix}\begin{pmatrix} e^{-i\frac{\varphi_Z}{2}} & 0 \\ 0 & e^{i\frac{\varphi_Z}{2}} \end{pmatrix} =$$

$$a + b(\vec{r})\begin{pmatrix} 0 & \frac{i}{2}\cos\varphi(\vec{r})e^{i(\Omega\tau + \varphi_Z)} \\ -\frac{i}{2}\cos\varphi(\vec{r})e^{-i(\Omega\tau + \varphi_Z)} & 0 \end{pmatrix} \qquad \text{(B.38)}$$

After decomposition of exponentials, density matrix becomes:

$$\rho\left(\vec{r}, t_{90^0} + t_1 + t_{90^0} + t_2 + t_{90^0} + \tau + t_z^{rg}\right) =$$

$$a + b(\vec{r})\begin{pmatrix} 0 & \frac{i}{2}\cos\varphi(\vec{r})\cos(\Omega\tau + \varphi_z) - \frac{1}{2}\cos\varphi(\vec{r})\sin(\Omega\tau + \varphi_z) \\ -\frac{i}{2}\cos\varphi(\vec{r})\cos(\Omega\tau + \varphi_z) - \frac{1}{2}\cos\varphi(\vec{r})\sin(\Omega\tau + \varphi_z) & 0 \end{pmatrix}$$

(B.39)

The matrix in brackets can be expressed as a sum of two matrices:

$$\rho\left(\vec{r}, t_{90^0} + t_1 + t_{90^0} + t_2 + t_{90^0} + \tau + t_z^{rg}\right) =$$

$$a + b(\vec{r})\left\{\begin{pmatrix} 0 & \frac{i}{2} \\ -\frac{i}{2} & 0 \end{pmatrix}\cos\varphi(\vec{r})\cos(\Omega\tau + \varphi_z) + \begin{pmatrix} 0 & -\frac{1}{2} \\ -\frac{1}{2} & 0 \end{pmatrix}\cos\varphi(\vec{r})\sin(\Omega\tau + \varphi_z)\right\} \quad (B.40)$$

This simplifies to:

$$\rho\left(\vec{r}, t_{90^0} + t_1 + t_{90^0} + t_2 + t_{90^0} + \tau + t_z^{rg}\right) =$$

$$a + b(\vec{r})\left\{-I_y\cos\varphi(\vec{r})\cos(\Omega\tau + \varphi_z) - I_x\cos\varphi(\vec{r})\cos(\Omega\tau + \varphi_z)\right\} \quad (B.41)$$

The NMR signal is a complex function, given by:

$$M(t) = M_x(t) + iM_y(t) \quad (B.42)$$

where

$$M_x(t) = Tr\left(\hat{\rho}(t)\hat{I}_x\right) \quad (B.43)$$

$$M_y(t) = Tr\left(\hat{\rho}(t)\hat{I}_y\right) \quad (B.44)$$

It is convenient, for a while, to write density matrix (C.40) in a simple form:

$$\rho(t) = \begin{pmatrix} 0 & \rho_{12} \\ \rho_{21} & 0 \end{pmatrix} \quad (B.45)$$

where $\rho_{12}$ and $\rho_{21}$ are the elements of density matrix, given by (C.40).

Hence,

$$M_x(t) = Tr\left\{\begin{pmatrix} 0 & \rho_{12} \\ \rho_{21} & 0 \end{pmatrix}\begin{pmatrix} 0 & \frac{1}{2} \\ \frac{1}{2} & 0 \end{pmatrix} + i\begin{pmatrix} 0 & \rho_{12} \\ \rho_{21} & 0 \end{pmatrix}\begin{pmatrix} 0 & -\frac{i}{2} \\ \frac{i}{2} & 0 \end{pmatrix}\right\} \tag{B.46}$$

After multiplication, it becomes:

$$M_x(t) = Tr\left\{\begin{pmatrix} \frac{1}{2}\rho_{12} & 0 \\ 0 & \frac{1}{2}\rho_{21} \end{pmatrix} + i\begin{pmatrix} \frac{i}{2}\rho_{12} & 0 \\ 0 & -\frac{i}{2}\rho_{21} \end{pmatrix}\right\} =$$

(B.47)

$$\frac{1}{2}(\rho_{12} + \rho_{21}) - \frac{1}{2}(\rho_{12} - \rho_{21}) = \rho_{21}$$

After substitution of matrix element from (C.40), NMR signal becomes:

$$M(t) = -\frac{i}{2}\cos\varphi(\vec{r})e^{i(\Omega\tau + \varphi_z)} \tag{B.48}$$

# Appendix C

## The gxysweep and gxycal macros

The data acquired with the use of macro gxysweep are processed by macro gxycal, which calibrates the strengths of linear *x*- and *y*- gradients by application of fitting programme **calibxy.c**, presented in Appendix F.

```
/* Macro gxysweep for setting up the x- and y-gradient strength calibration*/

/* G.A. Morris and V.V. Korostelev, March 2002 */
/****************************************************************/

$x1init=x1
$y1init=y1
$gxystep=100

if ($#>0) then $n=$1 else $n=16 endif
if ($#>1) then $gxystep=$2 endif

exists('gxystep','parameter','global'):$e

if $e=0 then
	create('gxystep','real','global')
	gxystep=$gxystep
endif
	write('line3','parameter gxystep set to %d', $gxystep)

if (x1+gxystep)>2047 then
	write('error','gxystep is too large; x1 will go out of range') abort
endif

if (x1-gxystep)<-2047 then
	write('error','gxystep is too large; x1 will go out of range') abort
endif
```

```
if (y1+gxystep)>2047 then
        write('error','gxystep is too large; y1 will go out of range') abort
endif

if (y1-gxystep)<-2047 then
        write('error','gxystep is too large; y1 will go out of range') abort
endif

/* Setting up an array for x1 and y1 shims*/

$i=1
$th=0.0
repeat
          x1[$i]=$x1init+gxystep*cos($th)
          y1[$i]=$y1init+gxystep*sin($th)
          $th=$th+3.141592654*2.0/$n
          $i=$i+1
until $i>$n

x1[$n+1]=$x1init
y1[$n+1]=$y1init

array='(x1, y1)'

load='y'
da

/* the end of gxysweep macro*/

/***********Macro gxycal************* */

/*  Step1: Measurement of the profile widths */

wft
dssh(1, arraydim)

/* set magnetogyric ratio in radian/(s*T)*/

if tn='H1'    then  $gamma=267522128.00 endif
if tn='H2'    then  $gamma=  41066279.10 endif
if tn='lk'    then  $gamma=  41066279.10 endif
if tn='none' then  $gamma=  41066279.10 endif
if tn='F19'  then  $gamma=251814800.00 endif
if ((tn<>'H1') and (tn<>'H2') and (tn<>'none') and (tn<>'F19') and (tn<>'lk'))
   write('line3','Nucleus not recognised')
   abort
endif

if $#>1 then
  write('error','Usage: gxycal or gxycal(threshold)')
```

```
  abort
endif

if $#=1 then
 $thr=$1
else
 $thr=0.05
endif

if arraydim<2 then
  write(‘error’,’gxycal expects an arrayed experiment’)
  abort
endif

exists(‘tubeid’,’parameter’,’global’):$tid
if $tid=0 then
  create(‘tubeid’,’real’,’global’)
  tubeid=4.2
  write(‘error’,’Tube inside diameter assumed to be 4.2 mm’)
endif

exists(‘gcalx’,’parameter’,’current’):$e1
if $e1=1 then destroy(‘gcalx’,’parameter’,’current’) endif

exists(‘gcaly’,’parameter’,’current’):$e2
if $e2=1 then destroy(‘gcaly’,’parameter’,’current’) endif

exists(‘gcalang’,’parameter’,’current’):$e3
if $e3=1 then destroy(‘gcalang’,’parameter’,’current’) endif

exists(‘gcalx’,’parameter’,’global’):$e4
if $e4=0 then create (‘gcalx’,’real,’global’) endif

exists(‘gcaly’,’parameter’,’global’):$e5
if $e5=0 then create (‘gcaly’,’real,’global’) endif

exists(‘gcalratio’,’parameter’,’global’):$e6
if $e6=0 then create (‘gcalratio’,’real,’global’) endif

exists(‘gcalang’,’parameter’,’global’):$e7
if $e7=0 then create (‘gcalang’,’real,’global’) endif

exists(‘gcalxerr’,’parameter’,’global’):$e8
if $e8=0 then create (‘gcalxerr’,’real,’global’) endif

exists(‘gcalyerr’,’parameter’,’global’):$e9
if $e9=0 then create (‘gcalyerr’,’real,’global’) endif

$fn=curexp+’/xydata’
write(‘reset’, $fn)
```

```
write('file', $fn, arraydim-1)
refpos='n' crl f

$I=1
repeat
echo($i)
        select($i)
        peak:r1,cr
        dres:$hw
        cz lifrq=cr+$hw*1.5, cr-$hw*1.5,0
        bc
        peak:$int,cr
        dres(cr,$thr):$w
        write('file',$fn,'%5d\t%5d\t%10.5f',x1[$i],y1[$i],$w)
        $i=$i+1
until ($i>arraydim-1)
clear

/* Step2: Fitting of the parameters gcalx, gcaly, gcalang, gcalxerr, gcalyerr */

calibxy: $wx, $wy, gcalang, gcalxerr, gcalyerr

/* Conversion of the fitted parameters into G/cm units */

gcalx=2*3.141592654*$wx*100000.0/($gamma*tubeid)
gcaly=2*3.141592654*$wy*100000.0/($gamma*tubeid)

full autoscale expl('regression','link')

/* the end of gxycal macro*/
```

# Appendix D

## The calibxy.c programme

The calibxy.c source code listed below was compiled as a standalone programme and is used for fitting of the parameters of *x*- and *y*- gradient strength calibration.

```
/* calibxy.c – G.A. Morris and V.V. Korostelev, University of Manchester, 2002 */

/**********************************************************************
The following functions have been obtained from 'Numerical Recipes in C', 2nd Edition.
mrqmin(), mrqcof(), covsrt(),gaussj(), nrerror(). Vladimir Korostelev – April 2002
**********************************************************************/

/* Declaration of the header files, definitions and global functions*/
#include  <stdio.h>
#include <math.h>
#include <stdlib.h>
#include <stddef.h>
#define MA 5
#define NR_END 1
#define ERROR 1
#define MAXLENGTH 250
#define Pi 3.141592654
#define twoPi 6.283185308
#define FREE_ARG char*
#define SWAP(a,b) {swap=(a); (a)=(b); (b)=swap;}

extern char curexpdir[];

static double root=0;
static double a[MA+1]={0,0,0,0,0,0}
static double gues[MA+1]={0,0.1,1.0,0.1,0,0}

static void nrerror(char error_text[]);
static void mrqmin(double x[], double y[], double sig[], int ndata, double a[],
        int ia[], int ma, double **covar, double **alpha, double *chisq,
```

```
            void (*funcs)(double, double[], double*, double[], int), double *alambda);
static void free_vector(double *v, long nl, long nh);
static double *vector(long nl, long nh);
static int *ivector(long nl, long nh);
static void free_ivector(int *v, long nl, long nh);
static double **matrix(long nrl, long nrh, long ncl, long nch);
static void free_matrix(double **m, long nrl, long nrh, long ncl, long nch);

extern char *realString();


/***************The main function***************/

int calibxy(int argc, char *argv[], int retc, char *retv[])
{
   FILE     *in, *out, *outreg;
   int        i, *ia, iter, itst, j, k, mfit=MA, np, npt;
   double   alambda, chisq, ochisq, *x, *y, *sig, **covar, **alpha,
              *dummydada, cwidth;
   char  fname[MAXLENGTH];
   char  in_file[MAXLENGTH]="/export/home/vnmr1/vnmrsys/3Dshimlib/xywidths";
   char  out_file[MAXLENGTH]="/export/home/vnmr1/vnmrsys/3Dshimlib/output";

     if ((in=fopen(in_file,"r"))==NULL)
     {
       Werrprintf("Error opening input file.\n");
        return (ERROR);
      }
      if ((out=fopen(out_file,"w"))==NULL)
     {
       Werrprintf("Error opening output file.\n");
        return (ERROR);
      }

strcpy(fname,curexpdir);
strcat(fname,"/regression.inp");

     if ((out=fopen(out_file,"w"))==NULL)
     {
       Werrprintf("Error opening input file.\n");
        return (ERROR);
      }
      if ((outreg=fopen(fname,"w"))==NULL)
      {
        Werrprintf("Error opening output file.\n");
         return(Error);
       }
        fprintf(out,"Output from calibxy\n");
        fscanf(in,"%d \n", &npt);
```

```
if ((npt>257)||(npt<2))
{
  Werrprintf(“Number of points %d is unreasonable\n”,npt);
   abort();
 }
 fprintf(outreg,”Output from fitXYprofiles\n”);
 fprintf(outreg,”profile widths in Hz\n”);
 fprintf(outreg,”1\t%d\n”,npt);
 fprintf(outreg,”NEXT\n”);

  ia=ivector(1,MA);
  dummydyda=vector(1,MA);
  x=vector(1,npt);
  y=vector(1,npt);
  sig=vector(1,npt);
  covar=matrix(1,MA,1,MA);
  alpha=matrix(1,MA,1,MA);
  i=1;

 while((fscanf(in,”%lf \n”, &y[i]))!=EOF)
        { i++; }
 gues[1]=y[1];

 for ( i=1;i<=MA;i++)
     { a[i]= gues[i]; }

 for (i=1; i<npt+1;i++)
    {
      x[i]=i*twoPi/npt);
      sig[i]=1.0;
     }
 modelfunc(x[1],a,&cwidth,dummydyda,MA);

  for(i=1;i<=mfit,i++)  ia[i]=1;
  for(i=1;i<=MA,i++)  a[i]=gues[i];

  alambda=-1;
  mrqmin(x,y,sig,npt,a,ia,MA,covar,alpha,&chisq,modelfunc,&alambda);

  k=1;
  itst=0;
  for(;;) { k++;
             ochisq=chisq;
             mrqmin(x,y,sig,npt,a,ia,MA,covar,alpha,&chisq,
                        modelfunc,&alambda);
              if (chisq>ochisq)
              itst=0;
              else if (fabs(ochisq-chisq)<0.1)
                       itst++;
              if(itst<4)
```

```
                    continue;
                    alambda=0.0;
                    mrqmin(x,y,sig,npt,a,ia,MA,covar,alpha,&chisq,
                              modelfunc,&alambda);
                    break;
                   }

/********************Calculation of line widths*********************/
   fprintf(out, “Calculated fitting parameters: \n”);
   fprintf(out,”%f\t%f\t%f\t%f\t%f\n”, a[1], a[2], a[3], a[4], a[5]);
   fprintf(out,”Rotation angle, degrees: \t Profile width, Hz: \n”);

   for (i=1; i<=npt; i++)
      {
        modelfunc(x[i], a, &cwidth, dummydyda, MA);
        fprintf(out,”%f\t\t%f\n”,x[i]*180.0/Pi,cwidth);
       }

   fprintf(outreg,”NEXT\n”);

   for (i=1; i<=npt; i++)
      {
        fprintf(outreg,”%f\t%f\n”,x[i]*180.0/Pi,y[i]);
      }

    if (retc>0)
      {
        retv[0]=realString((double) a[1]);
      }
    if (retc>1)
      {
        retv[1]=realString((double) a[2]);
       }
    if (retc>2)
      {
        retv[2]=realString((double) a[3]);
      }
    if (retc>3)
      {
        retv[3]=realString((double) a[4]);
      }
    if (retc>4)
      {
        retv[4]=realString((double) a[5]);
      }
        free_matrix(alpha,1,MA,1,MA);
        free_matrix(covar,1,MA,1,MA);
        free_vector(sig,1,npt);
        free_vector(x,1,npt);
        free_vector(y,1,npt);
```

```
        free_vector(dummydyda,1,npt);
        free_ivector(ia,1,MA);
        fclose(out);
        fclose(outreg);
        return 0;
}

 /******************End of main function******************/

/*Functions for dynamic allocation and release of memory for types:
                    *vector, *ivector, **matrix
****************************************************/
   static double *vector(long nl, long nh)
         {
           double *v;
           v=(double *)malloc((size_t)((nh-nl+1+NR_END)sizeof(double)))

         if (!v)
           {
             printf("allocation failure in vector() of indices %d and %d\n",nl,nh);
             (void) getchar();
             nrerror("allocation failure in vector()");
           }
             return v-nl+NR_END;
         }

  static int *ivector(long nl, long nh)
        {
          int *v;
          v=(int *)malloc((size_t)((nh-nl+1+NR_END)sizeof(int)))
          if (!v)
            {
              printf("allocation failure in vector() of indices %d and %d\n",nl,nh);
              (void) getchar();
              nrerror("allocation failure in ivector()");
             }
          return v-nl+NR_END;
        }

   static void free_vector(double *v, long nl, long nh)
         {
           free((FREE_ARG) (v+nl-NR_END));
         }

    static void free_ivector(int *v, long nl, long nh)
          {
            free((FREE_ARG) (v+nl-NR_END));
           }

    static double **matrix(long nrl, long nrh, long ncl, long nch)
```

```
          {
            long i, nrow=nrh-nrl+1, ncol=nch-ncl+1;
            double **m;
            m=(double **) malloc((size_t)((nrow+NR_END)*sizeof(double*)));
            if (!m) nrerror("allocation failure 1 in matrix()");
            m+=NR_END;
            m-=nrl;
            m[nrl]=(double *) malloc((size_t)((nrow*ncol+NR_END)*sizeof(double)));
            if (!m[nrl]) nrerror("allocation failure 2 in matrix()");
            m[nrl] +=NR_END;
            m[nrl] -=ncl;

            for (i=nrl+1;i<nrh;i++)
                m[i]=m[i-1]+ncol;

            return m;
          }

      static void free_matrix(double **m, long nrl, long nrh, long ncl, long nch)
            {
              free((FREE_ARG) (m[nrl]+ncl-NR_END));
              free((FREE_ARG) (m+nrl-NR_END));
            }

/* Function for Marquardt-Levenberg algorithm of nonlinear least squares fitting*/
/* has been obtained from 'Numerical Recipes in C', 2nd Edition                  */ /*Vladimir
Korostelev – April 2002*/

static void mrqmin(double x[], double y[], double sig[], int ndata, double a[],
             int ia[], int ma, double **covar, double **alpha, double *chisq,
             void (*funcs)(double, double[], double *, double [], int),
             double *alambda)
   {
      void covsrt (double **covar, int ma, int ia[], int mfit);
      void gaussj (double **a, int n, double **b, int m);

      void mrqcof(double x[], double y[], double sig[], int ndata, double a[],
            int ia[], int ma, double **alpha, double beta[], double *chisq,
            void (*funcs)(double, double [], double *, double [], int));
      int j,k,l,m;
      static int mfit;
      static double ochisq, *arty, *beta, *da, **oneda;

      if (*alambda<0.0)
        {
          atry=vector(1,ma);
          beta=vector(1,ma);
          da=vector(1,ma);

          for (mfit=0, j=1;j<=ma;j++)
```

```
      if (ia[j]) mfit++;

     oneda=matrix(1,mfit,1,1);
     *alambda=0.001;
     mrqcof(x,y,sig,ndata,a,ia,ma,alpha,beta,chisq,funcs);
     ochisq=(*chisq);

   for (j=1;j<=ma;j++)
        arty[j]=a[j];
  }

for j=0, l=1; l<=ma;l++)
   {
     if (ia[l])
       {
         for (j++, k=0, m=1; m<=ma; m++)
             {
               if (ia[m])
                 { k++;
                   covar [j][k]=alpha[j][k];
                  }
              }
             covar[j][j]=alpha[j][j]*(1.0+(*alambda));
             oneda[j][1]=beta[j];
       }
    }

   gaussj(covar, mfit, oneda, 1);

   for (j=1; j<=mfit; j++)
     {
       da[j]=oneda[j][1];
      }
       if (*alambda == 0.0)
         {
           covsrt(covar, ma, ia, mfit);
           free_matrix(oneda, 1, mfit, 1, 1);
           free_vector(da, 1, ma);
           free_vector(beta, 1, ma);
           free_vector(arty, 1, ma);
           return;
         }

   for (j=0, l=1; l<=ma; l++)
     {
       if (ia[l]) arty[l]=a[l]+da[++j];
      }

   mrqcof (x, y, sig, ndata, atry, ia, ma, covar,  da, chisq, funcs);
```

```
            if (*chisq<ochisq)
              {
                *alambda *=0.1;
                ochisq=(*chisq);

                for (j=0, l=1; l<=ma; l++)
                   {
                     if (ia[l])
                       {
                         for (j++, k=0, m=1; m<=ma; m++)
                            {
                              if (ia[m])
                                { k++;
                                  alpha[j][k]=covar[j][k];
                                 }
                            }

                         beta[j]=da[j];
                         a[l]=arty[l];
                       }
                     }
                 }
               else
                 { *alambda *=10.0;
                   *chisq=ochisq;
                  }
        }

static void mrqcof(double x[], double y[], double sig[], int ndata, double a[], int ia[],
                   int ma, double **alpha, double beta[], double *chisq,
                   void (*funcs) (double, double[], double *, double [], int))
{
   int I, j, k, l, m, mfit=0;
   double ymod, wt, sig2i, dy, *dyda;

   dyda=vector(1, ma);

   for (j=1; j<+ma; j++)
      {
        if (ia[j])
          {
            mfit++;
           }
       }
   for (j=1; j<=mfit; j++)
     { for (k=1; k,=j; k++)
           {
            alpha[j][k]=0.0;
            beta[j]=0.0;
            }
```

```
        *chisq=0.0;
         for (i=1; i<=ndata; i++)
            {
              (*funcs)(x[i], a, &ymod, dyda, ma);
              sig2i=1.0/(sig[i]*sig[i]);
              dy=y[i]-ymod;

              for (j=0, l=1; l<=ma; l++)
                 {
                  if (ia[l])
                    { wt=dyda[l]*sig2i;

                       for (j++, k=0, m=1; m<=l; m++)
                           { if (ia[m])
                                alpha[j][++k] +=wt*dyda[m];
                                beta[j] +=dy*wt;
                           }
                     }
                   }
                for (j=2; j<=mfit; j++)
                   {
                    for (k=1; k<j; k++)
                        {
                         alpha[k][j]=alpha[j][k];
                         }
                    free_vector(dyda, 1, ma);
}

static void covstr(double **covar, int ma, int ia[], int mfit)
{
   int i, j , k;
  double swap;

  for (i=mfit+1; i<=ma; i++)
    for (j=1; j<=i; j++)
        {
          covar[I][j]=covar[j][I]=0.0;
         }

          k=mfit;
    for (j=ma; j>=1; j--)
        {
          if (ia[j])
            {
              for (i=1; i<=ma; i++) swap(covar[i][k], covar[i][j])
              for (i=1; i<=ma; i++) swap(covar[k][i], covar[j][i])
              k--;
            }
         }
  }
```

```
static void gaussj(double **a, int n, double **b, int m)
{
   int *indxc, *indxr, *ipiv;
   int i, icol, irow, j, k, l, ll;
   double big, dum, pivinv, temp;
   dowble swap;

   indxc=ivector(1,n);
   indxr=ivector(1,n);
   ipiv=ivector(1,n);

   for (j=1; j<=n; j++)
      { ipiv[j]=0; }



   for (i=1; i<=n; i++)
      {
        big=0.0;

        for (j=1; j<=n; j++)
           if (ipiv[j] !=1)
              for (k=1; k<=n; k++)
                 {
                  if ( ipiv[k] ==0)
                    {
                      if (fabs(a[j][k])>=big)
                        {
                         big=fabs(a[j][k];
                         irow=j;
                         icol=k;
                        }
                    }
                    else if (ipiv[k] > 1)
                           {
                             nrerror("gaussj: Singular Matrix-1");
                           }
                 }
                 ++(ipiv[icol]);
                 if (irow !=icol)
                    {
                      for (l=1; l<=n; l++) SWAP(a[irow][l], a[icol][l])
                      for (l=1; l<=m; l++) SWAP(b[irow][l], b[icol][l])
                    }
                   indxr[i] = irow;
                   indxc[i] = icol;

                   if (a[icol][icol] == 0.0)
                      nrerror("gaussj: Singular Matrix-2");
```

```
                pivinv=1.0/a[icol][icol];
                a[icol][icol]=1.0;

                for (l=1; l<=n; l++)
                    a[icol][l] *=pivinv;

                for (l=1; l<=m; l++)
                    b[icol][l] *=pivinv;

                for (ll=1; ll<=n; ll++)
                    if (ll != icol)
                      {
                        dum=a[ll][icol];
                        a[ll][icol]=0.0;


                        for (l=1; l<=n; l++)
                            a[ll][l] -=a[icol][l]*dum;

                        for (l=1; l<=m; l++)
                            b[ll][l] -=b[icol][l]*dum;
                     }
          }

          for (l=n; l>=1; l--)
             {
               if (indxr[l] != indxc[l])
               for (k=1; k<=n; k++)
                 SWAP(a[k][indxr[l]], a[k][indxc[l]]);
              }
              free_ivector(ipiv, 1, n);
              free_ivector(indxr, 1, n);
              free_ivector(indxc, 1, n);
}

void nrerror(char error_text[])
{
  fprintf(stderr, "Numerical Recipes run-time error…\n");
  fprintf(stderr, "%s\n", error_text);
  fprintf(stderr, "…now exiting to system…\n");
  exit(1);
}

/* Model function used for fitting as described in Chapter 8*/
static void modelfunc(double x, double a[], double *y, double dyda[], int na)
{
  double ginc=100;
  root=sqrt(a[2]*a[2]*cos(a[3])*cos(a[3])*(a[5]+ginc*sin(x))*(a[5]+ginc*sin(x))+
         (a[4]+ginc*cos(x)+a[2]*sin(a[3])*(a[5]+ginc*sin(x)))*(a[4]+ginc*cos(x)+
```

```
          a[2]*sin(a[3])*(a[5]+ginc*sin(x))));
  *y=a[1]*root;
  dyda[1]=root;
  dyda[2]=(a[1]*(2*a[2]*cos(a[3])*cos(a[3])*(a[5]+ginc*sin(x))*(a[5]+
              ginc*sin(x))+2*sin(a[3])*(a[5]+ginc*sin(x))*(a[4]+ginc*cos(x)+
              a[2]*sin(a[3])*(a[5]+ginc*sin(x)))))/(2*root);
  dyda[3]=(a[1]*(-2*a[2]*a[2]*cos(a[3])*sin(a[3])*(a[5]+ginc*sin(x))*(a[5]+
              ginc*sin(x))+2*a[2]*cos(a[3])*(a[5]+ginc*sin(x))*(a[4]+
              ginc*cos(x)+a[2]*sin(a[3])*(a[5]+ginc*sin(x)))))/(2*root);
  dyda[4]=(a[1]*(a[4]+ginc*cos(x)+a[2]*sin(a[3])*(a[5]+ginc*sin(x))))/root;
  dyda[5]=(a[1]*(2*a[2]*a[2]*cos(a[3])*cos(a[3])*(a[5]+ginc*sin(x))+
             2*a[2]*sin(a[3])*(a[4]+ginc*cos(x)+a[2]*sin(a[3])*(a[5]+
             ginc*sin(x)))))/(2*root);
}
```

# Appendix E

## The gmapxyz.c pulse sequence

This is C programme for PFGSTE pulse sequence used in the field and shim mapping experiments. G.A. Morris and V.V. Korostelev , 2003.

Three pulse PFGSTE sequence for 3D gradient shimming uses x1 and y1 shim gradients for transverse linear gradient pulses and homospoil gradient pulses.

Local parameters:

| | |
|---|---|
| fov | transverse field of view in mm |
| gt1 | delay for xy gradient settling, 200ms |
| gt2 | delay between x and y switching, 10ms |
| gt3 | purge gradient pulse during del interval |
| d4 | dephasing delay before acquisition |
| tau | arrayed delay for field mapping |
| hspcorr | correction for homospoil rise time |
| sp1flag | flag, whose settings ‘y’ or ‘n’, allow switching of lock relay |

Global parameters:

| | |
|---|---|
| gcalx | x1 shim G/cm per DAC point |
| gcaly | y1 shim G/cm per DAC point |
| gcalang | error in degrees in y1 shim angle |

```
#include ,standard.h>
#include “acodes.h”

#define GLOBAL  0

static int ph1[16]={0,0,0,0,1,1,1,1,2,2,2,2,3,3,3,3};
static int ph2[4]={0,1,2,3};
static int ph3[16]={0,1,2,3,3,0,1,2,2,3,0,1,1,2,3,0};
```

```
pulsesequence()
{
     int index1, index2, ni, ni2;
     double gt1, gt2, gt3, tau, d5, del, range, hspcorr, shimset;
     double ginc, x1m, y1m, gamma;
     double gxlvl, gylvl, fov, gcalx, gcaly, gcalang;
     char sp1flag[MAXSTR], tn[MAXSTR];
     getstr(“sp1flag”, sp1flag);
     getstr(“tn”, tn);

if (P_getreal(GLOBAL, “shimset”, &shimset, 1)<0)
    printf(“shimset global parameter not found\n”);

if (P_getreal(GLOBAL, “gcalx”, &gcalx, 1)<0)
    printf(“gcalx global parameter not found\n”);

if (P_getreal(GLOBAL, “gcaly”, &gcaly, 1)<0)
    printf(“gcaly global parameter not found\n”);

if (P_getreal(GLOBAL, “gcalang”, &gcalang, 1)<0)
    printf(“gcalang global parameter not found\n”);

/*Load variables*/
            ni=getval(“ni”);
            ni2=getval(“ni2”);
            gt1=getval(“gt1”);
            gt2=getval(“gt2”);
            gt3=getval(“gt3”);
            del=getval(“del”);
            d4=getval(“d4”);
            tau=getval(“tau”);
            hspcorr=getval(“hspcorr”);
            x1m=getval(“x1”);
            y1m=getval(“y1”);
            fov=getval(“fov”);

            range=32767.0;
            if ((shimset<3.0) || ((shimset>9.5) && (shimset<11.5))) range = 2047;

            index1=(int)(d2*getval(“sw1”)+0.5);
            index2=(int)(d3*getval(“sw2”)+0.5);
            d5=d4+at/2;

/*Initialise value of magnetogyric ratio*/
/*  H1 mapping is assumed unless tn=’H2’ or ‘lk’ or ‘F19’ */
            gamma=267522128.0;
            if (tn[0] ==’F’) gamma=251814800.0;
            if (tn[0] ==’l’)  gamma=41066279.0;
            if (tn[1] ==’2’)  gamma=41066279.0;
```

```
        settable(t1, 16, ph1)
        settable(t2, 4, ph2)
        settable(t3, 16, ph3)

        ginc=200000.0*3.1416/(gamma*d5*fov); /* unit: G/cm */
        gxlvl=((index1-((ni-1)/2.0))*ginc/gcalx)-
              (index2-((ni2-1)/2.0))*ginc*tan(gcalang*3.14159/180.0)/gcalx;
        gylvl=(index2-((ni2-1)/2.0))*ginc/(gcaly*cos(gcalang*3.14159/180.0));

        if ((ni==0)||(ni==1))
        {
          gxlvl=0;
          gylvl=0;
         }

        if (((x1m+gxlvl)>range)||((x1m+gxlvl)<(-range)))
          {
            printf(" x shim is taken out of valid range\n");
            abort(1);
          }

/* Start of pulse sequence */

status(A);
rcvroff();
/*switch lock relay*/
if (sp1flag[0]=='y') sp1on();

/*delay before start of gradients and rf excitation*?
delay(d1-gt1);

/* x- and y-gradient pulses are set on*/
rgradient('x', gxlvl);
delay(gt2);
rgradient('y', gylvl);
delay(gt1-2.0*gt2);
   putcode(IHOMOSPOIL);
   putcode(TRUE);
delay(gt2);

/*first rf pulse*/
rgpulse(pw, t1, 0.00001, 0.00001);

/*delay between ferst and second rf pulses*/
delay(d5);

/*second rf pulse*/
rgpulse(pw, t2, 0.00001, 0.00001);
```

```
/*z-gradient pulse is set off*/
   putcode(IHOMOSPOIL);
   putcode(FALSE);
delay(gt2);

/*x- and y- gradient pulses are set off*/
rgradient('x',0.0);
delay(gt2);
rgradient('y',0.0);




status(B);

/*set purge gradient on*/
   putcode(IHOMOSPOIL);
   putcode(TRUE);
delay(gt3);

/*set purge gradient off*/
   putcode(IHOMOSPOIL);
   putcode(FALSE);
delay(del-2.0*gt2-gt3);

/*the third rf pulse*/
rgpulse(pw, zero, 0.00001, 0.00001);

setreceiver(t3);
rcvron();

/*tau interval is on*/
delay(tau);

/*set read-out gradient on*/
   putcode(IHOMOSPOIL);
   putcode(TRUE);
delay(d4+hspcorr);

/*signal acquisition*/
acquire(np,1.0/sw);

/*set read-out gradient off*/
   putcode(IHOMOSPOIL);
   putcode(FALSE);

/*switch lock relay*/
if (sp1flag[0]=='y') sp1off();
}
```